\pdfoutput=1
\documentclass[11pt]{article}

\usepackage{amsmath,amssymb,amsthm}
\usepackage{booktabs,longtable,array}
\allowdisplaybreaks
\usepackage{threeparttable}
\usepackage{caption}
\usepackage{booktabs}
\usepackage{multicol}
\usepackage{multirow}
\usepackage{graphicx}
\usepackage{float}
\usepackage{setspace}
\usepackage{geometry}
\usepackage{appendix}
\usepackage{natbib}
\newtheorem{theorem}{Theorem}
\newtheorem{proposition}{Proposition}
\newtheorem{corollary}{Corollary}
\newtheorem{lemma}{Lemma}
\newtheorem{assumption}{Assumption}
\usepackage[hidelinks,hypertexnames=false]{hyperref}
\usepackage[dvipsnames]{xcolor}

\usepackage{enumitem}
\usepackage{changepage}

\newcommand{\R}{\mathbb{R}}
\newcommand{\Var}{\mathrm{Var}}

\newcommand{\argmin}{\mathrm{arg\,min}}

\newtheorem{algorithm}{Algorithm}
\newcommand{\op}{\operatorname{op}}
\newcommand{\tr}{\operatorname{tr}}

\newtheorem{remarkx}{Remark}

\newcounter{appendixcounter}
\renewcommand{\theappendixcounter}{\Alph{appendixcounter}}

\renewcommand{\appendix}[1]{%
  \refstepcounter{appendixcounter}%
  \section*{Appendix \theappendixcounter: #1}%
  \addcontentsline{toc}{section}{Appendix \theappendixcounter: #1}%
}

\title{Semiparametric Inference in Panel Data with Interactive \\and High-dimensional Confounding\footnote{We thank Xun Lu, Anna Mikusheva, Whitney Newey, Zhentao Shi, Liangjun Su, and
seminar participants at the CUHK workshop, the 9th International Conference on Econometrics and Statistics, and the 41st Canadian
Econometrics Study Group for their useful comments and discussions. Any remaining errors are solely ours.}}
\author{Bin Chen\thanks{University of Rochester. Email: {binchen@rochester.edu} } \and Yukun Ma\thanks{University of Rochester. Email: {yma69@ur.rochester.edu}} \and Qiyang Yu\thanks{University of Rochester. Email: {qyu13@ur.rochester.edu} }}
\date{\today}

\begin{document}
\maketitle

\begin{abstract}
   This paper develops inference for a low-dimensional treatment coefficient in
   partially linear panel models with high-dimensional observed covariates and
   latent interactive confounding. Flexible adjustment for observed covariates
   can leave correlated factor components in treatment and outcome residuals,
   invalidating standard double/debiased machine learning (DML). We propose Factor-Augmented DML (FA-DML), which estimates treatment-side and reduced-form factor spaces from first-stage residuals. The estimator combines a feasible double projection with three-fold, unit-level cross-fitting. The double projection controls interactions between the two estimated factor spaces. The separate folds eliminate same-training-unit pairings in the leading disturbance components of the factor-estimation errors. Under fixed factor ranks, pervasive factor signals,
   strict exogeneity, and joint asymptotics in the number of units \(N\) and
   periods \(T\), FA-DML is first-order equivalent
   to an infeasible oracle, is asymptotically normal, and admits consistent
   variance estimation. In the baseline simulation design with strong interactive confounding, FA-DML confidence intervals have coverage close to the nominal level. Standard DML and a two-fold variant that estimates both factor spaces on the same units substantially undercover. In an application to U.S.\ states, FA-DML finds that a \(10\%\) increase in the real minimum wage is associated with a roughly \(1.1\%\) lower 50/10 wage ratio. The estimate is stable across control sets and first-stage learners.

\bigskip

\noindent \textit{JEL Codes}:
\textit{C10, C14, C23, C55}

\bigskip

\noindent
\textit{Keywords:} Debiased machine learning; High-dimensional panel data; Interactive fixed effects;
 Latent factor confounding; Semiparametric inference.
\bigskip

\end{abstract}
\section{Introduction}

Many panel applications have rich observed covariates, yet time-varying unobserved heterogeneity remains a concern. This paper asks how to conduct
inference on a low-dimensional treatment coefficient in a partially linear
panel model with both observed and latent interactive confounding. In this
setting, flexibly residualizing the treatment and outcome with respect to
observed covariates does not suffice. The resulting residuals may still contain
correlated interactive components generated by common shocks with
unit-specific loadings. The usual double/debiased machine learning (DML) score is then contaminated by latent
confounding. Additive fixed-effects transformations do not remove these
components, while classical interactive fixed-effects (IFE) methods typically impose
a parametric specification on the observed part of the model.

DML uses Neyman-orthogonal scores and
cross-fitting to make inference robust to regularized or nonparametric
first-stage estimation \citep{CCDDHNR18}. Existing panel extensions cover
difference-in-differences designs \citep{chang2020double} and partially linear
models with additive individual effects \citep{clarke2026double}. A complementary
factor-lasso literature extracts factors from a large set of observed controls
and combines them with sparse regression \citep{hansen2019factor}. These approaches do not directly address the residual interactive confounding considered here.

High-dimensional common-correlated-effects methods provide
a closer comparison. \citet{rucker2025estimation} construct a factor projection from cross-sectional covariate averages for desparsified Lasso inference, while \citet{chen2026dmlife} combine this
projection with DML and unit-level cross-fitting. Both require covariate averages to retain the relevant
factor directions, a condition that can fail when factors are absent
from the covariates or their loadings cancel across units. The structural projection approach of \citet{dong_gao_linton_peng} uses a related projection idea. It constructs internal instruments by projecting regressors away from a restricted nuisance-function space containing the conditional mean of the structural error. In our setting, the projected spaces are latent and must be estimated.

We propose Factor-Augmented DML (FA-DML). The estimator applies principal component analysis (PCA) to first-stage residuals to estimate the treatment and reduced-form factor spaces, then incorporates both estimated spaces into a double-projection score. It uses three disjoint folds of complete unit histories: one estimates the treatment nuisance function and factor space, a second estimates their reduced-form counterparts, and a third evaluates the score. Coefficient estimates are averaged over the six assignments of fold roles.

This paper makes three contributions. First, it extends DML to partially linear panels with interactive confounding and flexible  nuisance functions. The framework allows distinct treatment and outcome factor structures. We establish identification through a factor-orthogonal moment that removes the latent interactive components and is Neyman-orthogonal with respect to the nuisance functions. Under the maintained model, identification follows from strict exogeneity and a projected treatment-variation condition. It does not require the observed covariates to contain signals about the latent factors.

Second, it develops a score and sample-splitting construction that jointly control the additional errors introduced by residual-based factor estimation. Double projection makes the factor–factor remainder a product of treatment-side and reduced-form factor-estimation errors. A single estimated reduced-form projection can instead leave a first-order component or a quadratic self-product with nonzero mean. Estimating the two spaces on separate training folds removes same-unit disturbance pairings from their leading error components, while the evaluation fold separates both estimates from the final moment.

Third, under suitable regularity conditions, we establish oracle-equivalent inference with flexible first-stage learners. The feasible estimator has the same first-order asymptotic distribution as an infeasible estimator that knows the nuisance functions and factor spaces. We derive centered $\sqrt{NT}$-asymptotic normality and establish consistency of the proposed variance estimator as $N,T\to\infty$. We also develop an iterative nuclear-norm procedure for selecting the treatment and reduced-form factor dimensions. We prove its consistency and show that estimating these dimensions preserves the inference results.

The simulations separately assess factor augmentation, three-fold separation,
regularized nuisance estimation, and rank selection.
In the baseline simulations with strong interactive confounding, FA-DML delivers coverage close to the nominal level across linear and nonlinear nuisance specifications and across Lasso and Elastic Net first stages.
Standard DML and a two-fold double-projection variant substantially undercover. The proposed rank selector also recovers each factor dimension with high probability in the reported designs.

Our main application studies minimum wages and lower-tail wage inequality in a panel of \(50\) U.S.\ states from \(1979\) to \(2012\). With factor dimensions chosen by our selector, FA-DML implies that a \(10\%\) increase in the real minimum wage is associated with a roughly \(1.1\%\) lower 50/10 wage ratio. The estimate is stable across Lasso and random-forest first stages and across two control sets. The corresponding \(95\%\) confidence intervals exclude zero. With the original controls, pooled OLS, two-way fixed effects, and standard DML yield estimates close to zero, with confidence intervals that include zero. An additional application to branching deregulation  is presented in Appendix~\ref{app:banking}.

The remainder of the paper is organized as follows.
Section~\ref{sec_model} introduces the model and identification.
Section~\ref{Section_algorithm} develops the estimator, and
Section~\ref{Section_thm} presents the asymptotic theory and rank-selection
results. Section~\ref{sec:sim} reports the simulations.
Section~\ref{sec:ams} presents the minimum-wage application, and
Section~\ref{sec:conclusion} concludes. The appendix contains proofs,
implementation details, an additional empirical application, and sensitivity
analyses.
\subsection*{Notation}
{\color{black}For a matrix $A\in\mathbb R^{m\times n}$, let $s_j(A)$ denote
its $j$-th largest singular value. We write
$s_{\max}(A)=\|A\|_{\op}$ and $s_{\min}(A)$ for its largest
and smallest singular values, respectively. The operator,
Frobenius, and nuclear norms are
\[
\|A\|_{\op}=\sup_{\|x\|_2=1}\|Ax\|_2,
\qquad
\|A\|_F=\{\tr(A'A)\}^{1/2},
\qquad
\|A\|_*=\sum_{j=1}^{\min\{m,n\}}s_j(A),
\]
where $\|\cdot\|_2$ denotes the Euclidean norm.
For symmetric $A$, $\lambda_{\max}(A)$ and
$\lambda_{\min}(A)$ denote its largest and smallest
eigenvalues. We use $\tr(A)$ for the trace, $A'$ for
transpose, $I_d$ for the $d\times d$ identity matrix,
and $\mathbf 1\{\cdot\}$ for the indicator function.

We write $\xrightarrow{p}$ and $\xrightarrow{d}$ for
convergence in probability and distribution, respectively,
and $\to$ for convergence of deterministic sequences.
For a random sequence $Z_{NT}$ and a positive deterministic
sequence $a_{NT}$, $Z_{NT}=O_p(a_{NT})$ means that
$Z_{NT}/a_{NT}$ is bounded in probability, while
$Z_{NT}=o_p(a_{NT})$ means that this ratio converges
to zero in probability. In bounds, $C$ denotes a generic positive constant whose
value may vary from line to line. For positive deterministic
sequences, $a_{NT}\lesssim b_{NT}$ means that
$a_{NT}\le Cb_{NT}$ for all sufficiently large $N,T$,
and $a_{NT}\asymp b_{NT}$ means that both
$a_{NT}\lesssim b_{NT}$ and $b_{NT}\lesssim a_{NT}$ hold.
For nonnegative random sequences, with $Y_{NT}>0$ with
probability approaching one, $X_{NT}\lesssim_p Y_{NT}$
means that $X_{NT}/Y_{NT}=O_p(1)$. We write
$X_{NT}\asymp_p Y_{NT}$ when both
$X_{NT}\lesssim_p Y_{NT}$ and
$Y_{NT}\lesssim_p X_{NT}$ hold.
\ref{app:notation} collects the model-specific
notation for residuals, disturbances, loadings, and
first-stage errors.}

\section{Model and Identification}
\label{sec_model}

This section presents the model and a population moment that identifies the
target coefficient. Residualizing the outcome and treatment with respect to observed covariates may leave interactive components generated by common shocks and unit-specific loadings. The usual
partially linear DML moment can therefore fail. Under the exogeneity and
treatment-variation conditions stated below, projecting out the relevant factor
spaces yields a valid identifying moment.

\subsection{Model and Target Parameter}
\label{sub:partial_linear_panel}

We observe a balanced panel
\[
    \{(Y_{it},D_{it},X_{it}):i=1,\ldots,N,\ t=1,\ldots,T\},
\]
where \(i\) indexes units, \(t\) indexes periods, and \(N\) and \(T\) are the
corresponding panel dimensions. The variable \(Y_{it}\) is the outcome,
\(D_{it}\) is a scalar treatment, and \(X_{it}\in\mathcal X\) is a possibly
high-dimensional vector of observed covariates with support \(\mathcal X\).
Let
\[
    Y_i=(Y_{i1},\ldots,Y_{iT})',\qquad
    D_i=(D_{i1},\ldots,D_{iT})',\qquad
    X_i=(X_{i1},\ldots,X_{iT})'.
\]
The unknown functions \(f_0,g_0:\mathcal X\to\mathbb R\) capture how the
observed covariates enter the outcome and treatment equations.
We use \(f_0(X_i)\) and \(g_0(X_i)\) for the \(T\)-vectors obtained by
evaluating these functions period by period. The parameter of interest is the
common coefficient \(\theta_0\). Let \(r_Y^*\) and \(r_D\) denote the
outcome-side and treatment-side factor dimensions, respectively. Let
\(G_Y^0\in\mathbb R^{T\times r_Y^*}\) and
\(G_D^0\in\mathbb R^{T\times r_D}\) be full-column-rank matrices of
outcome-side and treatment-side common factors, respectively, and let
\(C_{Y,i}^*\in\mathbb R^{r_Y^*}\) and
\(C_{D,i}\in\mathbb R^{r_D}\) be their unit-specific loadings. Let
\(\varepsilon_i,\nu_i\in\mathbb R^T\) denote the corresponding idiosyncratic
disturbances. The primitive loading arrays are treated as fixed throughout.
We consider the partially linear panel system
\begin{align}
    Y_i
    &=D_i\theta_0+f_0(X_i)+G_Y^0C_{Y,i}^*+\varepsilon_i,
    \label{eq:modelY-main}\\
    D_i
    &=g_0(X_i)+G_D^0C_{D,i}+\nu_i.
    \label{eq:modelD-main}
\end{align}
Row \(t\) of each factor matrix contains the common-factor values in period
\(t\), while heterogeneity across units enters through the corresponding
loadings.\footnote{Equation~\eqref{eq:modelD-main} allows common forces to affect treatment differently across units even after adjustment for observed covariates. For example, states may respond
differently to national political developments or aggregate economic
conditions. These forces may also affect outcomes, generating latent
confounding. The additive factor structure is a maintained assumption
on treatment assignment. For binary treatment, with fixed loadings,
the conditional mean restriction below implies
\(
    \Pr(D_{it}=1\mid X_i,\mathcal F_T)
    =g_0(X_{it})+(G_D^0C_{D,i})_t\in[0,1].
\)
Thus binary treatment is permitted under a factor-augmented linear
probability specification, but a logit or probit model with a factor
index need not satisfy this condition.} A causal interpretation of \(\theta_0\) requires the outcome equation to be
structural and the strict-exogeneity condition stated below to hold.

Assume \begin{equation}
    \operatorname{rank}
    \begin{pmatrix}
        C_{D,1}' & C_{Y,1}^{*\prime}\\
        \vdots   & \vdots\\
        C_{D,N}' & C_{Y,N}^{*\prime}
    \end{pmatrix}
    =r_D+r_Y^*.
    \label{eq:joint-loading-rank}
\end{equation}
This rank condition allows dependence between the two sets of
loadings but rules out exact linear dependence among their
joint columns. The two equations may still share common factors.

Define $m_0=f_0+\theta_0g_0$, and let
$G^0\in\mathbb R^{T\times r_Y}$ be any full-column-rank basis for the
reduced-form factor space spanned by the interactive components
\[
    G_Y^0C_{Y,i}^*+\theta_0G_D^0C_{D,i},
    \qquad i=1,\ldots,N,
\]
and let $C_i\in\mathbb R^{r_Y}$ denote their coordinates in this basis.
Substituting the treatment equation into the outcome equation gives
\begin{equation}
    Y_i=m_0(X_i)+G^0C_i+\theta_0\nu_i+\varepsilon_i,
    \label{eq:reduced-form}
\end{equation}
where $G^0C_i=G_Y^0C_{Y,i}^*+\theta_0G_D^0C_{D,i}$.
The coordinates $C_i$ depend on the chosen basis, but $G^0C_i$ does not.
Under the joint loading-rank condition~\eqref{eq:joint-loading-rank},
\begin{equation}
    \operatorname{span}(G^0)
    =\operatorname{span}\!\left([G_Y^0,\theta_0G_D^0]\right),
    \qquad
    r_Y=\operatorname{rank}\!\left([G_Y^0,\theta_0G_D^0]\right).
    \label{eq:effective-factor-space}
\end{equation}
Thus $r_Y=\dim\operatorname{span}(G^0)$ is the reduced-form factor dimension
relevant to outcome-residual PCA; it need not equal the structural outcome
factor dimension $r_Y^*$. Our analysis permits any fixed $\theta_0$, including
zero, subject to the factor-strength and other regularity conditions in
Section~\ref{Section_thm}. The joint loading-rank condition alone does not
ensure consistent factor estimation.\footnote{For
$\theta_0\ne0$, the treatment factor space is contained in the reduced-form
space. At $\theta_0=0$, the latter equals the structural outcome factor space,
so this nesting need not hold. The results are established for each fixed $\theta_0$.}

\subsection{Why Covariate Residualization Is Not Enough}

To see why standard DML fails, first suppose that the observed nuisance
functions were known. Define the covariate-residualized treatment and outcome
by
\[
    v_{D,i}=D_i-g_0(X_i),
    \qquad
    v_{Y,i}=Y_i-m_0(X_i),
\]
and define the residual in the outcome equation at the true coefficient by
\[
    u_i=v_{Y,i}-\theta_0v_{D,i}.
\]
Equations~\eqref{eq:modelY-main}--\eqref{eq:reduced-form} imply
\begin{equation}
    v_{D,i}=G_D^0C_{D,i}+\nu_i,
    \qquad
    u_i=G_Y^0C_{Y,i}^*+\varepsilon_i.
    \label{eq:ordinary-residuals}
\end{equation}

The usual partially linear score uses \(v_{D,i}'u_i\) \citep{CCDDHNR18}. Even if every term
involving \(\nu_i\) or \(\varepsilon_i\) has conditional mean zero given the
common factors, this score contains the factor--factor term
\[
    C_{D,i}'G_D^{0\prime}G_Y^0C_{Y,i}^*.
\]
This term is the inner product of the treatment- and outcome-side factor components. When those components are not orthogonal, the score need not have mean zero. Flexible regression of \(Y_i\) and \(D_i\) on \(X_i\) does not solve
this problem because the two latent components in
\eqref{eq:ordinary-residuals} remain.

\subsection{Factor-Orthogonal Moment}

For a full-column-rank matrix \(Q\in\mathbb R^{T\times r}\), let
\[
    P_Q=Q(Q'Q)^{-1}Q',
    \qquad
    M_Q=I_T-P_Q,
\]
where \(I_T\) is the \(T\times T\) identity matrix. Thus \(P_Q\) projects onto
\(\operatorname{span}(Q)\), and \(M_Q\) projects onto its orthogonal
complement. Define the population factor-orthogonal operator\footnote{The matrix $A_0$ is a product of
residual-makers and need not be a symmetric orthogonal projector.
The order of the two factors is therefore maintained throughout.}
\begin{equation}
    A_0=M_{G_D^0}M_{G^0}.
    \label{eq:oracle-operator}
\end{equation}
By construction,
\[
    (G_D^0)'A_0=0,\qquad A_0G^0=0.
\]
Equation~\eqref{eq:effective-factor-space} also implies
\[
    A_0G_Y^0=0,\qquad \theta_0A_0G_D^0=0.
\]
These identities and \eqref{eq:ordinary-residuals} yield
\begin{equation}
    \begin{aligned}
    v_{D,i}'A_0u_i
    &=(G_D^0C_{D,i}+\nu_i)'A_0
      (G_Y^0C_{Y,i}^*+\varepsilon_i)\\
    &=\nu_i'A_0\varepsilon_i.
    \end{aligned}
    \label{eq_ortho}
\end{equation}

Let \(\mathcal F_T=\sigma(G_D^0,G_Y^0)\) denote the information generated by the
common factor paths.

\begin{assumption}[Strict exogeneity and projected treatment variation]
\label{ass:identification}
Conditional on \(\mathcal F_T\),
\begin{enumerate}[label=(\alph*)]
    \item The idiosyncratic disturbances satisfy the conditional mean restrictions
    \(E[\nu_i\mid X_i,\mathcal F_T]=0\) and
    \(E[\varepsilon_i\mid \nu_i,X_i,\mathcal F_T]=0\).
    \item The projected treatment variation satisfies
    \(E[\nu_i'A_0\nu_i\mid\mathcal F_T]>0\).
\end{enumerate}
\end{assumption}

Part (a) is a strict-exogeneity condition with respect to the complete observed
covariate path. It implies \(E[\nu_i'A_0\varepsilon_i\mid\mathcal F_T]=0\), the
score orthogonality used for identification. Part (b)
directly requires the quadratic-form denominator for the ordered
two-projection score to be positive. Since $A_0$ need not be symmetric,
this positivity does not follow merely from nonzero residual treatment
variation. Dynamic
panels and predetermined covariates require different moment conditions and are
not covered by this assumption.

Let \(W_i=(Y_i,D_i,X_i)\). For a candidate coefficient
\(\theta\in\mathbb R\), maps
\(g,m:\mathcal X^T\to\mathbb R^T\) and a matrix
\(A\in\mathbb R^{T\times T}\), define
\[
    \psi(W_i;\theta,g,m,A)
    =\{D_i-g(X_i)\}'A
      \bigl[Y_i-m(X_i)-\theta\{D_i-g(X_i)\}\bigr].
\]
The oracle score evaluated at the unprojected nuisance functions is
\begin{equation}
    \psi(W_i;\theta,g_0,m_0,A_0)
    =v_{D,i}'A_0\{v_{Y,i}-\theta v_{D,i}\}.
    \label{eq_score}
\end{equation}
The parameter \(\theta_0\) is identified by the conditional moment
\begin{equation}
E[\psi(W_i;\theta_0,g_0,m_0,A_0)\mid\mathcal F_T]=0.
    \label{eq:id-moment}
\end{equation}
Indeed,
\[
    v_{Y,i}-\theta v_{D,i}
    =u_i+(\theta_0-\theta)v_{D,i},
\]
so Assumption~\ref{ass:identification} and equation \eqref{eq_ortho} give
\begin{align*}
    E[\psi(W_i;\theta,g_0,m_0,A_0)\mid\mathcal F_T]
    &=(\theta_0-\theta)
      E[v_{D,i}'A_0v_{D,i}\mid\mathcal F_T]\\
    &=(\theta_0-\theta)
      E[\nu_i'A_0\nu_i\mid\mathcal F_T].
\end{align*}
The second equality uses
\[
    E[\nu_i'A_0G_D^0C_{D,i}\mid\mathcal F_T]=0,
\]
which follows from Assumption~\ref{ass:identification}(a) and the fixed-loading
convention; it does not require $A_0G_D^0=0$.
As \(E[\nu_i'A_0\nu_i\mid\mathcal F_T]\) is strictly positive, \(\theta_0\) is the unique
solution to \eqref{eq:id-moment}.

\subsection{Two Layers of Orthogonalization}

The factor projection addresses latent confounding. A second form of
orthogonalization protects the score from first-stage estimation error in the
observed nuisance functions. The feasible algorithm estimates factor-residualized
versions of these functions. Define
\[
    g_0^*(X_i)=M_{G_D^0}g_0(X_i),
    \qquad
    m_0^*(X_i)=M_{G^0}m_0(X_i).
\]
Although $g_0$ and $m_0$ act period by period, the projected
functions generally depend on the complete covariate path because the
projections mix periods. Under Assumption~\ref{ass:identification}, they
generate the same conditional population moment as the unprojected nuisance
functions: for every candidate coefficient $\theta$,
\begin{equation}
    E[\psi(W_i;\theta,g_0,m_0,A_0)\mid\mathcal F_T]
    =E[\psi(W_i;\theta,g_0^*,m_0^*,A_0)\mid\mathcal F_T].
    \label{eq:score-equivalence}
\end{equation}
At the true coefficient, the scores also coincide exactly:
\[
    \psi(W_i;\theta_0,g_0,m_0,A_0)
    =\psi(W_i;\theta_0,g_0^*,m_0^*,A_0).
\]
\ref{app:proof} gives the underlying algebra.

We state the orthogonality property for the projected nuisance functions. Let \(h_g,h_m:\mathcal X^T\to\mathbb R^T\) be square-integrable perturbation directions, which may depend on $\mathcal F_T$.
Under Assumption~\ref{ass:identification},
\begin{equation}
    \left.
    \frac{\partial}{\partial \tau}
    E\!\left[
        \psi(W_i;\theta_0,g_0^*+\tau h_g,m_0^*+\tau h_m,A_0)
        \mid\mathcal F_T
    \right]
    \right|_{\tau=0}
    =0.
    \label{eq:neyman_moment}
\end{equation}
Thus first-stage errors around \(g_0^*\) and \(m_0^*\) have no first-order
effect on the population moment. This Neyman orthogonality is distinct from
factor orthogonality: the former controls local first-stage error, whereas the
latter uses \(A_0\) to remove the latent components themselves.
Equality of the conditional population moments preserves
the identifying equation, while exact equality at $\theta_0$ preserves the
oracle score used in the asymptotic expansion. This allows the feasible
procedure to target the projected nuisance functions.

\section{Estimation}
\label{Section_algorithm}
This section develops the feasible FA-DML estimator. The
population-moment equivalence and the exact score identity
at $\theta_0$ established in Section~\ref{sec_model} permit the algorithm
to target the projected nuisance functions $g_0^*$ and $m_0^*$, together
with the latent factor spaces $\operatorname{span}(G_D^0)$ and
$\operatorname{span}(G^0)$. The estimator uses machine-learning (ML) estimates of the nuisance functions
and PCA estimates of the factor spaces.
Three-fold cross-fitting separates treatment-side training, reduced-form
training, and score evaluation.

For the analysis below, define the augmented loadings
\[
\bar C_{D,i}=C_{D,i}+(G_D^{0\prime}G_D^0)^{-1}G_D^{0\prime}g_0(X_i),
\qquad
\bar C_{Y,i}=C_i+(G^{0\prime}G^0)^{-1}G^{0\prime}m_0(X_i),
\]
and let \(\xi_i=\theta_0\nu_i+\varepsilon_i\). The projected residuals then
satisfy
\begin{equation}
    v_{D,i}^*=D_i-g_0^*(X_i)=G_D^0\bar C_{D,i}+\nu_i,
    \qquad
    v_{Y,i}^*=Y_i-m_0^*(X_i)=G^0\bar C_{Y,i}+\xi_i.
    \label{eq:projected-residual-representations}
\end{equation}
By \eqref{eq:effective-factor-space},
$\theta_0G_D^0\bar C_{D,i}\in\operatorname{span}(G^0)$. Hence there is a
unique $\bar C_i\in\mathbb R^{r_Y}$ satisfying
\[
    G^0\bar C_i
    =G^0\bar C_{Y,i}-\theta_0G_D^0\bar C_{D,i}.
\]
Thus the projected score residual at \(\theta_0\) is
\begin{equation}
    u_i^*=v_{Y,i}^*-\theta_0v_{D,i}^*
    =G^0\bar C_i+\varepsilon_i.
    \label{eq:projected-score-residual}
\end{equation}
The augmented loadings \(\bar C_{D,i}\) and \(\bar C_{Y,i}\) determine the
factor signal available to PCA, whereas \(\bar C_i\) enters the score
expansion.

\subsection{Three-Fold Double-Projection Estimator}
The sample is split into three folds of complete unit-level time series. For each
split, the treatment-side first stage, the reduced-form first stage, and score
evaluation are performed on mutually disjoint folds. Algorithm~\ref{algorithm_three_fold}
treats \(r_D\) and \(r_Y\) as given. Section~\ref{subsec:rank-selection}
establishes validity with estimated ranks, and the complete data-driven
rank-selection procedure appears in Algorithm~\ref{alg:rank_estimation} in
\ref{app:rank-selection-algorithm}.
The algorithm uses the same construction for all coefficient
values; it does not require a preliminary test on $\theta_0$ or impose an
ordering on the two factor dimensions.

\begin{algorithm}[Three-fold double-projection estimator]
\label{algorithm_three_fold}
Given panel data \(\{(Y_i,D_i,X_i):i=1,\dots,N\}\), factor dimensions \(r_D\)
and \(r_Y\), fold-specific initializations for \(\hat g\) and \(\hat m\), a
maximum number of iterations \(S\), and a tolerance
\(\epsilon_{\mathrm{tol}}>0\):

\begin{enumerate}
    \item Partition \(\{1,\dots,N\}\) into three disjoint folds
    \(\mathcal I_1,\mathcal I_2,\mathcal I_3\), and let
    \((\mathcal I_1^{(l)},\mathcal I_2^{(l)},\mathcal I_3^{(l)})\),
    \(l=1,\dots,6\), denote their six permutations.

    \item For each \(l=1,\dots,6\), proceed as follows.
    \begin{enumerate}
        \item On \(\mathcal I_1^{(l)}\), initialize \(\hat g\) using only this
        fold. Estimate \(g_0^*\) and \(\operatorname{span}(G_D^0)\) by alternating
        between two updates: (i) apply rank-\(r_D\) PCA to
        \(\{D_i-\hat g(X_i):i\in\mathcal I_1^{(l)}\}\) to obtain \(\hat G_D\);
        (ii) refit \(\hat g\) with the chosen first-stage method to the projected responses
        \(\{M_{\hat G_D}D_i:i\in\mathcal I_1^{(l)}\}\).\footnote{Because the
projector mixes periods, the learner must map the complete path \(X_i\) to a
\(T\)-vector, or use an equivalent stacked design. For linear or series learners, project the stacked design along with the
response. A learner using only \(X_{it}\) at row \(t\) requires the projected
target to remain a function of \(X_{it}\) alone.}
        Stop after \(S\) updates or when the root-mean-square change in fitted
        values on the fold is at most \(\epsilon_{\mathrm{tol}}\). After the
        final update, apply PCA once more to align the final nuisance and factor
        estimates; denote them by \(\hat g^{(l1)}\) and \(\hat G_D^{(l1)}\).

        \item On \(\mathcal I_2^{(l)}\), analogously initialize \(\hat m\)
        using only this fold and estimate \(m_0^*\) and
        \(\operatorname{span}(G^0)\) from the outcome residuals using rank
        \(r_Y\). Denote the outputs by \(\hat m^{(l2)}\) and \(\hat G^{(l2)}\).

	    \item On the evaluation fold \(\mathcal I_3^{(l)}\), form the feasible
        counterparts of \(v_{D,i}^*\) and \(v_{Y,i}^*\):
	    \[
	    \hat v_{D,i}=D_i-\hat g^{(l1)}(X_i),
	    \qquad
	    \hat v_{Y,i}=Y_i-\hat m^{(l2)}(X_i),
	    \qquad i\in\mathcal I_3^{(l)}.
	    \]
	    Let \(\hat A^{(l)}=M_{\hat G_D^{(l1)}}M_{\hat G^{(l2)}}\) and compute
        \begin{align*}
        \hat\theta^{(l)}
        &=
        \frac{\sum_{i\in\mathcal I_3^{(l)}}
        \hat v_{D,i}'\hat A^{(l)}\hat v_{Y,i}}
        {\sum_{i\in\mathcal I_3^{(l)}}
        \hat v_{D,i}'\hat A^{(l)}\hat v_{D,i}},\\
        \hat J_l
        &=\frac{1}{T|\mathcal I_3^{(l)}|}
        \sum_{i\in\mathcal I_3^{(l)}}
        \hat v_{D,i}'\hat A^{(l)}\hat v_{D,i},\\
        \hat\sigma_{\theta,l}^{2}
        &=\frac{\hat J_l^{-2}}{T|\mathcal I_3^{(l)}|}
        \sum_{i\in\mathcal I_3^{(l)}}
        \left\{\hat v_{D,i}'\hat A^{(l)}
        (\hat v_{Y,i}-\hat\theta^{(l)}\hat v_{D,i})\right\}^2.
        \end{align*}
        If
        \(
        \sum_{i\in\mathcal I_3^{(l)}}
        \hat v_{D,i}'\hat A^{(l)}\hat v_{D,i}=0
        \), set
        \(\hat\theta^{(l)}=\hat\sigma_{\theta,l}^{2}=0\) for that split.
        Proposition~\ref{proposition:denominator} shows that this convention is
        invoked with probability approaching zero.
    \end{enumerate}

    \item Average across the six permutations to symmetrize over fold roles:
    \[
    \hat\theta=\frac16\sum_{l=1}^6 \hat\theta^{(l)},
    \qquad
    \hat\sigma_\theta^2=\frac16\sum_{l=1}^6 \hat\sigma_{\theta,l}^2.
    \]
\end{enumerate}
\end{algorithm}

\subsection{Why Double Projection and Three-Fold Cross-Fitting?}
\label{subsection:2.3}
\label{subsection:2.4}

Double projection turns the factor--factor remainder into
a cross-product of estimation errors from the two factor
spaces. Separate training folds remove same-unit disturbance
pairings from its leading component, while the evaluation
fold separates both estimates from the final score.

Fix one permutation $l$, write
$n_3=|\mathcal I_3^{(l)}|$, and suppress fold superscripts.
Let $\hat A=M_{\hat G_D}M_{\hat G}$.
Consider the factor--factor term in the score expansion:
\begin{equation}
    \frac1{\sqrt{n_3T}}
    \sum_{i\in\mathcal I_3^{(l)}}
    \bar C_{D,i}'G_D^{0\prime}\hat A G^0\bar C_i.
    \label{eq:threefold-ff}
\end{equation}
Its oracle counterpart is zero, so this term arises
entirely from factor estimation.

Let $H_D$ and $H_G$ be the nonsingular rotation matrices
and write
\[
    \mathcal E_D=G_D^0H_D-\hat G_D,
    \qquad
    \mathcal E_G=G^0H_G-\hat G.
\]
Since $\hat G_D'\hat A=0$ and $\hat A\hat G=0$,
\begin{equation}
    G_D^{0\prime}\hat A G^0
    =
    H_D^{-\prime}
    \mathcal E_D'\hat A\mathcal E_G
    H_G^{-1}.
    \label{eq:doubleproj-cross-product}
\end{equation}
Thus double projection makes the factor--factor term
depend on a product of errors from the two estimated
spaces, without requiring nesting.

With only the outcome-side projector, the corresponding
matrix is
\[
    G_D^{0\prime}M_{\hat G}G^0
    =
    G_D^{0\prime}M_{\hat G}\mathcal E_GH_G^{-1}.
\]
Without nesting, this expression can retain a first-order
component in the reduced-form factor error. Under nesting, it becomes a
quadratic self-product of that error, which can contain same-training-unit
disturbance terms.

The cross-product in \eqref{eq:doubleproj-cross-product}
allows these pairings to be removed by separate training
folds. Recall that
$\xi_i=\theta_0\nu_i+\varepsilon_i$.
Up to the finite-dimensional coefficient matrices in
the PCA expansions, the leading disturbance product
with the oracle operator is
\[
    \frac1{n_1n_2}
    \sum_{j\in\mathcal I_1^{(l)}}
    \sum_{k\in\mathcal I_2^{(l)}}
    \bar C_{D,j}\bar C_{Y,k}'
    \nu_j'A_0\xi_k.
\]
For disjoint training folds, every pair has $j\ne k$.
Conditional on the factor paths and training covariates,
the summands have mean zero by cross-sectional independence
and the disturbance mean restrictions.

If both factor spaces were estimated on a common training
fold, the product would also contain same-unit terms
involving $\nu_j'A_0\xi_j$. Under the exogeneity conditions,
\[
    E[\nu_j'A_0\xi_j\mid \mathcal F_T]
    =
    \theta_0\operatorname{tr}\!\left(
        A_0E[\nu_j\nu_j'\mid \mathcal F_T]
    \right),
\]
which need not vanish. Separate training folds exclude these same-unit terms, while the third fold separates
both factor estimates from score evaluation.

\section{Theory}
\label{Section_thm}
This section
establishes the asymptotic validity of the three-fold double-projection
FA-DML estimator introduced in Section~\ref{Section_algorithm}.
We show that the feasible score, which uses estimated nuisance
functions and estimated factor spaces, is first-order equivalent to the infeasible
oracle score
\[
\frac{1}{\sqrt{NT}}\sum_{i=1}^N \nu_i'A_0\varepsilon_i.
\]
Consequently, the proposed estimator is centered at \(\theta_0\) and admits standard
\(\sqrt{NT}\)-inference.

\subsection{Oracle Decomposition}

Fix one split \(l\in\{1,\dots,6\}\), write
\(n_k=|\mathcal I_k^{(l)}|\) for \(k=1,2,3\), and suppress the superscript
\(l\) on fold-specific quantities whenever there is no ambiguity. We retain the population and augmented-loading notation
introduced in Sections~\ref{sec_model}
and~\ref{Section_algorithm}. Define
\[
    \delta_i^g=\hat g^{(l1)}(X_i)-g_0^*(X_i),
    \qquad
    \delta_i^m=\hat m^{(l2)}(X_i)-m_0^*(X_i),
\]
and write
\[
    \delta_i=\delta_i^m-\theta_0\delta_i^g,
    \qquad
    \Delta_A=\hat A-A_0.
\]
Equations~\eqref{eq:projected-residual-representations} and
\eqref{eq:projected-score-residual} then give
\[
    \hat v_{D,i}=G_D^0\bar C_{D,i}+\nu_i-\delta_i^g,
    \qquad
    \hat v_{Y,i}-\theta_0\hat v_{D,i}
    =G^0\bar C_i+\varepsilon_i-\delta_i.
\]

Throughout this section, we maintain the model and joint
loading-rank condition in Section~\ref{sec_model}.
The coefficient $\theta_0$ is fixed, and all nuisance functions and factor
spaces below refer to the corresponding data-generating process (DGP). Until
Section~\ref{subsec:rank-selection}, the ranks \(r_D\) and \(r_Y\) are
treated as fixed and known, as formalized in Assumption~\ref{ass:factor_fold}(a).
Because their defining projections depend on the factor paths, \(g_0^*\) and
\(m_0^*\) are \(\mathcal F_T\)-measurable random maps. Unless a
statement explicitly conditions on a specified sigma field, all \(O_p\) and
\(o_p\) orders refer to the joint law of the sample, the factor paths, the
random partition, and any auxiliary learner randomization.

The splitwise estimator can be written as
\begin{equation}\label{eq:theta_l}
\hat\theta^{(l)}-\theta_0
=
\frac{
\sum_{i\in\mathcal I_3^{(l)}} \hat v_{D,i}'\hat A(\hat v_{Y,i}-\theta_0\hat v_{D,i})
}{
\sum_{i\in\mathcal I_3^{(l)}} \hat v_{D,i}'\hat A\hat v_{D,i}
}.
\end{equation}
Expanding the numerator yields
{\small
\begin{align}\label{eq:theta_l_expand}
\hat\theta^{(l)}-\theta_0
=
\left(
\sum_{i\in\mathcal I_3^{(l)}} \hat v_{D,i}'\hat A\hat v_{D,i}
\right)^{-1}
\left(
\sum_{i\in\mathcal I_3^{(l)}} \nu_i'A_0\varepsilon_i
+
\sum_{i\in\mathcal I_3^{(l)}}R_{1i}
+
\sum_{i\in\mathcal I_3^{(l)}}R_{2i}
+
\sum_{i\in\mathcal I_3^{(l)}}R_{3i}
+
\sum_{i\in\mathcal I_3^{(l)}}R_{4i}
\right),
\end{align}
}
where
\begin{align*}
R_{1i}
&=
-\delta_i^{g\prime}A_0\varepsilon_i
-\nu_i'A_0\delta_i
+\delta_i^{g\prime}A_0\delta_i,
\\
R_{2i}
&=
\bar C_{D,i}'G_D^{0\prime}\hat A G^0\bar C_i
+
\bar C_{D,i}'G_D^{0\prime}\hat A \varepsilon_i
+
\nu_i'\hat A G^0\bar C_i,
\\
R_{3i}
&=
-\delta_i^{g\prime}\Delta_A\varepsilon_i
-\nu_i'\Delta_A\delta_i
+\delta_i^{g\prime}\Delta_A\delta_i
+\nu_i'\Delta_A\varepsilon_i,
\\
R_{4i}
&=
- \bar C_{D,i}'G_D^{0\prime}\hat A\delta_i
-\delta_i^{g\prime}\hat A G^0\bar C_i.
\end{align*}

The first term in \eqref{eq:theta_l_expand} is the oracle score. The remainder
terms \(R_{1i}\)–\(R_{4i}\) collect, respectively, the nuisance-error
terms under the oracle projector, the pure factor-projection terms, the
projector-perturbation terms, and the
factor--nuisance cross terms. The proof of Theorem~\ref{thm:CLT} shows that all four remainder sums are \(o_p(\sqrt{n_3T})\) under its stated assumptions and the three-fold double-projection construction.

\subsection{Regularity Conditions}
We state the assumptions in four groups. The first concerns first-stage learning
and remainder rates. The second concerns the factor structure and fold design.
The third concerns training-fold disturbances and residuals, and the fourth
concerns evaluation-fold moments.

Recall that \(\mathcal F_T=\sigma(G_D^0,G_Y^0)\) is the sigma field generated
by the factor paths. For each split $l$, let $\mathcal H_l$
be the sigma field generated by $\mathcal F_T$, the ordered random
partition, any auxiliary learner randomization, and the training data
\[
\{(Y_i,D_i,X_i):
i\in\mathcal I_1^{(l)}\cup\mathcal I_2^{(l)}\}.
\]
Any auxiliary learner randomization is independent of the sample
and partition conditional on $\mathcal F_T$.
The treatment learner $\hat g^{(l)}$ is trained on
$\mathcal I_1^{(l)}$, and the reduced-form learner $\hat m^{(l)}$
is trained on $\mathcal I_2^{(l)}$. Both are
$\mathcal H_l$-measurable.

For any $\mathcal H_l$-measurable map
$h:\mathcal X^T\to\mathbb R^T$, define the conditional prediction
norm and evaluation-fold empirical norm by
\begin{align*}
\|h\|_{P,2}
&=
\left\{
\frac{1}{T}
E\!\left[\|h(X_i)\|_2^2\mid\mathcal H_l\right]
\right\}^{1/2},
\qquad i\in\mathcal I_3^{(l)},\\
\|h\|_{n,l}
&=
\left\{
\frac{1}{Tn_3}
\sum_{i\in\mathcal I_3^{(l)}}\|h(X_i)\|_2^2
\right\}^{1/2}.
\end{align*}
By Assumption~\ref{ass:factor_fold}(d)--(e), together with
the independence of learner randomization, the first expression
does not depend on the choice of evaluation unit.
We suppress the dependence of the rate sequences on $(N,T)$.

\begin{assumption}[First-stage prediction rates]
\label{ass:1st_stage}
There exist positive deterministic sequences \(\bar r_{g}\) and
\(\bar r_{m}\), both converging to zero, such that the following
conditions hold uniformly over \(l\).
\begin{enumerate}[label=(\alph*)]
\item The feasible treatment learner satisfies
\[
\|\hat g^{(l1)}-g_0^*\|_{P,2}=O_p(\bar r_{g}).
\]
\item The feasible reduced-form learner satisfies
\[
\|\hat m^{(l2)}-m_0^*\|_{P,2}=O_p(\bar r_{m}).
\]
\item The prediction rates satisfy
\[
\sqrt{n_3T}\,\bar r_{g}
\bigl(\bar r_{m}+\bar r_{g}\bigr)=o(1).
\]
\end{enumerate}
\end{assumption}

Assumption~\ref{ass:1st_stage}(a)--(b) bounds prediction errors
relative to the projected targets $g_0^*$ and $m_0^*$. Part~(c) makes the quadratic nuisance-error term asymptotically
negligible. The proof of Proposition~\ref{proposition:1st_stage} transfers the conditional prediction
bounds to the evaluation fold and establishes the resulting bound on
\(R_{1i}\).

\begin{assumption}[First-stage remainder rates]
\label{ass:rate-main}
There exist positive deterministic sequences \(\eta_g\) and \(\eta_m\) such that the following conditions hold uniformly over $l$. Write
\(\eta=(\eta_g^2+\eta_m^2)^{1/2}\).
\begin{enumerate}[label=(\alph*)]
\item The loading-weighted first-stage errors satisfy
\begin{align*}
\left\|
\frac1{\sqrt{n_1T}}\sum_{i\in\mathcal I_1^{(l)}}
\delta_i^g\bar C_{D,i}'
\right\|_F
&=O_p(\eta_g),
&
\left\|
\frac1{\sqrt{n_2T}}\sum_{i\in\mathcal I_2^{(l)}}
\delta_i^m\bar C_{Y,i}'
\right\|_F
&=O_p(\eta_m),\\
\left\|
\frac1{\sqrt{n_3T}}\sum_{i\in\mathcal I_3^{(l)}}
\delta_i^g\bar C_i'
\right\|_F
&=O_p(\eta_g),
&
\left\|
\frac1{\sqrt{n_3T}}\sum_{i\in\mathcal I_3^{(l)}}
\delta_i\bar C_{D,i}'
\right\|_F
&=O_p(\eta).
\end{align*}
Let \(q_{NT}=T^{-1/2}+\sqrt T/N\). The associated rate sequences also satisfy
\[
\frac TN\eta_g^2=o(1),
\qquad
\frac TN\eta_m^2=o(1),
\qquad
\sqrt{\frac TN}\,\eta_g\eta=o(1),
\]
and
\[
q_{NT}(1+\eta_g+\eta_m)=o(1),
\qquad
\sqrt{\frac NT}\,q_{NT}^2=o(1).
\]
\item The evaluation-fold errors satisfy the loading-weighted squared-error
bounds
\begin{align*}
\frac1{n_3T}\sum_{i\in\mathcal I_3^{(l)}}
\|\delta_i^g\|_2^2\|\bar C_i\|_2^2
&=O_p(\eta_g^2),
&
\frac1{n_3T}\sum_{i\in\mathcal I_3^{(l)}}
\|\delta_i\|_2^2\|\bar C_{D,i}\|_2^2
&=O_p(\eta^2),
\end{align*}
and the fourth-moment bound
\[
\frac1{n_3T^2}\sum_{i\in\mathcal I_3^{(l)}}
\|\delta_i^g\|_2^4=O_p(1).
\]
\end{enumerate}
\end{assumption}

Assumption~\ref{ass:rate-main}(a) controls the loading-weighted
errors in the PCA expansions and their interactions with the score.
The rate restrictions bound the corresponding first-stage terms
and quadratic PCA remainders. Part~(b), needed only for variance
consistency, controls squared errors without uniformly bounding
individual loadings. These conditions complement the prediction
bounds in Assumption~\ref{ass:1st_stage}. \ref{app:decomposition}
discusses their roles and gives a Lasso illustration.

The basis $G^0$ is chosen measurably with respect to
$\mathcal F_T$, so the reduced-form coordinates $C_i$
are $\mathcal F_T$-measurable.
Cross-sectional sampling statements are conditional
on $\mathcal F_T$.

\begin{assumption}[Factor structure, fold design, and sampling]
\label{ass:factor_fold}
    The following conditions hold as $N,T\to\infty$.
    \begin{enumerate}[label=(\alph*)]
    \item The
    factor dimensions $r_D$ and $r_Y$ are positive integers that remain fixed.
    \item There exist constants \(0<c<C<\infty\) such that, with probability
    approaching one, all eigenvalues of
    \[
    \frac1N\sum_{i=1}^N \bar C_{D,i}\bar C_{D,i}',
    \quad
    \frac1N\sum_{i=1}^N \bar C_{Y,i}\bar C_{Y,i}',
    \quad
    \frac1T G_D^{0\prime}G_D^0,
    \quad
    \frac1T G^{0\prime}G^0
    \]
    lie in \([c,C]\).
\item The augmented loadings satisfy
\[
\frac1N\sum_{i=1}^N
\left(\|\bar C_{D,i}\|_2^4+\|\bar C_{Y,i}\|_2^4\right)=O_p(1).
\]
\item The ordered partition \((\mathcal I_1,\mathcal I_2,\mathcal I_3)\)
of \(\{1,\ldots,N\}\) is drawn uniformly at random from all such partitions
whose fold sizes differ by at most one. It is independent of the full
triangular array including the factor paths, primitive loadings, covariates,
and errors.
\item Conditional on \(\mathcal F_T\), the unit-level arrays
\(\{(X_i,\varepsilon_i,\nu_i)\}_{i=1}^N\) are independent across \(i\), and
the covariate paths \(X_i\) have a common conditional distribution. The
conditional distributions of \((\varepsilon_i,\nu_i)\) need not be identical
across units.
    \end{enumerate}
\end{assumption}

Assumption \ref{ass:factor_fold} collects the factor-structure and
sample-splitting requirements. Part (b) bounds the eigenvalues of the factor and
augmented-loading Gram matrices used by PCA without separately imposing
primitive-loading pervasiveness.
Part (c)
bounds average fourth moments, not the largest loading in the sample.

By the definition of $\bar C_i$,
\[
    \bar C_i
    =
    \bar C_{Y,i}
    -
    \theta_0
    (G^{0\prime}G^0)^{-1}G^{0\prime}G_D^0
    \bar C_{D,i}.
\]
Part (b) implies
\[
    \left\|
    (G^{0\prime}G^0)^{-1}G^{0\prime}G_D^0
    \right\|_{\op}=O_p(1).
\]
Since $\theta_0$ is fixed, part (c) therefore gives
\[
\frac1N\sum_{i=1}^N\|\bar C_i\|_2^4=O_p(1).
\]
Thus no separate moment restriction on \(\bar C_i\) is needed. Parts (d)--(e) specify balanced random folds and conditional cross-sectional independence. Cross-sectional dependence may enter through \(G_D^0\) and \(G^0\). Within-unit temporal dependence is allowed subject to the conditions below.

\begin{assumption}[Disturbance moments and training-fold PCA residuals]
\label{ass:training_main}
Parts~(a) and~(d) hold for each fold-role assignment $l$; parts~(b) and~(c) are stated over the full sample.
\begin{enumerate}[label=(\alph*)]
    \item The training-fold disturbances satisfy
    $E[\nu_{i}\mid \mathcal F_T,X_i]=0$ for \(i\in \mathcal I_1^{(l)}\) and
    $E[\xi_{i}\mid \mathcal F_T,X_i]=0$ for \(i\in \mathcal I_2^{(l)}\).
\item The corresponding conditional covariance operators are uniformly bounded:
\[
\sup_{j\leq N}
\left\|E[\nu_j\nu_j'\mid\mathcal F_T,X_j]\right\|_{\op}\le C,
\qquad
\textcolor{black}{\sup_{j\leq N}
\left\|E[\varepsilon_j\varepsilon_j'\mid\nu_j,\mathcal F_T,X_j]\right\|_{\op}\le C}
\]
almost surely.
\item The treatment disturbances have bounded full-sample average fourth moments:
\[
\frac1{NT^2}\sum_{j=1}^N\|\nu_j\|_2^4=O_p(1).
\]
\item Let \(V_1,\Delta_{g,1}\in\mathbb R^{n_1\times T}\) have rows
\(\nu_i'\) and \(\delta_i^{g\prime}\), respectively, indexed by
\(i\in\mathcal I_1^{(l)}\). Let
\(\Xi_2,\Delta_{m,2}\in\mathbb R^{n_2\times T}\) have rows
\(\xi_i'\) and \(\delta_i^{m\prime}\), respectively, indexed by
\(i\in\mathcal I_2^{(l)}\). The resulting residual matrices satisfy
\[
\|V_1-\Delta_{g,1}\|_{\op}
=O_p(\sqrt{n_1}+\sqrt T),
\qquad
\|\Xi_2-\Delta_{m,2}\|_{\op}
=O_p(\sqrt{n_2}+\sqrt T).
\]
\end{enumerate}
\end{assumption}
Assumption \ref{ass:training_main}(a)--(b) supplies the conditional means and
covariance-operator bounds for the PCA linear terms while allowing within-unit
serial dependence. Part (c) is used only for variance consistency. Part (d) permits data-dependent first-stage errors but rules out an additional factor-strength component in the residuals supplied to PCA. Together with the \(q_{NT}\)-rate conditions, part (d) makes the quadratic PCA remainder negligible.
Analogous oracle-residual bounds are standard in approximate factor models; see
\citet{bai2003inferential}.

For each split \(l\), define the evaluation-fold sigma field
\(\mathcal G_l=\sigma\!\bigl(\mathcal H_l,
\{X_i:i\in\mathcal I_3^{(l)}\}\bigr)\) and write $\Sigma_{\nu,i}=E[\nu_i\nu_i'\mid \mathcal G_l]$ and
$\Sigma_{\varepsilon,i}=E[\varepsilon_i\varepsilon_i'\mid \mathcal G_l]$. \textcolor{black}{By Assumptions~\ref{ass:factor_fold}(d)--(e) and
\ref{ass:training_main}(b), together with the independence
of auxiliary learner randomization, the evaluation-fold
conditional second moments satisfy, almost surely,
\[
\sup_{i\in\mathcal I_3^{(l)}}
\|\Sigma_{\nu,i}\|_{\op}\le C,
\qquad
\sup_{i\in\mathcal I_3^{(l)}}
\left\|
E[\varepsilon_i\varepsilon_i'
\mid\nu_i,\mathcal G_l]
\right\|_{\op}\le C.
\]
By iterated expectations, these bounds also imply
$\sup_{i\in\mathcal I_3^{(l)}}
\|\Sigma_{\varepsilon,i}\|_{\op}\le C$ almost surely.}

\begin{assumption}[Evaluation-fold moments]
\label{ass:eval_main}
The following conditions hold for each split $l$.
\begin{enumerate}[label=(\alph*)]
    \item The evaluation-fold disturbances satisfy the conditional mean restrictions
    \(E[\nu_i\mid \mathcal G_l]=0\) and
    \(E[\varepsilon_i\mid\nu_i,\mathcal G_l]=0\).
\item There exists \(\delta>0\) such that, almost surely,
\[
\frac1{n_3T^{1+\delta/2}}
\sum_{i\in\mathcal I_3^{(l)}}E\!\left[
|\nu_i'A_{0}\varepsilon_i|^{2+\delta}
\mid \mathcal G_l
\right]
\le C.
\]
    \item There exist constants \(\Omega_0>0\) and \(\kappa_0>0\), common to
    all splits and permutations, such that
\[
\frac{1}{n_3T}\sum_{i\in \mathcal I_3^{(l)}} E\!\left[(\nu_i'A_0\varepsilon_i)^2\mid \mathcal G_l\right] \xrightarrow{p} \Omega_0,
\]
and
\[
    \frac{1}{n_3T}
    \sum_{i\in\mathcal I_3^{(l)}}
    \nu_i'A_0\nu_i
    \xrightarrow{p}
    \kappa_0.
\]

\end{enumerate}
\end{assumption}

Assumption \ref{ass:eval_main} imposes high-level conditional-moment conditions
on the held-out evaluation fold, conditional on the sigma field
\(\mathcal G_l\), which collects the population factor space, the training-fold
information and the evaluation-fold covariates. Part (a) is the score-relevant
conditional exogeneity restriction; it does not require conditional independence
between \(\nu_i\) and \(\varepsilon_i\).
Part
(b) supplies the Lyapunov condition for the conditional central limit theorem (CLT).
Part (c) specifies the positive limits of the normalized
score variance and the ordered-score denominator, which determine
\(\Omega_0\) and \(\kappa_0\).

Assumption \ref{ass:eval_main} does not cover generic dynamic-panel
specifications with predetermined regressors. Extending the FA-DML framework to dynamic panels would
require a separate treatment of predetermined regressors, the resulting
additional score terms, and the interaction between dynamic endogeneity and
factor estimation. We leave this extension for future research.

\begin{proposition} \label{proposition:1st_stage}
    Suppose Assumptions \ref{ass:1st_stage}, \ref{ass:factor_fold}(d)--(e), \ref{ass:training_main}(b), and \ref{ass:eval_main}(a) hold. Then, uniformly over $l$,
    \begin{align*}
        \frac{1}{\sqrt{Tn_3}}
        \sum_{i\in \mathcal{I}_3^{(l)}}R_{1i}
        =
        O_p\!\left(
        \bar{r}_{g}
        +
        \bar{r}_{m}
        +
        \sqrt{Tn_3}\,
        \bar{r}_{g}
        \left(
        \bar{r}_{m}+\bar{r}_{g}
        \right)
        \right)
        =
        o_p(1).
    \end{align*}
\end{proposition}

The proof of Proposition~\ref{proposition:1st_stage} shows that the prediction
rates carry over to the evaluation fold and make the
nuisance-error remainder negligible.
The first-stage learners may be chosen flexibly, provided their
final feasible estimates satisfy the stated prediction,
loading-weighted, and residual conditions.

\subsection{Main Inference Results}
We first state the asymptotic distribution of the FA-DML estimator when the factor dimensions $r_D$ and $r_Y$ are treated as fixed and known. Rank selection is discussed in Section~\ref{subsec:rank-selection}.
Assumption~\ref{ass:identification} gives the population identification
conditions.  For the inference results below,
exogeneity and asymptotic nondegeneracy are imposed directly
through the fold-specific conditions in
Assumptions~\ref{ass:training_main} and~\ref{ass:eval_main}.

\begin{theorem}[Asymptotic normality of the estimator]
\label{thm:CLT}
Suppose Assumption \ref{ass:1st_stage}, Assumption
\ref{ass:rate-main}(a), Assumption \ref{ass:factor_fold}, Assumption
\ref{ass:training_main}(a),(b),(d), and Assumption \ref{ass:eval_main} hold.
Then
\[
\sqrt{NT}\,(\hat\theta-\theta_0)
\;\xrightarrow{d}\;
N\!\left(0,\sigma_\theta^2\right),
\]
where
\(
\sigma_\theta^2=\kappa_0^{-2}\Omega_0
\),
and $\kappa_0$ and $\Omega_0$ are defined in Assumption~\ref{ass:eval_main}(c).
\end{theorem}

Theorem \ref{thm:CLT} follows from the expansion
\[
\sqrt{NT}(\hat\theta-\theta_0)
=
\kappa_0^{-1}
\frac{1}{\sqrt{NT}}\sum_{i=1}^N \nu_i'A_0\varepsilon_i
+o_p(1).
\]
This expansion establishes first-order equivalence between
FA-DML and an infeasible oracle estimator that knows the
nuisance functions and factor spaces. Under the stated rate conditions, Neyman orthogonality and cross-fitting make
the nuisance-estimation remainders negligible. Under the same conditions, double projection and separate
training folds do the same for the factor-estimation remainders.

The resulting CLT permits valid inference without
analytical bias correction. This contrasts with the
interactive-effects least-squares framework of
\citet{bai2009panel}, where heteroskedasticity and serial
correlation can yield asymptotic bias.

\begin{theorem}[Consistency of the variance estimator]
\label{thm:var-consistency}
Under the conditions of Theorem \ref{thm:CLT}, if Assumption
\ref{ass:rate-main}(b) and Assumption \ref{ass:training_main}(c) also hold,
\[
\hat\sigma_\theta^2 \xrightarrow{p} \sigma_\theta^2,
\]
where \(\sigma_\theta^2\) is defined in Theorem \ref{thm:CLT}.

\end{theorem}
Theorem \ref{thm:var-consistency} complements Theorem \ref{thm:CLT} by showing
that inference is fully feasible. The estimator \(\hat\sigma_\theta^2\) uses estimated nuisance functions and
estimated factor spaces. Consistency of this plug-in estimator follows from
the orthogonalization argument, together with the loading-weighted squared-error
bounds in Assumption~\ref{ass:rate-main}(b) and the fourth-moment condition in
Assumption~\ref{ass:training_main}(c). In particular, the
foldwise quadratic forms used in \(\hat\sigma_\theta^2\) are asymptotically
equivalent to their infeasible counterparts based on \(\nu_i\), \(\varepsilon_i\),
and \(A_0\). Hence the numerator consistently estimates \(\Omega_0\), while the
denominator consistently estimates \(\kappa_0\). Together, Theorems
\ref{thm:CLT} and \ref{thm:var-consistency} justify confidence intervals of the
form
\[
\hat\theta \pm z_{1-\alpha/2}\frac{\hat\sigma_\theta}{\sqrt{NT}},
\]
where \(z_{1-\alpha/2}\) is the \((1-\alpha/2)\)-quantile of the standard normal
distribution.

\subsection{Rank Selection}
\label{subsec:rank-selection}

Theorems~\ref{thm:CLT} and~\ref{thm:var-consistency} treat
$r_D$ and $r_Y$ as known. We estimate these dimensions using
an alternating procedure that combines first-stage learning
with nuclear-norm penalization. The complete procedure appears
in Algorithm~\ref{alg:rank_estimation} in
\ref{app:rank-selection-algorithm}. Here we define
its terminal outputs and establish rank consistency and
the validity of subsequent inference.

Let \(D\) and \(Y\) denote the \(N\times T\) matrices whose \(i\)-th rows are
\(D_i'\) and \(Y_i'\), respectively, and write \(X=(X_1,\ldots,X_N)\). For any
vector-valued learner \(h\), let \(h(X)\) denote the \(N\times T\) matrix whose
\(i\)-th row is \(h(X_i)'\). The primitive decompositions used for rank
selection are
\[
    D=g_0(X)+\Theta_D+V,
    \qquad
    \Theta_D=C_DG_D^{0\prime},
\]
and
\[
    Y=m_0(X)+\Theta_Y+\Xi,
    \qquad
    \Theta_Y=CG^{0\prime},
\]
where the \(i\)-th rows of \(C_D\), \(C\), \(V\), and \(\Xi\) are
\(C_{D,i}'\), \(C_i'\), \(\nu_i'\), and \(\xi_i'\), respectively.

Let \(\hat g^{(nn)}\) and \(\hat m^{(nn)}\) denote the terminal treatment and
reduced-form learners used by the rank selector;
\ref{app:rank-selection-algorithm} gives their complete construction. Define
the effective residual perturbations
\[
    \widetilde V_D
    =
    V-\{\hat g^{(nn)}(X)-g_0(X)\},
\]
and
\[
    \widetilde \Xi_Y
    =
    \Xi-\{\hat m^{(nn)}(X)-m_0(X)\}.
\]

For positive tuning sequences \(\rho_D,\rho_Y\), define the terminal low-rank
estimates by
\[
\hat\Theta_D
=
\argmin_{\Theta\in\mathbb R^{N\times T}}
\left\{
\frac12\|D-\hat g^{(nn)}(X)-\Theta\|_F^2
+
\rho_D\|\Theta\|_*
\right\}.
\]
Define \(\hat\Theta_Y\) by the same criterion using \(Y\),
\(\hat m^{(nn)}\), and \(\rho_Y\).
For a fixed \(\mu\in(-1/2,1/2)\) and \(e\in\{D,Y\}\), define
\[
\hat r_e
=
\begin{cases}
0, & s_1(\hat\Theta_e)=0,\\[2mm]
\displaystyle
\sum_{j=1}^{\min\{N,T\}}
\mathbf 1
\left\{
s_j(\hat\Theta_e)
>
\rho_e^{1/2-\mu}s_1(\hat\Theta_e)^{1/2+\mu}
\right\},
&s_1(\hat\Theta_e)>0.
\end{cases}
\]
Once
the ranks are selected, Algorithm~\ref{algorithm_three_fold} continues to use
the projected first-stage targets and double-projection score from
Section~\ref{Section_algorithm}.

\begin{assumption}[Low-rank signal and final residual order]
\label{assp:rank_consistency}
\leavevmode\par
\begin{enumerate}[label=(\alph*)]
\item For each \(e\in\{D,Y\}\),
there exist constants
\(0<c_e<C_e<\infty\) such that
\[
P\!\left(
\operatorname{rank}(\Theta_e)=r_e,
\quad
c_e\sqrt{NT}\le s_{r_e}(\Theta_e)
\le s_1(\Theta_e)\le C_e\sqrt{NT}
\right)\to1.
\]
\item The positive, possibly random tuning sequences \(\rho_D\) and \(\rho_Y\)
satisfy
\[
\frac{\|\widetilde V_D\|_{\op}}{\rho_D}=O_p(1),
\qquad
\frac{\|\widetilde \Xi_Y\|_{\op}}{\rho_Y}=O_p(1),
\]
and
\[
    \rho_D=o_p(\sqrt{NT}),
    \qquad
    \rho_Y=o_p(\sqrt{NT}).
\]
\end{enumerate}
\end{assumption}
Assumption \ref{assp:rank_consistency}(a) imposes the signal separation used only
for rank selection. It follows, for example, if the minimum and
maximum eigenvalues of \(N^{-1}C_D'C_D\), \(N^{-1}C'C\),
\(T^{-1}G_D^{0\prime}G_D^0\), and \(T^{-1}G^{0\prime}G^0\) are bounded away
from zero and infinity with probability approaching one. Those primitive Gram
conditions are sufficient, not separately required by the rank theorem. Part
(b) requires the effective residual perturbation to
be no larger in stochastic order than \(\rho_D\) or \(\rho_Y\), while these
sequences remain asymptotically smaller than the \(\sqrt{NT}\)-order signal.
Unlike a hard domination inequality, this is a simple sufficient separation
condition for the proof. It also rules out a final learner that absorbs a leading low-rank component.
Its relation to first-stage prediction rates is discussed in
\ref{app:decomposition}.

\begin{theorem}[Consistency of rank selection]
\label{thm:rank_consistency_hl}
Suppose Assumption \ref{ass:factor_fold}(a) and Assumption
\ref{assp:rank_consistency} hold.
Then, for any fixed \(\mu\in(-1/2,1/2)\),
\[
P(\hat r_D=r_D)\to 1,
\quad \text{and} \quad
P(\hat r_Y=r_Y)\to 1.
\]
\end{theorem}

Theorem \ref{thm:rank_consistency_hl} formalizes the singular-value separation
behind the rank selector. For the treatment equation, the final residual matrix
satisfies
\[
D-\hat g^{(nn)}(X)=\Theta_D+\widetilde V_D.
\]
Assumption \ref{assp:rank_consistency} implies
\(\|\widetilde V_D\|_{\op}=o_p(\sqrt{NT})\), while the nonzero singular values of
\(\Theta_D\) are of order \(\sqrt{NT}\). If the residual perturbation is
\(O_p(\rho_D)\) and \(\rho_D=o_p(\sqrt{NT})\), the nuclear-norm step estimates
\(\Theta_D\) with operator-norm error of order \(\rho_D\). Hence the
estimated singular values corresponding to the true factors remain of order
\(\sqrt{NT}\), whereas the remaining singular values are of order at most
\(\rho_D\). The threshold
\(
\rho_D^{1/2-\mu}s_1(\hat\Theta_D)^{1/2+\mu}
\)
lies between these two orders, yielding \(P(\hat r_D=r_D)\to 1\). The same
argument applies to the outcome equation.

This argument parallels the singular-value separation logic used in
nuclear-norm penalized low-rank estimation and factor-number selection
\citep[e.g.,][]{koltchinskiilounicitsybakov2011,bai2002determining,ahn2013}.
Here the residual matrices
also contain errors from the estimated nuisance functions.
The terminal residual bounds ensure that these errors
preserve the separation needed for rank recovery.

Let \(\widehat\theta_{\widehat r}\) and
\(\widehat\sigma_{\theta,\widehat r}^2\) be obtained by Algorithm \ref{algorithm_three_fold} with
\((r_D,r_Y)\) replaced by the rank estimators \((\hat r_D,\hat r_Y)\)
from Algorithm \ref{alg:rank_estimation}. Use a fixed measurable convention for
SVD and eigenspace ties; when a selected rank is zero, take its projection to
be zero and its residual-maker to be \(I_T\). If a selected rank exceeds a
corresponding training-fold matrix dimension, or if \(\widehat J_l=0\) for any
split, set
\(\widehat\theta_{\widehat r}=0\) and
\(\widehat\sigma_{\theta,\widehat r}^2=0\). Under the assumptions below, this fallback occurs with probability tending to zero.

\begin{corollary}[Inference with estimated ranks]
\label{cor:estimated-ranks}
Under the assumptions of Theorems \ref{thm:CLT}--\ref{thm:rank_consistency_hl},
\[
\sqrt{NT}\,(\widehat\theta_{\widehat r}-\theta_0)\ \xrightarrow{d}\ N(0,\sigma_\theta^2),
\quad \text{and} \quad
\widehat\sigma_{\theta,\widehat r}^2\ \xrightarrow{p}\ \sigma_\theta^2,
\]
where \(\sigma_\theta^2\) is defined in Theorem~\ref{thm:CLT}.
\end{corollary}
Corollary \ref{cor:estimated-ranks} shows that preliminary rank selection does
not affect first-order inference when the ranks are selected consistently. On
the correct-rank event, the estimated-rank procedure coincides with the
oracle-rank procedure. Because this event has probability approaching one, the
limiting distribution and variance consistency carry over to the estimated-rank
procedure.

\section{Simulation Study} \label{sec:sim}
We use Monte Carlo experiments to assess coverage of
FA-DML with known and data-selected factor dimensions. We compare FA-DML with a two-fold variant, standard DML without factor augmentation, and an OLS first-stage benchmark. These comparisons examine the finite-sample roles of sample separation, factor adjustment, and regularization. We also report the rank-selection frequencies.

\subsection{Simulation Design and Implementation}\label{subsec:sim-design}
We consider panels with
\((N,T)\in\{100,200\}\times\{50,100,200,300\}\). The DGP
follows the model in Section~\ref{sub:partial_linear_panel} and takes the form
\[
Y_i=D_i\theta_0+f_0(X_i)+G_Y^0C_{Y,i}^*+\varepsilon_i,
\qquad
D_i=g_0(X_i)+G_D^0C_{D,i}+\nu_i,
\]
where \(\theta_0=0.5\), \(X_i\in\R^{T\times p}\), and \(p=30\). Let
\(\Sigma_d\) denote the Toeplitz covariance matrix of dimension \(d\), with
\((s,t)\) entry \(0.5^{|s-t|}\), and let
\(\beta=(1,0.5,\ldots,0.5^{p-1})'\). We consider linear nuisance functions,
\(f_0(X_i)=g_0(X_i)=X_i\beta\), and quadratic nuisance functions,
\(f_0(X_i)=g_0(X_i)=X_i\beta+(X_i\circ X_i)\beta\), where \(\circ\)
is the Hadamard product.

We let the treatment and structural
outcome factor spaces share two of their three directions. Specifically, we set
\[
    G_D^{0} = \big[\, G^{c} \;\; G^{d}\, \big],
    \qquad
    G_Y^{0} = \big[\, G^{c} \;\; G^{y}\, \big],
\]
where \(G^c\in\R^{T\times2}\) is common and
\(G^d,G^y\in\R^T\) are equation-specific.
In this baseline design with \(\theta_0=0.5\),
\(r_D=r_Y^*=3\), while the reduced-form factor space is
spanned by \(G^0=[G^c,G^y,G^d]\), with \(r_Y=4\).
We draw
\(G^0=\Sigma_T^{1/2}\widetilde G\Sigma_4^{1/2}\),
where \(\widetilde G\) has i.i.d.\ \(N(0,0.75^2)\) entries.

We write
\(C_{D,i}=({\Lambda_{D,i}^c}',B_{D,i})'\) and
\(C_{Y,i}^*=({\Lambda_{Y,i}^c}',B_{Y,i})'\). We stack the loadings by unit
and draw
\[
    \Lambda_{Y}^{c} = \Sigma_N^{1/2}\, \tilde\Lambda_{Y}^{c},
    \qquad
    \Lambda_{D}^{c} = \Lambda_{Y}^{c} + \Sigma_N^{1/2}\, \tilde\Lambda_{D}^{c},
    \qquad
    B_Y = \Sigma_N^{1/2}\, \tilde B_Y,
    \qquad
    B_D = \Sigma_N^{1/2}\, \tilde B_D,
\]
where \(\widetilde\Lambda_Y^c\) has i.i.d.\ \(N(0,1)\) entries,
\(\widetilde B_Y,\widetilde B_D\) have i.i.d.\ \(N(0,2)\) entries, and
\(\widetilde\Lambda_D^c\) has i.i.d.\ \(N(0,0.25)\) entries. These arrays
are mutually independent and independent of \(G^0\). Sharing
\(\Lambda_Y^c\) in \(\Lambda_D^c\) induces treatment--outcome confounding
along the common factor directions.

The \(k\)-th covariate path, \(X_{ik}\in\R^T\), combines the two shared
factor directions and an idiosyncratic Gaussian component according to
\[
    X_{ik} = \frac{1}{2} \sum_{q=1}^{2} 0.5^{\,k-1}\, G^c_q
    + \Sigma_T^{1/2}\, U_{x,ik},
    \qquad k = 1, \dots, p,
\]
where \(G^c_q\) is the \(q\)-th column of \(G^c\), and the \(U_{x,ik}\) have
i.i.d.\ \(N(0,0.75)\) entries and are independent of the factors and loadings.
This makes \(X_i\) correlated with both factor spaces but leaves a shared
interactive component after covariate residualization.

Finally, the disturbances are conditionally heteroskedastic across units,
\[
    \varepsilon_i = \sigma_{\varepsilon, i}\, \tilde\varepsilon_i,
    \qquad
    \nu_i = \sigma_{\nu, i}\, \Sigma_T^{1/2}\, \tilde\nu_i,
\]
where \(\sigma_{\varepsilon,i}^2,\sigma_{\nu,i}^2\overset{\mathrm{i.i.d.}}{\sim}
\mathrm{Unif}[0.5,2]\) are drawn once for each \((N,T)\) design and held fixed
across Monte Carlo replications. The entries of \(\widetilde\varepsilon_i\) are
i.i.d.\ \(N(0,1)\), whereas those of \(\widetilde\nu_i\) are i.i.d.\
\(N(0,0.75)\). These variables are independent across units and of all other
primitives. Thus \(\nu_i\) is serially correlated, whereas
\(\varepsilon_i\) is not. Cross-sectional dependence operates through the
common factors and \(\Sigma_N\)-correlated loadings.

All first stages use the same additive cubic dictionary containing the first,
second, and third powers of each covariate, without cross-products. Its
dimension is \(\widetilde p=90\). Both nuisance specifications lie in the span
of this dictionary before factor projection. The quadratic design therefore
introduces nonlinearity without dictionary approximation error. Each regularized
specification is implemented with both Lasso and Elastic Net. OLS provides
an unregularized first-stage benchmark.

For both Lasso and Elastic Net, we select the penalty coefficient separately at each regularized regression update using three-fold cross-validation within the relevant training sample. For Elastic Net, we fix the mixing parameter at \(\omega=1/2\). \ref{app:additional-sim} provides additional details of the simulation setting.

We implement three-fold FA-DML using Algorithm~\ref{algorithm_three_fold}
with either known or selected ranks. Complete unit histories are randomly partitioned into three folds of sizes
differing by at most one. Estimates are averaged over all six assignments of
the treatment-training, outcome-training, and score-evaluation roles. The known-rank implementation
uses \((r_D,r_Y)=(3,4)\) but still estimates the factor spaces and nuisance
functions. In the selected-rank implementation, the iterative nuclear-norm procedure
selects the two ranks on the full sample once per replication. These ranks
are then held fixed across the six fold-role assignments.

To examine the role of sample splitting, we also consider a known-rank
two-fold variant. Units are split into two equal folds. Both nuisance
functions and factor spaces are estimated on one fold, and the
double-projection score is evaluated on the other. The estimates are
averaged after exchanging the fold roles. Relative to known-rank three-fold
FA-DML, this changes both training-sample separation and size, so the
comparison does not isolate separation at equal training-sample sizes.
Standard DML uses the same two-fold arrangement but omits factor estimation
and sets \(\widehat A=I_T\). This comparison holds the splitting scheme fixed.

To examine first-stage regularization, the OLS first-stage benchmark replaces
both regularized nuisance fits with OLS using the same cubic dictionary.
It retains the true ranks, factor augmentation, double-projection score,
and three-fold cross-fitting of known-rank FA-DML. Additional implementation
details are given in \ref{app:additional-sim}.

\subsection{Inference Results}\label{subsec:sim-main-results}
\begin{figure}[!htb]
    \centering
    \includegraphics[width=\linewidth]{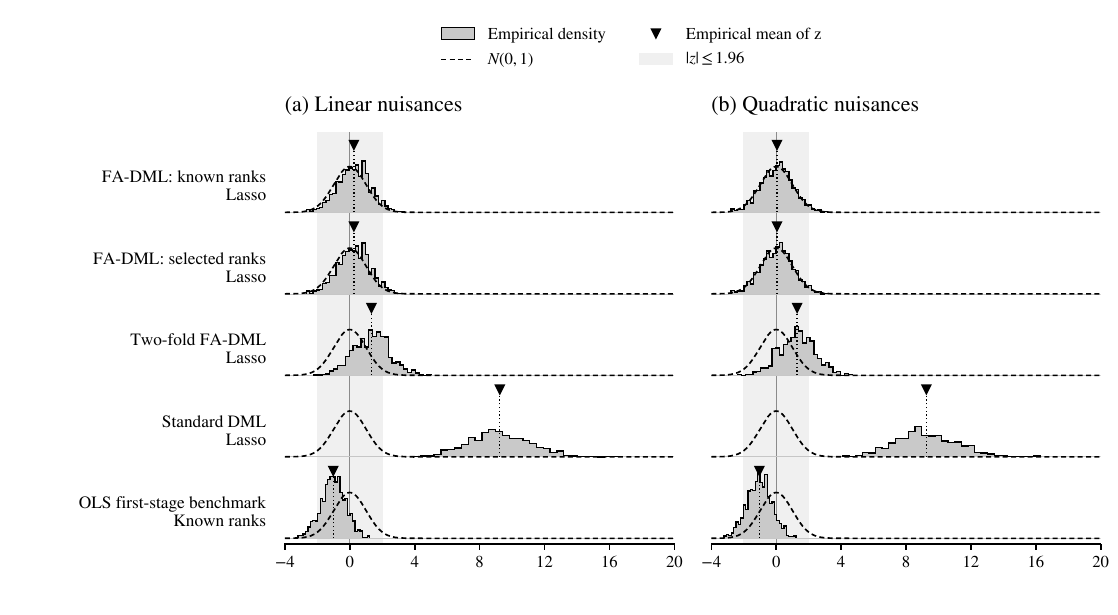}
    \begingroup
    \setstretch{1.0}
    \footnotesize
    \color{black}
    \captionsetup{font={footnotesize,stretch=1,color=black},labelfont={color=black}}
    \caption{Empirical distributions of
    \(z=\sqrt{NT}(\hat\theta-\theta_0)/\hat\sigma_\theta\) at
    \((N,T)=(100,100)\), with \(\theta_0=0.5\), based on \(1{,}000\) Monte
    Carlo replications. The estimated standard error is
    \(\hat\sigma_\theta/\sqrt{NT}\). Panels (a) and (b) use linear and
    quadratic nuisance functions, respectively. The first four rows use
    Lasso. Known-rank FA-DML, its two-fold variant, and the OLS first-stage
    benchmark use the true ranks \((r_D,r_Y)=(3,4)\). Selected-rank FA-DML
    selects both ranks once per replication on the full sample and keeps them
    fixed across its three-fold fits. The OLS benchmark retains three-fold
    cross-fitting and factor augmentation but uses unregularized first stages.
    Standard DML uses two folds and omits factor augmentation.
    Histograms have \(30\) bins. Dashed curves show the standard normal
    density. Triangles and dotted lines mark the mean of \(z\), and
    shading denotes \(|z|\le1.96\). Horizontal scales are identical, and heights
    above each row's baseline use a common density scale.
    Figure~\ref{fig:sim_clt_full} adds Elastic Net.}
    \label{fig:sim_studentized_main}
    \endgroup
\end{figure}

Figure~\ref{fig:sim_studentized_main} compares the studentized statistic
\(z=\sqrt{NT}(\hat\theta-\theta_0)/\hat\sigma_\theta\) in the
design with \((N,T)=(100,100)\). The comparison includes known-rank and
selected-rank three-fold FA-DML, the known-rank two-fold variant, standard
DML, and the OLS first-stage benchmark. The first four methods use Lasso, and all five are evaluated on the same simulated panels within each nuisance
design. All distributions use the same horizontal and density scales.
Figure~\ref{fig:sim_clt_full} in
\ref{app:additional-sim} adds the Elastic Net implementations.

For known-rank FA-DML, the studentized distribution is close to the
standard normal benchmark, with a modest rightward shift under linear
nuisance functions and a mean close to zero under quadratic nuisance functions.
The selected-rank distribution nearly
overlaps it in both panels. At this sample size the selected pair is correct
in \(98.4\%\) of linear and \(98.5\%\) of quadratic replications.

The distribution for the two-fold variant is shifted to the right. This
pattern is consistent with the
same-training-fold interaction discussed in Section~\ref{subsection:2.3},
although the comparison does not identify that interaction separately from
the change in training-sample sizes. The distribution for standard DML is
displaced far to the right in this design, which retains interactive
confounding after covariate adjustment. For the OLS first-stage benchmark,
the studentized statistic is shifted to the left of zero and is less
dispersed than the standard normal under
both nuisance specifications. The benchmark's undercoverage at this
sample size is thus
accompanied by a leftward displacement of the studentized statistic.  The complete figure in the appendix shows similar qualitative
patterns with Elastic Net.

Table~\ref{tab:sim_cr} reports coverage of nominal $95\%$
confidence intervals across the sample-size grid. Coverage
for known-rank and selected-rank FA-DML ranges from
$93.4\%$ to $95.9\%$ in the linear design and from
$92.6\%$ to $95.8\%$ in the quadratic design. Undercoverage
in the quadratic design is concentrated in shorter panels,
particularly with Elastic Net. Selected-rank coverage differs
from known-rank coverage by at most $0.7$ percentage points.
At $(N,T)=(100,100)$, the corresponding Lasso coverages are
$94.0\%$ and $93.9\%$ in the linear design and $95.3\%$
and $95.2\%$ in the quadratic design.

The two-fold variant undercovers substantially.
Its coverage
improves with $N$ at each fixed $T$ and generally declines
as $T$ increases at fixed $N$. Standard DML has zero
empirical coverage throughout the reported grid. OLS coverage
increases with $T$, but undercoverage in shorter panels can
give way to overcoverage: at $N=100$, coverage rises from
$65.8\%$ to $99.4\%$ in the linear design and from
$63.2\%$ to $99.5\%$ in the quadratic design as $T$
increases from $50$ to $300$. Across this baseline grid,
regularized three-fold FA-DML is closer to nominal coverage
than the benchmark methods.

\begin{table}[H]
    \centering
        \captionsetup{font={footnotesize,stretch=1},labelfont={color=black},justification=justified,
        singlelinecheck=false,width=\textwidth,skip=6pt}
    \setlength{\tabcolsep}{3.6pt}
    \renewcommand{\arraystretch}{1.04}
    \footnotesize
    \begin{tabular}{cc*{9}{c}}
    \toprule
    \multirow{2}{*}{\(N\)} & \multirow{2}{*}{\(T\)}
                  & \multicolumn{4}{c}{Lasso}
                  & \multicolumn{4}{c}{Elastic Net}
                  & \multicolumn{1}{c}{\multirow{2}{*}{OLS}} \\
    \cmidrule(lr){3-6}\cmidrule(lr){7-10}
          &
          & \multicolumn{1}{c}{Known ranks}
          & \multicolumn{1}{c}{Selected ranks}
          & \multicolumn{1}{c}{Two-fold}
          & \multicolumn{1}{c}{DML}
          & \multicolumn{1}{c}{Known ranks}
          & \multicolumn{1}{c}{Selected ranks}
          & \multicolumn{1}{c}{Two-fold}
          & \multicolumn{1}{c}{DML}
          & \\
    \midrule
    \multicolumn{11}{l}{\textit{Panel A. Linear nuisance functions}} \\
    100 & 50  & 95.5 & 95.1 & 79.6 & 0.0 & 95.2 & 95.4 & 82.7 & 0.0 & 65.8 \\
        & 100 & 94.0 & 93.9 & 71.9 & 0.0 & 95.1 & 95.1 & 74.9 & 0.0 & 88.6 \\
        & 200 & 94.6 & 94.6 & 61.7 & 0.0 & 94.7 & 94.7 & 63.0 & 0.0 & 96.4 \\
        & 300 & 94.0 & 94.0 & 51.7 & 0.0 & 94.3 & 94.3 & 54.8 & 0.0 & 99.4 \\
    \addlinespace[2pt]
    200 & 50  & 94.3 & 94.3 & 86.5 & 0.0 & 93.4 & 93.6 & 85.9 & 0.0 & 74.7 \\
        & 100 & 95.0 & 95.0 & 85.1 & 0.0 & 94.1 & 94.1 & 86.3 & 0.0 & 85.0 \\
        & 200 & 95.9 & 95.9 & 81.4 & 0.0 & 95.5 & 95.5 & 81.6 & 0.0 & 93.4 \\
        & 300 & 94.4 & 94.4 & 73.0 & 0.0 & 95.0 & 95.0 & 72.6 & 0.0 & 95.9 \\
    \midrule
    \multicolumn{11}{l}{\textit{Panel B. Quadratic nuisance functions}} \\
    100 & 50  & 95.7 & 95.8 & 81.5 & 0.0 & 93.9 & 93.4 & 82.8 & 0.0 & 63.2 \\
        & 100 & 95.3 & 95.2 & 72.7 & 0.0 & 93.8 & 93.7 & 75.0 & 0.0 & 87.8 \\
        & 200 & 94.6 & 94.6 & 63.4 & 0.0 & 95.2 & 95.2 & 64.2 & 0.0 & 96.9 \\
        & 300 & 94.4 & 94.4 & 52.5 & 0.0 & 95.3 & 95.2 & 54.9 & 0.0 & 99.5 \\
    \addlinespace[2pt]
    200 & 50  & 93.8 & 94.5 & 86.3 & 0.0 & 92.6 & 92.7 & 86.9 & 0.0 & 73.7 \\
        & 100 & 94.9 & 94.9 & 85.8 & 0.0 & 93.8 & 93.8 & 85.8 & 0.0 & 85.1 \\
        & 200 & 95.5 & 95.5 & 81.8 & 0.0 & 95.7 & 95.7 & 81.4 & 0.0 & 93.3 \\
        & 300 & 94.9 & 94.9 & 72.2 & 0.0 & 95.1 & 95.1 & 73.4 & 0.0 & 96.7 \\
    \bottomrule
    \end{tabular}
    \caption[Coverage of nominal \(95\%\) confidence intervals.]{Coverage of
    nominal \(95\%\) confidence intervals. Entries are percentages of
    \(1{,}000\) replications covering \(\theta_0=0.5\). Known ranks and Selected
    ranks denote three-fold FA-DML with true ranks \((r_D,r_Y)=(3,4)\) and
    data-selected ranks, respectively. Both estimate factor spaces and nuisance functions.
    Two-fold uses known ranks and common training units for both nuisance
    functions and factor spaces. DML uses two-fold cross-fitting without
    factor augmentation. OLS uses unregularized first stages within known-rank
    FA-DML\@. All methods use the same \(90\)-term cubic dictionary. The Monte
    Carlo standard error at \(95\%\) coverage is approximately \(0.69\)
    percentage points.}
    \label{tab:sim_cr}
\end{table}

\subsection{Rank-Selection Performance}\label{subsec:sim-rank-results}

We evaluate the rank selector using the
same DGP and sample-size grid.  Table~\ref{tab:sim_rank} reports the marginal frequencies
of correctly selecting $r_D=3$ and $r_Y=4$. Each rank is recovered in at least \(95.0\%\) of
replications in every design, in at least \(98.4\%\) when \(T\ge100\), and in
all replications when \(T\ge200\). Performance is similar across Lasso and
Elastic Net and across the two nuisance specifications.

\begin{table}[!htb]
  \centering
      \captionsetup{font={footnotesize,stretch=1},labelfont={color=black},justification=justified,
      singlelinecheck=false,width=\textwidth,skip=6pt}
  \setlength{\tabcolsep}{8pt}
  \renewcommand{\arraystretch}{1.05}
  \small
  \begin{tabular}{ccrrrr}
    \toprule
          &       & \multicolumn{2}{c}{Lasso}
                  & \multicolumn{2}{c}{Elastic Net} \\
    \cmidrule(lr){3-4}\cmidrule(lr){5-6}
    \(N\) & \(T\)
          & \multicolumn{1}{c}{Treatment}
          & \multicolumn{1}{c}{Outcome}
          & \multicolumn{1}{c}{Treatment}
          & \multicolumn{1}{c}{Outcome} \\
    \midrule
    \multicolumn{6}{l}{\textit{Panel A. Linear nuisance functions}} \\
    100 & 50  & 95.0 & 95.4 & 95.4 & 95.5 \\
        & 100 & 98.4 & 99.9 & 98.5 & 99.9 \\
        & 200 & 100.0 & 100.0 & 100.0 & 100.0 \\
        & 300 & 100.0 & 100.0 & 100.0 & 100.0 \\
    \addlinespace[2pt]
    200 & 50  & 99.6 & 97.2 & 99.6 & 97.3 \\
        & 100 & 100.0 & 100.0 & 100.0 & 100.0 \\
        & 200 & 100.0 & 100.0 & 100.0 & 100.0 \\
        & 300 & 100.0 & 100.0 & 100.0 & 100.0 \\
    \midrule
    \multicolumn{6}{l}{\textit{Panel B. Quadratic nuisance functions}} \\
    100 & 50  & 95.3 & 95.5 & 95.5 & 95.4 \\
        & 100 & 98.5 & 100.0 & 98.6 & 100.0 \\
        & 200 & 100.0 & 100.0 & 100.0 & 100.0 \\
        & 300 & 100.0 & 100.0 & 100.0 & 100.0 \\
    \addlinespace[2pt]
    200 & 50  & 99.6 & 97.2 & 99.6 & 97.3 \\
        & 100 & 100.0 & 100.0 & 100.0 & 100.0 \\
        & 200 & 100.0 & 100.0 & 100.0 & 100.0 \\
        & 300 & 100.0 & 100.0 & 100.0 & 100.0 \\
    \bottomrule
  \end{tabular}
  \caption[Correct factor-rank selection.]{Correct factor-rank selection.
  Entries report marginal, not joint, recovery frequencies  over
  \(1{,}000\) replications. Treatment and Outcome correspond to \(r_D=3\)
  and \(r_Y=4\), respectively. Both nuisance specifications use the same
  \(90\)-term cubic dictionary and
  \(\rho_D=\rho_Y=\max\{\sqrt N,\sqrt T\}\). }
  \label{tab:sim_rank}
\end{table}

\section{Minimum Wages and Lower-Tail Wage Inequality}
\label{sec:empirical}
\label{sec:ams}
We study whether the negative relationship between minimum
wages and lower-tail wage inequality persists after flexible
adjustment for observed workforce composition and latent
interactive heterogeneity. States may respond differently
to common economic shocks, and these responses may be related
to both minimum-wage policy and wage inequality. FA-DML
accommodates this heterogeneity beyond additive state and
year effects. Appendix~\ref{app:banking} presents an additional
application to branching deregulation and income inequality.

\citet{autor2016contribution} find that minimum wages compress the lower tail
of the U.S. wage distribution. We revisit this relationship using a panel of
all \(50\) states from \(1979\) to \(2012\), with \(1{,}700\) state-year
observations. We use their state-level wage percentiles and minimum wages, constructed from
the Current Population Survey Merged Outgoing Rotation Groups.
 Our outcome is
\[
Y_{it}=\log w_{it}(50)-\log w_{it}(10),
\]
where \(w_{it}(q)\) is the \(q\)-th percentile of wages in state \(i\) and
year \(t\). The treatment \(D_{it}\) is the log real effective state minimum
wage. Thus \(\theta_0\) is the elasticity of the 50/10 wage ratio with respect to the real minimum wage.

Unlike \citet{autor2016contribution}, we do not measure the treatment as the
minimum wage's ``bite'' relative to the state median. Because their outcome is also expressed relative to the median, sampling error in the median enters both sides of the regression, a division-bias problem \citep{borjas1980division} that they address with instrumental variables. Using the minimum-wage level avoids this shared component, so our linear specification targets a different parameter from their nonlinear specification in the bite.

The original control dictionary contains four state-year characteristics:
mean age, female employment share, non-white employment share, and the log sum of the hours-adjusted CPS earnings weights. The expanded dictionary contains
\(32\) features obtained from a degree-four polynomial
expansion after collinear terms are removed. These controls allow observed
composition to affect wages flexibly. Interactive effects allow common shocks, such as business-cycle fluctuations, to affect states differently. Such heterogeneity may be correlated with both minimum-wage policy and wage inequality and is not absorbed by additive state and year effects.

We estimate the first stages by Lasso and by random forests with each
dictionary. The Lasso penalty is a multiple of \(\sqrt{\log p/(NT)}\), with the multiplier chosen from \([0.01,10]\) by three-fold cross-validation. The random forest uses \(500\) trees of maximum depth three, a minimum leaf size of \(30\), and \(\sqrt p\) candidate features per split. FA-DML follows
Algorithm~\ref{algorithm_three_fold}, repeated over \(15\) random sample splits and aggregated with the median adjustment of \citet[Section~3.4]{CCDDHNR18}.

The benchmarks are pooled OLS, two-way fixed effects (TWFE),
standard DML, and the interactive fixed-effects estimator
of \citet{bai2009panel} (OLS-IFE). Standard DML uses two-fold
cross-fitting over states and cross-validated Lasso on
standardized controls. OLS-IFE uses the iterative estimator
of \citet{bai2009panel} with non-clustered standard
errors. For OLS-IFE, $r$ counts structural outcome factors. The FA-DML outcome rank
instead refers to the reduced-form factor space. Their rank grids are
therefore not directly comparable.

Applying Algorithm~\ref{alg:rank_estimation} with Lasso to the original
dictionary, with the calibration in Appendix~\ref{subsection:robustness_rank_miniwage}, selects
\((\hat r_D,\hat r_Y)=(2,3)\). We use this pair as the primary specification
for both dictionaries and learners and treat other dimensions as sensitivity
checks. Table~\ref{tab:ams_results} reports
the results.

At the selected ranks, the Lasso and random-forest estimates are \(-0.1103\) and \(-0.1167\) with the original controls and \(-0.1088\) and \(-0.1120\) with the expanded controls. All four \(95\%\) intervals exclude zero. A \(10\%\) increase in the real minimum wage is thus associated with a roughly \(1.1\%\) lower 50/10 wage ratio, and this estimate is stable across dictionaries and learners.

The benchmark estimates respond differently to the control
dictionary. Expanding the controls changes pooled OLS from
$-0.0213$ to $-0.1327$ and TWFE from $-0.0249$ to
$-0.0647$; their intervals include zero with the original
controls but exclude it with the expanded dictionary.
Standard DML estimates remain close to zero and imprecise
under both dictionaries. OLS-IFE point estimates remain negative
and decrease in magnitude as $r$ increases. Their confidence intervals
exclude zero at all ranks considered with the original controls. With the expanded controls, they exclude zero for $r=2,\ldots,5$ but include zero at
$r=6$.

\begin{table}[htbp!]
  \centering
  \captionsetup{font={footnotesize,stretch=1},justification=justified,
      singlelinecheck=false,width=\textwidth,skip=6pt}
  \fontsize{8}{12}\selectfont
    \begin{tabular}{lccccccc}
    \toprule
          & \multicolumn{3}{c}{Original controls} & & \multicolumn{3}{c}{Expanded controls} \\
    \cmidrule(lr){2-4}\cmidrule(lr){6-8}
          & Estimate   & s.e.    & 95\% CI                 & & Estimate   & s.e.     & 95\% CI \\
    \midrule
        \multicolumn{8}{c}{Panel A. Benchmark estimators} \\
    OLS                & $-0.0213$ & $0.0390$ & $(-0.098,\,+0.055)$ & & $-0.1327$ & $0.0327$ & $(-0.197,\,-0.069)$ \\
    OLS (TWFE)         & $-0.0249$ & $0.0318$ & $(-0.087,\,+0.037)$ & & $-0.0647$ & $0.0260$ & $(-0.116,\,-0.014)$ \\
    DML                & $-0.0128$ & $0.0403$ & $(-0.092,\,+0.066)$ & & $-0.0091$ & $0.0401$ & $(-0.088,\,+0.070)$ \\
    \midrule
    \multicolumn{8}{c}{Panel B. OLS-IFE } \\
    $r=2$ & $-0.1063$ & $0.0093$ & $(-0.125,\,-0.088)$ & & $-0.0688$ & $0.0116$ & $(-0.091,\,-0.046)$ \\
    $r=3$ & $-0.0908$ & $0.0110$ & $(-0.112,\,-0.069)$ & & $-0.0547$ & $0.0126$ & $(-0.079,\,-0.030)$ \\
    $r=4$ & $-0.0876$ & $0.0103$ & $(-0.108,\,-0.067)$ & & $-0.0479$ & $0.0121$ & $(-0.072,\,-0.024)$ \\
    $r=5$ & $-0.0611$ & $0.0108$ & $(-0.082,\,-0.040)$ & & $-0.0326$ & $0.0117$ & $(-0.056,\,-0.010)$ \\
    $r=6$ & $-0.0562$ & $0.0101$ & $(-0.076,\,-0.036)$ & & $-0.0174$ & $0.0112$ & $(-0.039,\,+0.005)$ \\
    \midrule
    \multicolumn{8}{c}{Panel C. FA-DML (Lasso)} \\
    Selected $(2,3)^{\dagger}$ & $-0.1103$ & $0.0391$ & $(-0.187,\,-0.034)$ & & $-0.1088$ & $0.0391$ & $(-0.185,\,-0.032)$ \\
    \addlinespace[2pt]
    $r=2$ & $-0.1155$ & $0.0425$ & $(-0.199,\,-0.032)$ & & $-0.1152$ & $0.0407$ & $(-0.195,\,-0.035)$ \\
    $r=3$ & $-0.1398$ & $0.0482$ & $(-0.234,\,-0.045)$ & & $-0.1296$ & $0.0471$ & $(-0.222,\,-0.037)$ \\
    $r=4$ & $-0.1612$ & $0.0604$ & $(-0.280,\,-0.043)$ & & $-0.1520$ & $0.0572$ & $(-0.264,\,-0.040)$ \\
    $r=5$ & $-0.1484$ & $0.0625$ & $(-0.271,\,-0.026)$ & & $-0.1487$ & $0.0606$ & $(-0.267,\,-0.030)$ \\
    $r=6$ & $-0.1590$ & $0.0639$ & $(-0.284,\,-0.034)$ & & $-0.1652$ & $0.0715$ & $(-0.305,\,-0.025)$ \\
    \midrule
    \multicolumn{8}{c}{Panel D. FA-DML (Random Forest)} \\
    Selected $(2,3)^{\dagger}$ & $-0.1167$ & $0.0406$ & $(-0.196,\,-0.037)$ & & $-0.1120$ & $0.0403$ & $(-0.191,\,-0.033)$ \\
    \addlinespace[2pt]
    $r=2$ & $-0.1201$ & $0.0423$ & $(-0.203,\,-0.037)$ & & $-0.1177$ & $0.0433$ & $(-0.203,\,-0.033)$ \\
    $r=3$ & $-0.1382$ & $0.0494$ & $(-0.235,\,-0.041)$ & & $-0.1405$ & $0.0490$ & $(-0.237,\,-0.044)$ \\
    $r=4$ & $-0.1608$ & $0.0601$ & $(-0.279,\,-0.043)$ & & $-0.1610$ & $0.0600$ & $(-0.279,\,-0.043)$ \\
    $r=5$ & $-0.1496$ & $0.0645$ & $(-0.276,\,-0.023)$ & & $-0.1486$ & $0.0641$ & $(-0.274,\,-0.023)$ \\
    $r=6$ & $-0.1606$ & $0.0645$ & $(-0.287,\,-0.034)$ & & $-0.1600$ & $0.0642$ & $(-0.286,\,-0.034)$ \\
    \bottomrule
    \end{tabular}
    \caption[Minimum wages and lower-tail wage inequality.]{Minimum wages and lower-tail wage inequality. The outcome is the log 50/10 wage
ratio and the treatment is the log real effective minimum wage (\(N=50\),
\(T=34\)). Entries are estimates, standard errors, and \(95\%\) Wald intervals
with the original (\(p=4\)) or expanded (\(p=32\)) dictionary. OLS-IFE uses
\(r\) outcome factors. The dagger marks the selected FA-DML pair
\((r_D,r_Y)=(2,3)\), and the other FA-DML rows set \(r_D=r_Y=r\). FA-DML
estimates use the \(15\)-split median adjustment. Standard errors are clustered
by state, except for OLS-IFE. Table~\ref{tab:ams_results_appendix} reports the full grid of
dimension pairs.
}
\label{tab:ams_results}
\end{table}

For FA-DML, this finding also extends beyond the selected
rank pair $(2,3)$. The $20$ equal-rank specifications with
$r_D=r_Y=r\in\{2,\ldots,6\}$ yield estimates between
$-0.1652$ and $-0.1152$. Across the full grid in
Appendix~\ref{subsection:robustness_rank_miniwage}, all $60$
estimates lie between $-0.1652$ and $-0.1003$, and every
reported interval excludes zero. Results are similar across
learners: for each rank pair and control dictionary, the
Lasso and random-forest estimates differ by at most $0.0116$.

Overall, the results indicate that higher minimum wages
are associated with lower wage inequality at the bottom
of the distribution, even after flexible adjustment for
observed workforce composition and latent interactive
heterogeneity.

\section{Conclusion}
\label{sec:conclusion}

Learning latent factor spaces from panel residuals introduces
estimation errors that standard DML orthogonality does not
automatically control. FA-DML addresses these errors through
double projection and separate training folds for the treatment
and reduced-form equations. Double projection makes the
factor--factor interaction a product of errors from the two
estimated spaces. Separate training folds remove same-unit
pairings in their leading disturbance components. Under the
stated conditions, the
estimator delivers oracle-equivalent
inference without analytical bias correction. We also develop a consistent
rank selector and show that estimating the factor
dimensions preserves inference results.

A natural extension is to dynamic panels with predetermined
regressors. Such an extension would require adapting the score
and sample-splitting scheme to account for dependence between
current disturbances and future covariates. Extending the
analysis to weak factors is another direction for future work.

\bibliographystyle{apalike}
\bibliography{Bibliography}
\clearpage
\pagenumbering{arabic}

\setcounter{section}{0}
\setcounter{subsection}{0}
\setcounter{lemma}{0}

\renewcommand{\thesection}{Appendix \Alph{section}}
\renewcommand{\thesubsection}{\Alph{section}.\arabic{subsection}}
\renewcommand{\thelemma}{\Alph{section}.\arabic{lemma}}

\makeatletter
\@addtoreset{lemma}{section}
\makeatother
\renewcommand{\theHsection}{appendix.\Alph{section}}
\renewcommand{\theHsubsection}{appendix.\Alph{section}.\arabic{subsection}}
\renewcommand{\theHlemma}{appendix.\Alph{section}.\arabic{lemma}}
\begin{center}
{\large Supplementary Material of ``Semiparametric Inference in Panel Data with Interactive
and High-dimensional Confounding''}
\end{center}
\ref{app:notation} collects the notation for residuals, disturbances,
loadings, and first-stage errors.
\ref{app:rank-selection-algorithm} gives the complete data-driven
rank-selection procedure. \ref{app:proof} collects the proofs of the main
theoretical results. \ref{app:decomposition} discusses the
first-stage prediction and remainder conditions and provides a
Lasso rate illustration.
\ref{app:additional-sim} records additional simulation
implementation details and evidence. \ref{app:additional_empi} reports
additional evidence for the minimum-wage and branching-deregulation
applications.

\section{Notation Guide}
\label{app:notation}

Table~\ref{tab:notation-guide} summarizes the residuals,
disturbances, loadings, and first-stage errors used in the analysis.
The residuals retain latent factor components, whereas the
idiosyncratic disturbances do not. Projecting the nuisance
functions changes the associated loadings but leaves these
disturbances unchanged.

For a fixed fold-role assignment $l$, $\hat g^{(l)}$ and
$\hat m^{(l)}$ denote the treatment and reduced-form learners.
We suppress $l$ on the first-stage errors, which are measured
relative to the projected targets $g_0^*$ and $m_0^*$.

\begingroup
\small
\renewcommand{\arraystretch}{1.15}
\setlength{\LTleft}{0pt}
\setlength{\LTright}{0pt}
\setlength{\LTcapwidth}{\textwidth}
\begin{longtable}{@{}p{0.16\textwidth}p{0.78\textwidth}@{}}
\caption{Notation for residuals, loadings, and first-stage errors}
\label{tab:notation-guide}\\
\toprule
Symbol & Definition and interpretation \\
\midrule
\endfirsthead
\multicolumn{2}{l}{\tablename~\thetable\ (continued)}\\
\toprule
Symbol & Definition and interpretation \\
\midrule
\endhead
\midrule
\multicolumn{2}{r}{Continued on next page}\\
\endfoot
\bottomrule
\endlastfoot

\multicolumn{2}{@{}l}{\textit{Residuals and disturbances}}\\*
\addlinespace
$v_{D,i}$
& $D_i-g_0(X_i)=G_D^0C_{D,i}+\nu_i$.
Treatment residual after observed-covariate adjustment.\\
\addlinespace
$v_{Y,i}$
& $Y_i-m_0(X_i)=G^0C_i+\xi_i$, where $m_0=f_0+\theta_0g_0$.
Reduced-form outcome residual after observed-covariate adjustment.\\
\addlinespace
$u_i$
& $v_{Y,i}-\theta_0v_{D,i}=G_Y^0C_{Y,i}^*+\varepsilon_i$.
Structural outcome residual, including the latent factor component.\\
\addlinespace
$\nu_i$
& Idiosyncratic disturbance in the treatment equation.\\
\addlinespace
$\varepsilon_i$
& Idiosyncratic disturbance in the structural outcome equation.\\
\addlinespace
$\xi_i$
& $\theta_0\nu_i+\varepsilon_i$.
Idiosyncratic disturbance in the reduced-form outcome equation.\\
\addlinespace[0.8em]

\multicolumn{2}{@{}l}{\textit{Projected targets and residuals}}\\*
\addlinespace
$g_0^*(X_i)$
& $M_{G_D^0}g_0(X_i)$.
Projected target of the treatment learner.\\
\addlinespace
$m_0^*(X_i)$
& $M_{G^0}m_0(X_i)$.
Projected target of the reduced-form learner.\\
\addlinespace
$v_{D,i}^*$
& $D_i-g_0^*(X_i)=G_D^0\bar C_{D,i}+\nu_i$.
Treatment residual relative to the projected nuisance target.\\
\addlinespace
$v_{Y,i}^*$
& $Y_i-m_0^*(X_i)=G^0\bar C_{Y,i}+\xi_i$.
Reduced-form residual relative to the projected nuisance target.\\
\addlinespace
$u_i^*$
& $v_{Y,i}^*-\theta_0v_{D,i}^*=G^0\bar C_i+\varepsilon_i$.
Structural score residual relative to the projected nuisance targets.\\
\addlinespace[0.8em]

\multicolumn{2}{@{}l}{\textit{Loadings}}\\*
\addlinespace
$C_{D,i}$
& Loading on $G_D^0$ in the treatment equation.\\
\addlinespace
$C_{Y,i}^*$
& Loading on $G_Y^0$ in the structural outcome equation.\\
\addlinespace
$C_i$
& $G^0C_i=G_Y^0C_{Y,i}^*+\theta_0G_D^0C_{D,i}$.
Loading in the reduced-form outcome equation.\\
\addlinespace
$\bar C_{D,i}$
& $C_{D,i}+(G_D^{0\prime}G_D^0)^{-1}G_D^{0\prime}g_0(X_i)$.
Augmented treatment loading.\\
\addlinespace
$\bar C_{Y,i}$
& $C_i+(G^{0\prime}G^0)^{-1}G^{0\prime}m_0(X_i)$.
Augmented reduced-form loading.\\
\addlinespace
$\bar C_i$
& $G^0\bar C_i=G^0\bar C_{Y,i}-\theta_0G_D^0\bar C_{D,i}$.
Loading in the projected structural score residual.\\
\addlinespace[0.8em]

\multicolumn{2}{@{}l}{\textit{First-stage errors}}\\*
\addlinespace
$\delta_i^g$
& $\hat g^{(l)}(X_i)-g_0^*(X_i)$.
Treatment first-stage error.\\*
\addlinespace
$\delta_i^m$
& $\hat m^{(l)}(X_i)-m_0^*(X_i)$.
Reduced-form first-stage error.\\*
\addlinespace
$\delta_i$
& $\delta_i^m-\theta_0\delta_i^g$.
Combined first-stage error in the structural score residual.\\
\end{longtable}
\endgroup

\clearpage

\section{Data-Driven Rank-Selection Procedure}
\label{app:rank-selection-algorithm}

Section~\ref{subsec:rank-selection} defines the terminal rank estimators used in
the theoretical results. This appendix provides the complete alternating
procedure that constructs their terminal learners and low-rank estimates.
The selector uses nuclear-norm penalization, a standard convex relaxation of
rank regularization used in low-rank matrix estimation and matrix completion
\citep{candesrecht2009,rechtfazelparrilo2010,
mazumderhastietibshirani2010,koltchinskiilounicitsybakov2011}, including in
econometric panel settings
\citep{athey2021matrix,chernozhukov2023,moonweidner2026}.

For \(M\in\mathbb R^{N\times T}\), let \(q=\min\{N,T\}\) and write its singular
value decomposition as
\[
M=U\operatorname{diag}(s_1,\ldots,s_q)V',
\qquad
s_1\ge\cdots\ge s_q\ge0.
\]
For \(\zeta>0\), define the singular-value thresholding operator
\[
\mathrm{SVT}(M,\zeta)
=
U\operatorname{diag}\{(s_1-\zeta)_+,\ldots,(s_q-\zeta)_+\}V',
\]
where \((a)_+=\max\{a,0\}\).

Let \(D\) and \(Y\) denote the \(N\times T\) matrices whose \(i\)-th rows are
\(D_i'\) and \(Y_i'\), respectively, and write \(X=(X_1,\ldots,X_N)\). For any
vector-valued learner \(h\), let \(h(X)\) denote the \(N\times T\) matrix whose
\(i\)-th row is \(h(X_i)'\).

\begin{algorithm}[Rank selection for the treatment equation]
\label{alg:rank_estimation}

Given \(D\), covariates \(X\), an initial learner
\(\hat g^{(\mathrm{init})}\), a tuning parameter \(\rho_D>0\), a fixed
threshold parameter \(\mu\in(-1/2,1/2)\), a maximum number of iterations \(K\),
and the tolerance level \(\epsilon_{\mathrm{tol}}\), proceed as follows.

\begin{enumerate}
\item Initialize \(\hat g^{(0)}=\hat g^{(\mathrm{init})}\).

\item For \(k=0,1,\ldots,K-1\), repeat the following steps until convergence:
\begin{enumerate}
\item Form the residual matrix
\[
\hat v_D^{(k)}=D-\hat g^{(k)}(X).
\]

\item Compute the nuclear-norm penalized low-rank estimate
\[
\hat\Theta_D^{(k)}
=
\argmin_{\Theta\in\mathbb R^{N\times T}}
\left\{
\frac12\|\hat v_D^{(k)}-\Theta\|_F^2
+
\rho_D\|\Theta\|_*
\right\}.
\]
Equivalently,
\[
\hat\Theta_D^{(k)}
=
\mathrm{SVT}\bigl(\hat v_D^{(k)},\rho_D\bigr).
\]

\item Update the nonparametric component by applying the chosen learner to the
pseudo-outcome
\[
D-\hat\Theta_D^{(k)}
\]
with covariates \(X\). Denote the updated learner by \(\hat g^{(k+1)}\).

\item Stop if \(k+1=K\) or if
\[
    \frac{1}{\sqrt{NT}}
    \bigl\|\hat g^{(k+1)}(X)-\hat g^{(k)}(X)\bigr\|_F
    \leq \epsilon_{\mathrm{tol}}.
\]
Otherwise continue to the next iteration.
\end{enumerate}

\item Let \(\hat g^{(nn)}\) denote the final learner from the nuclear-norm
iteration. Recompute the low-rank component using this final learner:
\[
\hat\Theta_D
=
\argmin_{\Theta\in\mathbb R^{N\times T}}
\left\{
\frac12\|D-\hat g^{(nn)}(X)-\Theta\|_F^2
+
\rho_D\|\Theta\|_*
\right\}.
\]

\item Define
\[
\hat r_D
=
\begin{cases}
0, & s_1(\hat\Theta_D)=0,\\[2mm]
\displaystyle
\sum_{j=1}^{\min\{N,T\}}
\mathbf 1
\left\{
s_j(\hat\Theta_D)
>
\rho_D^{1/2-\mu}s_1(\hat\Theta_D)^{1/2+\mu}
\right\},
&s_1(\hat\Theta_D)>0.
\end{cases}
\]
Here \(\mathbf 1\{\cdot\}\) denotes the indicator function.
\end{enumerate}

Estimate the reduced-form rank $r_Y$ analogously
by replacing $(D,\hat g^{(\mathrm{init})},\rho_D)$ with
$(Y,\hat m^{(\mathrm{init})},\rho_Y)$.
\end{algorithm}
The two estimated dimensions \((\hat r_D,\hat r_Y)\) are supplied to
Algorithm~\ref{algorithm_three_fold}. Assumption~\ref{assp:rank_consistency}
and Theorem~\ref{thm:rank_consistency_hl} give sufficient conditions for their
consistency, while Corollary~\ref{cor:estimated-ranks} establishes inference
for the resulting fully feasible FA-DML estimator.

\section{Proofs of Theorems}\label{app:proof}
The proof of Theorems \ref{thm:CLT} and \ref{thm:var-consistency} proceeds in
four steps. Proposition \ref{proposition:1st_stage} controls the standard
nuisance remainder. Lemma \ref{lem:pca_expansion_dml} gives an exact
linearization of the two PCA estimators, and Lemma
\ref{lem:double_projection_bound_sharp} uses the disjoint training folds to
control their interaction. Propositions \ref{proposition:R2}--\ref{proposition:R4}
then establish negligibility of the factor-projection remainders, while
Proposition \ref{proposition:denominator} controls the denominator. The
rank-selection result is proved last.

\noindent\textbf{Auxiliary identities for Sections 2 and 3.}\par\smallskip
Let $L$ be the $N\times(r_D+r_Y^*)$ matrix with $i$-th row
$(C_{D,i}',C_{Y,i}^{*\prime})$. The matrix of reduced-form interactive
components is
\[
    [\theta_0G_D^0,G_Y^0]L'.
\]
Since $L'$ has full row rank by \eqref{eq:joint-loading-rank}, this product
and $[\theta_0G_D^0,G_Y^0]$ have the same column space, establishing
\eqref{eq:effective-factor-space}. Consequently,
\[
    (G_D^0)'A_0=0,\qquad
    A_0G^0=0,\qquad
    A_0G_Y^0=0,\qquad
    \theta_0A_0G_D^0=0.
\]
These identities give \eqref{eq_ortho}.

To verify \eqref{eq:score-equivalence}, define
\[
    a_i=P_{G_D^0}g_0(X_i),
    \qquad
    b_i=P_{G^0}m_0(X_i).
\]
Then $D_i-g_0^*(X_i)=v_{D,i}+a_i$ and
$Y_i-m_0^*(X_i)=v_{Y,i}+b_i$. Using $a_i'A_0=0$,
$A_0b_i=0$, and $(G_D^0)'A_0=0$, we obtain, for every $\theta$,
\[
\begin{aligned}
&\psi(W_i;\theta,g_0^*,m_0^*,A_0)
 -\psi(W_i;\theta,g_0,m_0,A_0)\\
&\qquad=-\theta\,\nu_i'A_0a_i.
\end{aligned}
\]
Its conditional expectation given $\mathcal F_T$ is zero by
Assumption~\ref{ass:identification}(a), proving
\eqref{eq:score-equivalence}. At $\theta=\theta_0$, the difference vanishes
exactly because $\theta_0A_0a_i=0$.

Finally, write $v_{D,i}^*=D_i-g_0^*(X_i)$ and
$u_i^*=Y_i-m_0^*(X_i)-\theta_0v_{D,i}^*$. The projected residual
representations imply
\[
    A_0u_i^*=A_0\varepsilon_i,
    \qquad
    v_{D,i}^{*\prime}A_0=\nu_i'A_0.
\]
Therefore,
\[
\begin{aligned}
&\left.
\frac{\partial}{\partial\tau}
E[\psi(W_i;\theta_0,g_0^*+\tau h_g,m_0^*+\tau h_m,A_0)
  \mid\mathcal F_T]
\right|_{\tau=0}\\
&\quad=
E[-h_g(X_i)'A_0\varepsilon_i
  -\nu_i'A_0h_m(X_i)
  +\theta_0\nu_i'A_0h_g(X_i)
  \mid\mathcal F_T]\\
&\quad=0,
\end{aligned}
\]
establishing \eqref{eq:neyman_moment}.

\noindent\textbf{Proof of Proposition \ref{proposition:1st_stage}.}
\begin{proof}

Fix a split \(l\). Since the number of splits is fixed, it suffices to
prove the claim for an arbitrary \(l\). Assumption~\ref{ass:1st_stage}(a)--(b)
directly gives
\[
\|\hat g^{(l1)}-g_0^*\|_{P,2}=O_p(\bar r_{g}),
\qquad
\|\hat m^{(l2)}-m_0^*\|_{P,2}=O_p(\bar r_{m}).
\]

We next transfer the population prediction rates to the evaluation-fold
empirical norms. Let \(h\) be any \(\mathcal H_l\)-measurable map from
\(\mathcal X^T\) to \(\mathbb R^T\) with \(\|h\|_{P,2}<\infty\). By
Assumption~\ref{ass:factor_fold}(d)--(e) and the independence
of the auxiliary learner randomization specified in
Section~\ref{Section_thm}, conditional on \(\mathcal F_T\) and the random
partition, the evaluation-fold covariates are i.i.d.\ across units and
independent of the training-fold data and learner randomization. Hence
\[
E\!\left[\|h\|_{n,l}^2 \mid \mathcal H_l\right]
=
E\!\left[
\frac{1}{Tn_3}
\sum_{i\in \mathcal I_3^{(l)}} \|h(X_i)\|_2^2
\ \middle|\ \mathcal H_l
\right]
=
    \frac{1}{T}E\!\left[
    \|h(\widetilde X)\|_2^2\mid\mathcal H_l
    \right]
=
\|h\|_{P,2}^2.
\]

For any positive deterministic sequence \(r_{NT}\) and constants \(K,M>0\),
conditional Markov's inequality and the preceding identity imply
\[
P\!\left(\|h\|_{n,l}>Mr_{NT}\right)
\le
P\!\left(\|h\|_{P,2}>Kr_{NT}\right)+\frac{K^2}{M^2}.
\]
If \(\|h\|_{P,2}=O_p(r_{NT})\), choosing first \(K\) and then \(M\)
sufficiently large therefore gives
\[
\|h\|_{n,l}=O_p(r_{NT}).
\]

Taking
\[
h=\hat g^{(l1)}-g_0^*
\qquad\text{and}\qquad
h=\hat m^{(l2)}-m_0^*,
\]
and using the population prediction bounds established above, we obtain
\[
\|\hat g^{(l1)}-g_0^*\|_{n,l}
=
O_p(\bar r_{g}),
\qquad
\|\hat m^{(l2)}-m_0^*\|_{n,l}
=
O_p(\bar r_{m}).
\]
Since $\delta_i=\delta_i^m-\theta_0\delta_i^g$, and \(\theta_0\) is fixed,
\[
\left(
\frac1{Tn_3}\sum_{i\in\mathcal I_3^{(l)}}\|\delta_i\|_2^2
\right)^{1/2}
=
O_p(\bar r_{m}+\bar r_{g}).
\]

Recall that
\[
R_{1i}
=
-\delta_i^{g\prime}A_0\varepsilon_i
-\nu_i'A_0\delta_i
+\delta_i^{g\prime}A_0\delta_i.
\]
Both \(\delta_i^g\) and \(\delta_i\) are \(\mathcal G_l\)-measurable.
By Assumptions~\ref{ass:factor_fold}(e), \ref{ass:training_main}(b), and
\ref{ass:eval_main}(a), conditional on \(\mathcal G_l\), the
relevant cross-covariance terms vanish and
\[
\Var\!\left(
\frac1{\sqrt{Tn_3}}
\sum_{i\in\mathcal I_3^{(l)}}\delta_i^{g\prime}A_0\varepsilon_i
\middle| \mathcal G_l
\right)
\le
\frac{C}{Tn_3}
\sum_{i\in\mathcal I_3^{(l)}}\|\delta_i^g\|_2^2
=
O_p(\bar r_{g}^2).
\]
Thus, by conditional Chebyshev's inequality,
\[
\frac1{\sqrt{Tn_3}}
\sum_{i\in\mathcal I_3^{(l)}}\delta_i^{g\prime}A_0\varepsilon_i
=
O_p(\bar r_{g}).
\]
The same argument gives
\[
\frac1{\sqrt{Tn_3}}
\sum_{i\in\mathcal I_3^{(l)}}\nu_i'A_0\delta_i
=
O_p(\bar r_{m}+\bar r_{g}).
\]
Finally, by Cauchy--Schwarz and \(\|A_0\|_{\op}\le 1\),
\[
\left|
\frac1{\sqrt{Tn_3}}
\sum_{i\in\mathcal I_3^{(l)}}\delta_i^{g\prime}A_0\delta_i
\right|
\le
\sqrt{Tn_3}
\left(
\frac1{Tn_3}\sum_i\|\delta_i^g\|_2^2
\right)^{1/2}
\left(
\frac1{Tn_3}\sum_i\|\delta_i\|_2^2
\right)^{1/2},
\]
and hence
\[
\frac1{\sqrt{Tn_3}}
\sum_{i\in\mathcal I_3^{(l)}}\delta_i^{g\prime}A_0\delta_i
=
O_p\!\left(
\sqrt{Tn_3}\,
\bar r_{g}(\bar r_{m}+\bar r_{g})
\right).
\]
Combining the three displays,
\[
\frac1{\sqrt{Tn_3}}
\sum_{i\in\mathcal I_3^{(l)}}R_{1i}
=
O_p\!\left(
\bar r_{g}
+
\bar r_{m}
+
\sqrt{Tn_3}\,
\bar r_{g}(\bar r_{m}+\bar r_{g})
\right).
\]
Since \(\bar r_{g},\bar r_{m}\to0\), Assumption~\ref{ass:1st_stage}(c) gives
\[
\frac1{\sqrt{Tn_3}}
\sum_{i\in\mathcal I_3^{(l)}}R_{1i}
=o_p(1).
\]
Since \(l\) was arbitrary and the number of splits is fixed, the same
conclusion holds uniformly over \(l\).
\end{proof}
\noindent\textbf{Proof of Theorem \ref{thm:CLT}}
\begin{proof}
    By Propositions \ref{proposition:1st_stage},
    \ref{proposition:denominator}, and \ref{proposition:R2}--\ref{proposition:R4},
    for each fixed split \(l\),
\[
\hat\theta^{(l)}-\theta_0
=
\frac{\kappa_0^{-1}}{n_3T}
\sum_{i\in \mathcal I_3^{(l)}} \nu_i'A_0\varepsilon_i
+
o_p\!\big((NT)^{-1/2}\big).
\]
Let \(n_j=|\mathcal I_j|\) for the three original folds. Under the six
permutations, each fold appears as the evaluation fold twice. Therefore
\[
\hat\theta-\theta_0
=
\frac{\kappa_0^{-1}}{3T}
\sum_{j=1}^3\frac{1}{n_j}
\sum_{i\in \mathcal I_j}\nu_i'A_0\varepsilon_i
+
o_p\!\big((NT)^{-1/2}\big).
\]
Assumption~\ref{ass:factor_fold}(d) gives
\(n_j=N/3+O(1)\), and hence
\[
    \frac{1}{3n_j}=\frac{1}{N}+O(N^{-2}).
\]
The evaluation-fold moment conditions imply
\(\sum_{i\in\mathcal I_j}\nu_i'A_0\varepsilon_i
=O_p(\sqrt{n_jT})\). It follows that
\[
\hat\theta-\theta_0
=
\frac{\kappa_0^{-1}}{NT}
\sum_{i=1}^N \nu_i'A_0\varepsilon_i
+
o_p\!\big((NT)^{-1/2}\big).
\]
It remains to verify the CLT for the full-sample leading term. Put
\(W_i=\nu_i'A_0\varepsilon_i\). For each original fold, choose a permutation in
which that fold is the evaluation fold. Since
\(\mathcal F_T\subseteq\mathcal G_l\), Assumption
\ref{ass:eval_main}(a) and iterated expectations give
\(E[W_i\mid\mathcal F_T]=0\). Applying the fold-average condition in
Assumption~\ref{ass:eval_main}(b) to the three original folds, taking
conditional expectations down to \(\mathcal F_T\), and combining the folds
gives
\[
\frac1{NT^{1+\delta/2}}
\sum_{i=1}^N E[|W_i|^{2+\delta}\mid\mathcal F_T]
\le C.
\]
Assumption \ref{ass:factor_fold}(e) gives conditional independence across
units.  Moreover, Assumption \ref{ass:eval_main}(c), applied to the three
original folds, implies
\[
\frac1{NT}\sum_{i=1}^N E[W_i^2\mid\mathcal F_T]
\xrightarrow{p}\Omega_0.
\]
To see the last implication, take conditional expectations of the three
foldwise variance limits down to \(\mathcal F_T\) and combine them with weights
\(n_j/N\); the moment bound above supplies uniform integrability.  Finally,
\[
\frac1{(NT)^{1+\delta/2}}
\sum_{i=1}^N E[|W_i|^{2+\delta}\mid\mathcal F_T]
\le C N^{-\delta/2}\to0.
\]
The conditional Lyapunov CLT therefore yields
\[
\frac1{\sqrt{NT}}\sum_{i=1}^N W_i
\xrightarrow{d}N(0,\Omega_0),
\]
and hence
\[
\sqrt{NT}\,(\hat\theta-\theta_0)
\;\xrightarrow{d}\;
N\!\left(0,\kappa_0^{-2}\Omega_0\right).
\]
\end{proof}

\noindent\textbf{Proof of Theorem \ref{thm:var-consistency}}
\begin{proof}
By Propositions~\ref{proposition:denominator} and
\ref{proposition:num_variance},
\[
\frac{1}{T|\mathcal I_3^{(l)}|}
\sum_{i\in\mathcal I_3^{(l)}}
\hat v_{D,i}'\hat A\hat v_{D,i} \xrightarrow{p} \kappa_0,
\qquad
\frac{1}{T|\mathcal I_3^{(l)}|} \sum_{i \in \mathcal I_3^{(l)}} \left( \hat v_{D,i}'  \hat A (\hat v_{Y,i}-\hat{\theta}^{(l)}\hat{v}_{D,i})\right)^2 \xrightarrow{p} \Omega_0.
\]
Hence the foldwise variance estimator satisfies
\[
\hat\sigma_{\theta,l}^{2}\xrightarrow{p}
\sigma_\theta^2.
\]
It follows immediately that
\[
\hat\sigma_{\theta}^{2}
\xrightarrow{p}
\sigma_\theta^2.
\]
\end{proof}
\begin{proposition}\label{proposition:denominator}
Under the conditions of Theorem \ref{thm:CLT}, the denominator in
\eqref{eq:theta_l} satisfies
\[
\frac{1}{Tn_3}\sum_{i\in \mathcal I_3^{(l)}}
\hat v_{D,i}' \hat A \hat v_{D,i}=\kappa_0+o_p(1),
\]
where \(\kappa_0\) is the constant defined in Assumption
\ref{ass:eval_main}(c).
 \end{proposition}
 \begin{proof}
     Recall
\[
\hat v_{D,i}=v_{D,i}^*-\delta_i^g,
\qquad
\hat A=A_0+\Delta_A.
\]
Plugging them into $\hat v_{D,i}'\hat A\hat v_{D,i}$, we have
\begin{align*}
\frac1{n_3T}\sum_{i\in\mathcal I_3}
\hat v_{D,i}'\hat A\hat v_{D,i}
&=
\frac1{n_3T}\sum_{i\in\mathcal I_3}
v_{D,i}^{*\prime}A_0v_{D,i}^*\\
&\quad-
\frac1{n_3T}\sum_{i\in\mathcal I_3}
\left(
\delta_i^{g\prime}A_0v_{D,i}^*
+
v_{D,i}^{*\prime}A_0\delta_i^g
\right)\\
&\quad+
\frac1{n_3T}\sum_{i\in\mathcal I_3}
\delta_i^{g\prime}A_0\delta_i^g\\
&\quad+
\frac1{n_3T}\sum_{i\in\mathcal I_3}
v_{D,i}^{*\prime}\Delta_Av_{D,i}^*\\
&\quad-
\frac1{n_3T}\sum_{i\in\mathcal I_3}
\left(
\delta_i^{g\prime}\Delta_Av_{D,i}^*
+
v_{D,i}^{*\prime}\Delta_A\delta_i^g
\right)\\
&\quad+
\frac1{n_3T}\sum_{i\in\mathcal I_3}
\delta_i^{g\prime}\Delta_A\delta_i^g\\
&=D_0+D_1+D_2+D_3+D_4+D_5.
\end{align*}

We first note that Assumptions~\ref{ass:factor_fold}(d)--(e) and
\ref{ass:training_main}(b), together with conditional Markov's inequality, imply
\[
    \frac1{n_3T}\sum_{i\in\mathcal I_3}
    \|\nu_i\|_2^2=O_p(1).
\]
The factor Gram bound and the fold-average loading
moment bound therefore give
\begin{align*}
    \frac1{n_3T}\sum_{i\in\mathcal I_3}
    \|v_{D,i}^*\|_2^2
    &\le
    \frac{2\|G_D^0\|_{\op}^2}{T}
    \frac1{n_3}\sum_{i\in\mathcal I_3}
    \|\bar C_{D,i}\|_2^2
    +
    \frac2{n_3T}\sum_{i\in\mathcal I_3}
    \|\nu_i\|_2^2\\
    &=O_p(1).
\end{align*}
Moreover, Proposition~\ref{proposition:1st_stage} yields
\[
    \frac1{n_3T}\sum_{i\in\mathcal I_3}
    \|\delta_i^g\|_2^2
    =O_p(\bar r_{g}^2).
\]

Since $G_D^{0\prime}A_0=0$, the leading term satisfies
\[
    D_0
    =
    \frac1{n_3T}\sum_{i\in\mathcal I_3}
    \nu_i'A_0\nu_i
    +
    \frac1{n_3T}\sum_{i\in\mathcal I_3}
    \nu_i'A_0G_D^0\bar C_{D,i}.
\]
The second term is retained because $A_0G_D^0$ need not
vanish; its negligibility follows from the conditional moment bounds below.
The first sum converges in probability to $\kappa_0$
by Assumption~\ref{ass:eval_main}(c).
For the second sum, Assumptions~\ref{ass:factor_fold}(e),
\ref{ass:training_main}(b), and \ref{ass:eval_main}(a) give conditional
mean zero and
\begin{align*}
&\operatorname{Var}\!\left(
    \frac1{n_3T}\sum_{i\in\mathcal I_3}
    \nu_i'A_0G_D^0\bar C_{D,i}
    \,\middle|\,\mathcal G_l
\right)\\
&\qquad\le
    \frac{C\|G_D^0\|_{\op}^2}{n_3^2T^2}
    \sum_{i\in\mathcal I_3}\|\bar C_{D,i}\|_2^2
    =
    O_p\!\left(\frac1{n_3T}\right),
\end{align*}
where $\|A_0\|_{\op}\le1$.
Conditional Chebyshev's inequality thus implies
$D_0=\kappa_0+o_p(1)$.

For $D_1$, Cauchy--Schwarz gives
\begin{align*}
    |D_1|
    &\le
    2
    \left(
        \frac1{n_3T}\sum_{i\in\mathcal I_3}
        \|\delta_i^g\|_2^2
    \right)^{1/2}
    \left(
        \frac1{n_3T}\sum_{i\in\mathcal I_3}
        \|v_{D,i}^*\|_2^2
    \right)^{1/2}\\
    &=O_p(\bar r_{g})=o_p(1).
\end{align*}
Similarly,
\[
    |D_2|
    \le
    \frac1{n_3T}\sum_{i\in\mathcal I_3}
    \|\delta_i^g\|_2^2
    =
    O_p(\bar r_{g}^2)
    =o_p(1).
\]

Lemma~\ref{lem:DeltaA_rate} gives
$\|\Delta_A\|_{\op}=o_p(1)$. Consequently,
\[
    |D_3|
    \le
    \|\Delta_A\|_{\op}
    \frac1{n_3T}\sum_{i\in\mathcal I_3}
    \|v_{D,i}^*\|_2^2
    =o_p(1),
\]
\begin{align*}
    |D_4|
    &\le
    2\|\Delta_A\|_{\op}
    \left(
        \frac1{n_3T}\sum_{i\in\mathcal I_3}
        \|\delta_i^g\|_2^2
    \right)^{1/2}
    \left(
        \frac1{n_3T}\sum_{i\in\mathcal I_3}
        \|v_{D,i}^*\|_2^2
    \right)^{1/2}\\
    &=o_p(1),
\end{align*}
and
\[
    |D_5|
    \le
    \|\Delta_A\|_{\op}
    \frac1{n_3T}\sum_{i\in\mathcal I_3}
    \|\delta_i^g\|_2^2
    =o_p(1).
\]
Combining these bounds proves the desired result.
\end{proof}
\begin{lemma}[Exact PCA linearization]\label{lem:pca_expansion_dml}
Under Assumption \ref{ass:rate-main}(a), Assumption
\ref{ass:factor_fold}, and Assumption
\ref{ass:training_main}(a),(b),(d), there is an event
\(\mathcal E_{NT}^{\mathrm{PCA}}\), with
\(P(\mathcal E_{NT}^{\mathrm{PCA}})\to1\), on which there are nonsingular
alignment matrices \(H_D,H_G\) and matrices \(B_{D,n},B_{G,n}\). Extend these
proof-specific matrices to the complement of
\(\mathcal E_{NT}^{\mathrm{PCA}}\) by setting \(H_D,H_G\) equal to identity
matrices and \(B_{D,n},B_{G,n}\) equal to zero. Define
\(\mathcal E_D=G_D^0H_D-\widehat G_D\) and
\(\mathcal E_G=G^0H_G-\widehat G\) on the full probability space. Off
\(\mathcal E_{NT}^{\mathrm{PCA}}\), define the remainder matrices as the
residuals in \eqref{eq:EDt}, so that the decomposition continues to hold. Then
\[
\|H_D\|_{\op}+\|H_D^{-1}\|_{\op}
+\|H_G\|_{\op}+\|H_G^{-1}\|_{\op}
+\|B_{D,n}\|_{\op}+\|B_{G,n}\|_{\op}=O_p(1),
\]
and
\begin{equation}\label{eq:EDt}
\mathcal E_D=L_{D\nu}+L_{Dg}+\mathcal R_D,
\qquad
\mathcal E_G=L_{G\xi}+L_{Gm}+\mathcal R_G.
\end{equation}
The linear terms are
\[
    L_{D\nu}
    =-\frac1{n_1}V_1'\bar C_D B_{D,n}',
    \qquad
    L_{Dg}
    =\frac1{n_1}\Delta_{g,1}'\bar C_D B_{D,n}',
\]
and
\[
    L_{G\xi}
    =-\frac1{n_2}\Xi_2'\bar C_Y B_{G,n}',
    \qquad
    L_{Gm}
    =\frac1{n_2}\Delta_{m,2}'\bar C_Y B_{G,n}',
\]
where $\bar C_D$ and $\bar C_Y$ have rows
$\bar C_{D,j}'$ and $\bar C_{Y,j}'$ on their respective
training folds.
Row by row,
\[
L_{D\nu,t}'
=-B_{D,n}\frac1{n_1}\sum_{j\in\mathcal I_1}
\bar C_{D,j}\nu_{jt},
\qquad
L_{Dg,t}'
=B_{D,n}\frac1{n_1}\sum_{j\in\mathcal I_1}
\bar C_{D,j}\delta_{jt}^g,
\]
with analogous definitions of \(L_{G\xi,t}\) and \(L_{Gm,t}\), replacing
\((n_1,\bar C_{D,j},\nu_{jt},\delta_{jt}^g)\) by
\((n_2,\bar C_{Y,j},\xi_{jt},\delta_{jt}^m)\). Moreover,
\begin{align}
\|L_{D\nu}\|_F&=O_p\!\left(\sqrt{T/n_1}\right), &
\|L_{Dg}\|_F&=O_p\!\left(\sqrt{T/n_1}\,\eta_g\right),
\label{eq:pca-linear-rates-D}\\
\|L_{G\xi}\|_F&=O_p\!\left(\sqrt{T/n_2}\right), &
\|L_{Gm}\|_F&=O_p\!\left(\sqrt{T/n_2}\,\eta_m\right),
\label{eq:pca-linear-rates-G}\\
\|\mathcal R_D\|_F&=O_p\!\left(T^{-1/2}+\sqrt T/n_1\right), &
\|\mathcal R_G\|_F&=O_p\!\left(T^{-1/2}+\sqrt T/n_2\right).
\label{eq:pca-quadratic-rates}
\end{align}
Consequently, defining
\(q_k=T^{-1/2}+\sqrt T/n_k\),
\begin{align}
\|\mathcal E_D\|_F
&=O_p\!\left(\sqrt{T/n_1}\,(1+\eta_g)+q_1\right),
&
\frac1T\|\mathcal E_D\|_F^2
&=O_p\!\left(\frac{(1+\eta_g)^2}{n_1}+\frac{q_1^2}{T}\right),
\label{eq:pca-total-rate-D}\\
\|\mathcal E_G\|_F
&=O_p\!\left(\sqrt{T/n_2}\,(1+\eta_m)+q_2\right),
&
\frac1T\|\mathcal E_G\|_F^2
&=O_p\!\left(\frac{(1+\eta_m)^2}{n_2}+\frac{q_2^2}{T}\right).
\label{eq:pca-total-rate-G}
\end{align}
\end{lemma}

\begin{proof}
We prove the treatment-side result. Let \(\bar C_D\) be the
\(n_1\times r_D\) matrix with rows \(\bar C_{D,j}'\), let
\(U_D=V_1-\Delta_{g,1}\), and write the matrix supplied to PCA as
\[
Z_D=\bar C_DG_D^{0\prime}+U_D.
\]
The inverse-based identities below are first read on the event that the
relevant sample eigenvalues are positive. The argument below shows that this
event, and the event on which the resulting alignment is nonsingular, have
probability approaching one.
Let \(\widehat{\mathcal V}_D\) contain the largest \(r_D\) eigenvalues of
\(Z_D'Z_D/(n_1T)\). The PCA normalization and eigen-equation give
\[
T^{-1}\widehat G_D'\widehat G_D=I_{r_D},
\qquad
\frac1{n_1T}Z_D'Z_D\widehat G_D
=\widehat G_D\widehat{\mathcal V}_D.
\]
Expanding this identity and collecting the two terms in the span of \(G_D^0\)
yields the exact equality
\begin{equation}\label{eq:pca-exact-identity}
\widehat G_D
=G_D^0H_D+\frac1{n_1}U_D'\bar C_DK_D
+\frac1{n_1T}U_D'U_D\widehat G_D
\widehat{\mathcal V}_D^{-1},
\end{equation}
where
\[
K_D=\left(\frac1T G_D^{0\prime}\widehat G_D\right)
\widehat{\mathcal V}_D^{-1}
\]
and
\[
H_D=
\left(\frac1{n_1}\bar C_D'\bar C_D\right)
\left(\frac1T G_D^{0\prime}\widehat G_D\right)
\widehat{\mathcal V}_D^{-1}
+\frac1{n_1T}\bar C_D'U_D\widehat G_D
\widehat{\mathcal V}_D^{-1}.
\]
We first make the pervasiveness and alignment steps explicit. Conditional on
the realized augmented loadings, a balanced fold is a simple random sample
without replacement. Since the factor dimensions are fixed and the
full-sample fourth moments of the augmented loadings are bounded in probability, the finite-population
variance formula gives
\[
\left\|
\frac1{n_1}\bar C_D'\bar C_D
-\frac1N\sum_{i=1}^N\bar C_{D,i}\bar C_{D,i}'
\right\|_{\op}=O_p(n_1^{-1/2}).
\]
The same sampling argument and conditional Markov's inequality imply, for
each \(k=1,2,3\),
\[
\frac1{n_k}\sum_{i\in\mathcal I_k^{(l)}}
\left(\|\bar C_{D,i}\|_2^4+\|\bar C_{Y,i}\|_2^4
+\|\bar C_i\|_2^4\right)=O_p(1).
\]
Assumption \ref{ass:factor_fold}(b)--(d) therefore implies fold-specific
pervasiveness. Assumption \ref{ass:training_main}(d) gives
\(\|U_D\|_{\op}=O_p(\sqrt{n_1}+\sqrt T)=o_p(\sqrt{n_1T})\). Consequently,
\[
\left\|
\frac1{n_1T}Z_D'Z_D
-\frac1T G_D^0
\left(\frac1{n_1}\bar C_D'\bar C_D\right)G_D^{0\prime}
\right\|_{\op}=o_p(1).
\]
Weyl's inequality and the singular-subspace perturbation inequality now show
that \(\widehat{\mathcal V}_D\),
\(\widehat{\mathcal V}_D^{-1}\), and \(K_D\) are bounded in probability and
that the singular values of
\(T^{-1}G_D^{0\prime}\widehat G_D\) are bounded away from zero.  In the
definition of \(H_D\), the second summand is \(o_p(1)\), because
\[
\frac1{n_1T}
\|\bar C_D\|_{\op}\|U_D\|_{\op}\|\widehat G_D\|_{\op}=o_p(1).
\]
The first summand is a product of fixed-dimensional matrices whose smallest
singular values are bounded away from zero. Hence \(H_D\) is nonsingular with
\(\|H_D\|_{\op}+\|H_D^{-1}\|_{\op}=O_p(1)\).

Since \(U_D=V_1-\Delta_{g,1}\), subtracting
\eqref{eq:pca-exact-identity} from \(G_D^0H_D\) gives
\eqref{eq:EDt} with \(B_{D,n}=K_D'\) and
\[
\mathcal R_D
=-\frac1{n_1T}U_D'U_D\widehat G_D
\widehat{\mathcal V}_D^{-1}.
\]
Conditional cross-sectional independence, conditional mean zero, the
covariance-operator bound, and the fold-average loading moment bound give
\[
E\!\left[
\left\|\frac1{n_1}V_1'\bar C_D\right\|_F^2
\middle|\mathcal F_T,\{X_j:j\in\mathcal I_1\}
\right]
\le
\frac{CT}{n_1^2}\sum_{j\in\mathcal I_1}\|\bar C_{D,j}\|_2^2
=O_p\!\left(\frac{T}{n_1}\right).
\]
The last order follows from the fold-average loading moment bound. Applying
conditional Markov's inequality on sets where the normalized loading average
is bounded, and then letting the localization constant diverge, proves the
first bound in \eqref{eq:pca-linear-rates-D}. The second follows from
\[
\left\|\frac1{n_1}\Delta_{g,1}'\bar C_D\right\|_F
=\sqrt{\frac{T}{n_1}}
\left\|\frac1{\sqrt{n_1T}}
\Delta_{g,1}'\bar C_D\right\|_F
=O_p\!\left(\sqrt{T/n_1}\,\eta_g\right).
\]
Finally, \(\|\widehat G_D\|_F=\sqrt{Tr_D}\), so
\[
\|\mathcal R_D\|_F
\le
\frac{\|U_D\|_{\op}^2}{n_1T}
\|\widehat G_D\|_F
\|\widehat{\mathcal V}_D^{-1}\|_{\op}
=O_p\!\left(T^{-1/2}+\sqrt T/n_1\right).
\]
The condition \(q_{NT}(1+\eta_g+\eta_m)=o(1)\), together with balanced folds,
makes this bound \(o_p(1)\). Combining the three
components of \(\mathcal E_D\), dividing by \(\sqrt T\), and squaring gives
the stated aggregate rate. The reduced-form result follows identically from
\(Z_G=\bar C_YG^{0\prime}+\Xi_2-\Delta_{m,2}\), using the
\(\bar C_{Y,i}\)-weighted rate in Assumption \ref{ass:rate-main}.
\end{proof}

\begin{lemma}\label{lem:DeltaA_rate}
Under Assumption \ref{ass:rate-main}(a), Assumption
\ref{ass:factor_fold}, and Assumption
\ref{ass:training_main}(a),(b),(d), let
\[
\alpha_{NT}
=
n_1^{-1/2}+n_2^{-1/2}
+\frac{\eta_g}{\sqrt{n_1}}+\frac{\eta_m}{\sqrt{n_2}}
+T^{-1}+n_1^{-1}+n_2^{-1}.
\]
Then
\[
\|\Delta_A\|_{\op}=O_p(\alpha_{NT})=o_p(1),
\qquad
\|\Delta_A\|_F=O_p(\alpha_{NT})=o_p(1).
\]
\end{lemma}

\begin{proof}
By the definitions of \(\widehat A\) and \(A_0\), adding and
subtracting \(M_{\widehat G_D}M_{G^0}\) gives
\[
\Delta_A
=
M_{\widehat G_D}(M_{\widehat G}-M_{G^0})
+
(M_{\widehat G_D}-M_{G_D^0})M_{G^0}.
\]
It therefore suffices to control the two estimated factor spaces. By
Lemma \ref{lem:pca_expansion_dml},
\[
\frac{\|\mathcal E_D\|_F}{\sqrt T}
=
O_p\!\left(
n_1^{-1/2}+\frac{\eta_g}{\sqrt{n_1}}+T^{-1}+n_1^{-1}
\right),
\]
and the analogous bound holds for \(\mathcal E_G\). Pervasiveness and the
bounded, nonsingular alignments imply
\[
s_{\min}(G_D^0H_D)\asymp_p\sqrt T,
\qquad
s_{\min}(G^0H_G)\asymp_p\sqrt T.
\]
The singular-subspace perturbation inequality consequently gives
\[
\|P_{\widehat G_D}-P_{G_D^0}\|_{\op}
\lesssim_p \frac{\|\mathcal E_D\|_{\op}}{\sqrt T},
\qquad
\|P_{\widehat G}-P_{G^0}\|_{\op}
\lesssim_p \frac{\|\mathcal E_G\|_{\op}}{\sqrt T}.
\]
Combining the last three displays proves the operator-norm bound. Finally,
\(\Delta_A\) has rank at most \(2r_D+2r_Y\). Because the factor dimensions
are fixed,
\[
\|\Delta_A\|_F
\le \sqrt{2r_D+2r_Y}\,\|\Delta_A\|_{\op}
=O_p(\alpha_{NT}),
\]
which completes the proof.
\end{proof}

\begin{proposition}\label{proposition:leading}
Under Assumptions \ref{ass:factor_fold}(d)--(e) and
\ref{ass:eval_main}(a)--(c), the first term in
\eqref{eq:theta_l_expand} satisfies
\[
\frac{1}{\sqrt{n_3T}}\sum_{i\in \mathcal I_3^{(l)}}
\nu_{i}'A_0\varepsilon_i\xrightarrow{d} N(0,\Omega_0).
\]
\end{proposition}
\begin{proof}
Define
\[
W_i=\nu_i'A_0\varepsilon_i.
\]
Then
\[
\frac1{\sqrt{n_3T}}\sum_{i\in \mathcal I_3^{(l)}}\nu_{i}'A_0\varepsilon_i
=
\frac1{\sqrt{n_3}}\sum_{i\in \mathcal I_3^{(l)}}\frac{W_i}{\sqrt T}.
\]

Conditional on \(\mathcal G_l\), the variables \(\{W_i : i\in \mathcal I_3^{(l)}\}\) are independent and mean zero. Also,
\[
\frac1{n_3}\sum_{i\in \mathcal I_3^{(l)}}
\Var\!\left(\frac{W_i}{\sqrt T}\,\middle|\,\mathcal G_l\right)
=
\frac1{Tn_3}\sum_{i\in \mathcal I_3^{(l)}}
E\!\left(W_i^2\mid \mathcal G_l\right)
\xrightarrow{p} \Omega_0.
\]

Moreover, by the moment condition,
\[
\frac1{n_3^{1+\delta/2}}
\sum_{i\in \mathcal I_3^{(l)}}
E\!\left[
\left|\frac{W_i}{\sqrt T}\right|^{2+\delta}
\,\middle|\,
\mathcal G_l
\right]
\le
C n_3^{-\delta/2}
\to 0.
\]

Therefore the conditional Lyapunov condition holds, and the conditional
Lyapunov CLT yields
\[
\frac1{\sqrt{n_3T}}\sum_{i\in \mathcal I_3^{(l)}}\nu_{i}'A_0\varepsilon_i
\;\xrightarrow{d}\;
N(0,\Omega_0).
\]
\end{proof}

\begin{proposition}\label{proposition:R2}
Under the conditions of Theorem \ref{thm:CLT},
\[
\frac{1}{\sqrt{n_3T}}\sum_{i\in \mathcal{I}_3^{(l)}}R_{2i}=o_p(1).
\]
\end{proposition}

 \begin{proof}
Write
\begin{align*}
  \frac1{\sqrt{n_3T}}\sum_{i\in \mathcal I_3^{(l)}} R_{2i}
  &=\frac1{\sqrt{n_3T}}
\sum_{i\in \mathcal I_3^{(l)}} \bar C_{D,i}'G_D^{0\prime}\hat A G^0 \bar C_i\\
&+\frac1{\sqrt{n_3T}}
\sum_{i\in \mathcal I_3^{(l)}} \bar C_{D,i}'G_D^{0\prime}\hat A \varepsilon_i\\
&+\frac1{\sqrt{n_3T}}
\sum_{i\in \mathcal I_3^{(l)}} \nu_i'\hat A G^0 \bar C_i\\
&=S_{21}+S_{22}+S_{23}.
\end{align*}

For \(S_{22}\), note that conditional on \(\mathcal G_l\), the matrix \(\hat A\) is fixed, and the vectors
\(\{\varepsilon_i:i\in \mathcal I_3^{(l)}\}\) are independent with conditional mean
zero. Hence
\[
E(S_{22}\mid \mathcal G_l)=0.
\]
For the conditional variance, using Assumptions \ref{ass:factor_fold}(d)--(e),
\ref{ass:training_main}(b), and \ref{ass:eval_main}(a),
\begin{align*}
\Var(S_{22}\mid \mathcal G_l)
&=
\frac1{n_3T}
\sum_{i\in \mathcal I_3^{(l)}}
\bar C_{D,i}'G_D^{0\prime}\hat A\Sigma_{\varepsilon,i}\hat A'G_D^0 \bar C_{D,i} \\
&\le
\frac{C}{n_3T}
\sum_{i\in \mathcal I_3^{(l)}}
\|\bar C_{D,i}\|^2\,\|G_D^{0\prime}\hat A\|_F^2 \\
&\le C\left(\frac1{n_3}\sum_{i\in \mathcal I_3^{(l)}}\|\bar C_{D,i}\|^2\right)
\left(\frac1T\|\mathcal E_D\|_F^2\right)\\
&=O_p\!\left(
\frac{(1+\eta_g)^2}{n_1}+\frac{q_1^2}{T}
\right),
\end{align*}
where we have used the fact that
\[
\|G_D^{0\prime}\hat A\|_F
\le
\|H_D^{-1}\|_{\op}\,\|\mathcal E_D\|_F\,\|\hat A\|_{\op}
\le
C\|\mathcal E_D\|_F.
\]
Together with Lemma~\ref{lem:pca_expansion_dml}, this bound and conditional
Chebyshev's inequality give
\[
S_{22}
=
O_p\!\left(
\frac{1+\eta_g}{\sqrt{n_1}}+\frac{q_1}{\sqrt T}
\right)
=
o_p(1).
\]
Here \(q_1=O(q_{NT})\) by balanced folds. Moreover,
\(\eta_g/\sqrt{n_1}=O(\sqrt{T/N}\,\eta_g/\sqrt T)=o(1)\), so the last
equality does not require \(\eta_g=o(1)\).

The term \(S_{23}\) is handled similarly. We have
\[
S_{23}
=
\frac1{\sqrt{n_3T}}
\sum_{i\in \mathcal I_3^{(l)}} \nu_i'\hat A G^0 \bar C_i.
\]
Conditional on \(\mathcal G_l\), the matrix \(\hat A G^0\) is fixed, and
\(\{\nu_i:i\in \mathcal I_3^{(l)}\}\) are independent with conditional mean zero.
Therefore
\[
E(S_{23}\mid \mathcal G_l)=0.
\]
Using Assumptions \ref{ass:factor_fold}(d)--(e) and \ref{ass:training_main}(b),
\begin{align*}
\Var(S_{23}\mid \mathcal G_l)
&=
\frac1{n_3T}
\sum_{i\in \mathcal I_3^{(l)}}
\bar C_i'G^{0\prime}\hat A'\Sigma_{\nu,i}\hat A G^0 \bar C_i \\
&\le
\frac{C}{n_3T}
\sum_{i\in \mathcal I_3^{(l)}}
\|\bar C_i\|^2\,\|\hat A G^0\|_F^2 \\
&\le C\left(\frac1{n_3}\sum_{i\in \mathcal I_3^{(l)}}\|\bar C_i\|^2\right)
\left(\frac1T\|\mathcal E_G\|_F^2\right)\\
&=O_p\!\left(
\frac{(1+\eta_m)^2}{n_2}+\frac{q_2^2}{T}
\right),
\end{align*}
where we have used the fact that
\[
\|\hat A G^0\|_F
\le
\|\hat A\|_{\op}\,\|\mathcal E_G\|_F\,\|H_G^{-1}\|_{\op}
\le
C\|\mathcal E_G\|_F.
\]
Together with Lemma~\ref{lem:pca_expansion_dml}, this bound and conditional
Chebyshev's inequality yield
\[
S_{23}
=
O_p\!\left(
\frac{1+\eta_m}{\sqrt{n_2}}+\frac{q_2}{\sqrt T}
\right)
=
o_p(1).
\]

It remains to bound \(S_{21}\). Using
\[
G_D^{0\prime}\hat A G^0
=
H_D^{-\prime}\mathcal E_D'\hat A \mathcal E_G H_G^{-1},
\]
we obtain
\[
S_{21}
=
\sqrt{\frac{n_3}{T}}\,
\tr\!\left[
\left(\frac1{n_3}\sum_{i\in \mathcal I_3^{(l)}} \bar C_i \bar C_{D,i}'\right)
H_D^{-\prime}\mathcal E_D'\hat A \mathcal E_G H_G^{-1}
\right].
\]
Hence
\[
|S_{21}|
\le
\left\|
\frac1{n_3}\sum_{i\in \mathcal I_3^{(l)}} \bar C_i \bar C_{D,i}'
\right\|_F
\|H_D^{-1}\|_{\op}
\|H_G^{-1}\|_{\op}
\sqrt{\frac{n_3}{T}}\,
\|\mathcal E_D'\hat A \mathcal E_G\|_{F}.
\]
The first factor is \(O_p(1)\) by the fold-average fourth-moment condition.
Because the factor dimensions are fixed, Frobenius and operator norms are
equivalent for the last matrix, and Lemma
\ref{lem:double_projection_bound_sharp} makes the last factor \(o_p(1)\). Therefore
\[
S_{21}=o_p(1).
\]

Combining the three bounds gives
\[
\frac1{\sqrt{n_3T}}\sum_{i\in \mathcal I_3^{(l)}} R_{2i} = o_p(1).
\]
\end{proof}

\begin{lemma}[Cross-fitted double-projection bound]
\label{lem:double_projection_bound_sharp}
Under Assumption \ref{ass:rate-main}(a), Assumption
\ref{ass:factor_fold}, and Assumption
\ref{ass:training_main}(a),(b),(d),
\[
\sqrt{\frac{n_3}{T}}\,
\|\mathcal E_D'\widehat A\mathcal E_G\|_{\op}
=o_p(1).
\]
\end{lemma}

\begin{proof}
Write
\[
L_D=L_{D\nu}+L_{Dg},
\qquad
L_G=L_{G\xi}+L_{Gm}.
\]
The exact linearization in Lemma \ref{lem:pca_expansion_dml} gives
\[
\mathcal E_D=L_D+\mathcal R_D,
\qquad
\mathcal E_G=L_G+\mathcal R_G.
\]
We first control \(L_D'\widehat A L_G\). Decompose
\(\widehat A=A_0+\Delta_A\). For the structural--structural component, let
\(\widetilde L_{D\nu}\) and \(\widetilde L_{G\xi}\) denote the matrices
defined as \(L_{D\nu}\) and \(L_{G\xi}\), but without the bounded
finite-dimensional matrices \(B_{D,n}\) and \(B_{G,n}\). Conditional on the
factor paths and training-fold covariates, cross-sectional independence and
the conditional mean restrictions imply
\begin{align*}
E\!\left[
\|\widetilde L_{D\nu}'A_0\widetilde L_{G\xi}\|_F^2
\,\middle|\,
\mathcal F_T,\{X_j:j\in\mathcal I_1\cup\mathcal I_2\}
\right]
&\le \frac{CT}{n_1n_2}
\left(\frac1{n_1}\sum_{j\in\mathcal I_1}\|\bar C_{D,j}\|_2^2\right)
\left(\frac1{n_2}\sum_{k\in\mathcal I_2}\|\bar C_{Y,k}\|_2^2\right)\\
&=O_p\!\left(\frac{T}{n_1n_2}\right).
\end{align*}
Indeed, after expanding the squared Frobenius norm, all terms with an
unmatched training index have conditional mean zero; the remaining
\(n_1n_2\) terms, before multiplication by their loading weights, are bounded by
\[
E[(\nu_j'A_0\xi_k)^2\mid\mathcal F_T,X_j,X_k]
\le C\|A_0\|_F^2\le CT.
\]
The last stochastic order uses the two fold-average loading moment bounds.
Conditional Markov's inequality after localization, together with the fact
that the \(B\)-matrices are \(O_p(1)\), therefore gives
\[
\sqrt{\frac{n_3}{T}}\,
\|L_{D\nu}'A_0L_{G\xi}\|_{\op}
=O_p\!\left(\sqrt{\frac{n_3}{n_1n_2}}\right)
=o_p(1).
\]

The remaining terms in \(L_D'A_0L_G\) contain at least one first-stage
component. Equations \eqref{eq:pca-linear-rates-D} and
\eqref{eq:pca-linear-rates-G} give, for example,
\[
\sqrt{\frac{n_3}{T}}\,
\|L_{D\nu}'A_0L_{Gm}\|_{\op}
=
O_p\!\left(
\sqrt{\frac{n_3T}{n_1n_2}}\,\eta_m
\right)
=o_p(1),
\]
and symmetrically for \(L_{Dg}'A_0L_{G\xi}\). The component
\(L_{Dg}'A_0L_{Gm}\) has the same bound with
\(\eta_m\) replaced by \(\eta_g\eta_m\). The two structural--generated
terms are \(o_p(1)\) by the one-sided restrictions in Assumption
\ref{ass:rate-main}; the generated--generated term is \(o_p(1)\) by its joint
restriction \(\sqrt{T/N}\,\eta_g\eta=o(1)\).

For the projector-perturbation component, Lemma \ref{lem:DeltaA_rate} gives
\[
\sqrt{\frac{n_3}{T}}\,
\|L_D'\Delta_A L_G\|_{\op}
=
O_p\!\left(
\sqrt{\frac TN}\,(1+\eta_g)(1+\eta_m)\alpha_{NT}
\right)
=o_p(1).
\]
To verify the last equality without imposing bounded \(\eta_g\) or \(\eta_m\),
put \(a=\sqrt{T/N}\), \(b=N^{-1/2}\), and
\(c=T^{-1}+N^{-1}\). Balanced folds give
\(\alpha_{NT}\lesssim b(1+\eta_g+\eta_m)+c\). Moreover,
\(b,c=O(q_{NT})\), \(ab\le q_{NT}\), and \(ac=o(1)\). Expanding the finite
product above, its remaining monomials are products of the vanishing quantities
\[
a\eta_g,\quad a\eta_m,\quad a\eta_g\eta,
\quad\text{and}\quad q_{NT}(1+\eta_g+\eta_m),
\]
so the displayed order is \(o_p(1)\).

It remains to control terms containing a quadratic PCA remainder. Put
\[
q_k=T^{-1/2}+\frac{\sqrt T}{n_k},
\qquad k=1,2.
\]
Lemma \ref{lem:pca_expansion_dml} implies
\begin{align*}
\|\mathcal R_D\|_F&=O_p(q_1),
\\
\|\mathcal R_G\|_F&=O_p(q_2),\\
\|L_D\|_F&=O_p\!\left(\sqrt{T/n_1}\,(1+\eta_g)\right),
\\
\|L_G\|_F&=O_p\!\left(\sqrt{T/n_2}\,(1+\eta_m)\right).
\end{align*}
Because \(\|\widehat A\|_{\op}\le1\),
\[
\sqrt{\frac{n_3}{T}}
\left\{
\|\mathcal R_D'\widehat A L_G\|_{\op}
+
\|L_D'\widehat A\mathcal R_G\|_{\op}
\right\}
=O_p\!\left(q_1(1+\eta_m)+q_2(1+\eta_g)\right)
=o_p(1),
\]
where \(q_k=O(q_{NT})\) for balanced folds.
Finally,
\[
\sqrt{\frac{n_3}{T}}\,
\|\mathcal R_D'\widehat A\mathcal R_G\|_{\op}
=
O_p\!\left(\sqrt{\frac NT}\,q_1q_2\right)
=o_p(1),
\]
where the last rate follows from
\(\sqrt{N/T}\,q_{NT}^2=o(1)\).
Combining the preceding bounds proves the result.
\end{proof}
\begin{proposition}\label{proposition:R3}
Under the conditions of Theorem \ref{thm:CLT},
\[
\frac{1}{\sqrt{n_3T}}\sum_{i\in \mathcal I_3^{(l)}}R_{3i}=o_p(1).
\]
       \end{proposition}
\begin{proof}
Recall the decomposition:
\[
R_{3i}
=
-\delta_i^{g\prime}\Delta_A\varepsilon_i
-\nu_i'\Delta_A\delta_i
+\delta_i^{g\prime}\Delta_A\delta_i
+\nu_i'\Delta_A\varepsilon_i.
\]
We bound each term separately.

For the first term, conditional on $\mathcal{G}_l$, both $\delta_i^g$ and $\Delta_A$ are fixed, and $\varepsilon_i$ has mean zero. Following the same conditional-variance argument used in Proposition \ref{proposition:1st_stage}, we obtain
\[
\frac1{\sqrt{n_3T}}
\sum_{i\in \mathcal I_3^{(l)}}\delta_i^{g\prime}\Delta_A\varepsilon_i
=
O_p(\|\Delta_A\|_{\op}\bar r_{g}).
\]
Applying the corresponding covariance bound for \(\nu_i\) to the second term
yields
\[
\frac1{\sqrt{n_3T}}
\sum_{i\in \mathcal I_3^{(l)}}\nu_i'\Delta_A\delta_i
=
O_p\!\left(\|\Delta_A\|_{\op}
(\bar r_{m}+\bar r_{g})\right).
\]
For the third term, we apply the Cauchy--Schwarz inequality:
\begin{align*}
\left| \frac1{\sqrt{n_3T}} \sum_{i\in \mathcal I_3^{(l)}}\delta_i^{g\prime}\Delta_A\delta_i \right|
&\le
\frac{\|\Delta_A\|_{\op}}{\sqrt{n_3T}} \sum_{i\in \mathcal I_3^{(l)}} \|\delta_i^g\|\,\|\delta_i\|.
\end{align*}
Since \(\delta_i=\delta_i^m-\theta_0\delta_i^g\), Proposition
\ref{proposition:1st_stage} implies
\[
\frac1{\sqrt{n_3T}}
\sum_{i\in \mathcal I_3^{(l)}}\|\delta_i^g\|\,\|\delta_i\|
=
o_p(1).
\]
Together with Lemma \ref{lem:DeltaA_rate}, this yields
\[
\frac1{\sqrt{n_3T}}
\sum_{i\in \mathcal I_3^{(l)}}\delta_i^{g\prime}\Delta_A\delta_i
=
o_p(1).
\]
Finally, by Lemma \ref{lem:DeltaA_cross}, the fourth term satisfies:
\[
\frac1{\sqrt{n_3T}}
\sum_{i\in \mathcal I_3^{(l)}}\nu_i'\Delta_A\varepsilon_i
=o_p(1).
\]
Combining these bounds, we have:
\[
\frac1{\sqrt{n_3T}}
\sum_{i\in \mathcal I_3^{(l)}}R_{3i}
=
o_p(1).
\]
\end{proof}
\begin{lemma}\label{lem:DeltaA_cross}
Under the conditions of Theorem \ref{thm:CLT},
\[
\frac1{\sqrt{n_3T}}\sum_{i\in \mathcal I_3^{(l)}}\nu_i'\Delta_A\varepsilon_i=o_p(1).
\]
\end{lemma}

\begin{proof}
Note that
\[
E\!\left[ \frac1{\sqrt{n_3T}}\sum_{i\in \mathcal I_3^{(l)}}\nu_i'\Delta_A\varepsilon_i \;\middle|\; \mathcal{G}_l \right] = 0.
\]
For the conditional variance, iterated expectations and Assumptions
\ref{ass:factor_fold}(d)--(e), \ref{ass:training_main}(b), and
\ref{ass:eval_main}(a) yield
\begin{align*}
    \Var\!\left( \frac1{\sqrt{n_3T}}\sum_{i\in \mathcal I_3^{(l)}}\nu_i'\Delta_A\varepsilon_i \;\middle|\; \mathcal{G}_l \right)
&=\frac{1}{n_3T} \sum_{i\in \mathcal I_3^{(l)}} E\!\left[ \tr(\Delta_A \varepsilon_i \varepsilon_i' \Delta_A' \nu_i \nu_i') \;\middle|\;\mathcal{G}_l \right]\\
&\le \frac{C}{n_3T} \sum_{i\in \mathcal I_3^{(l)}}
E[\nu_i'\Delta_A\Delta_A'\nu_i\mid\mathcal G_l]\\
&\le \frac{C^2}{T}\|\Delta_A\|_F^2=o_p(1),
\end{align*}
where we have used Lemma \ref{lem:DeltaA_rate}. By Chebyshev's inequality, we conclude:
\[
\frac1{\sqrt{n_3T}}\sum_{i\in \mathcal I_3^{(l)}}\nu_i'\Delta_A\varepsilon_i = o_p(1).
\]
\end{proof}
\begin{proposition}\label{proposition:R4}
Under Assumption \ref{ass:rate-main}(a), Assumption
\ref{ass:factor_fold}, and Assumption
\ref{ass:training_main}(a),(b),(d),
\[
\frac{1}{\sqrt{n_3T}}\sum_{i\in \mathcal I_3^{(l)}}R_{4i}=o_p(1).
\]
\end{proposition}
\begin{proof}
Because \(\widehat G_D'\widehat A=0\) and
\(\widehat A\widehat G=0\),
\[
G_D^{0\prime}\widehat A
=H_D^{-\prime}\mathcal E_D'\widehat A,
\qquad
\widehat A G^0
=\widehat A\mathcal E_GH_G^{-1}.
\]
Define
\[
M_f=\frac1{\sqrt{n_3T}}
\sum_{i\in\mathcal I_3^{(l)}}\delta_i\bar C_{D,i}',
\qquad
M_g=\frac1{\sqrt{n_3T}}
\sum_{i\in\mathcal I_3^{(l)}}\delta_i^g\bar C_i'.
\]
The loading-weighted rates in Assumption \ref{ass:rate-main} imply
\[
\|M_f\|_F=O_p(\eta),
\qquad
\|M_g\|_F=O_p(\eta_g),
\qquad
\eta^2=\eta_g^2+\eta_m^2.
\]
Summing the two components of \(R_{4i}\) and using cyclicity of the trace gives
\begin{equation}\label{eq:R4}
\frac1{\sqrt{n_3T}}\sum_{i\in\mathcal I_3^{(l)}}R_{4i}
=
-\operatorname{tr}\!\left(
\mathcal E_D'\widehat A M_fH_D^{-\prime}
\right)
-\operatorname{tr}\!\left(
\mathcal E_GH_G^{-1}M_g'\widehat A
\right).
\end{equation}
By the trace Cauchy--Schwarz inequality,
\[
\left|
\operatorname{tr}\!\left(
\mathcal E_D'\widehat A M_fH_D^{-\prime}
\right)
\right|
\le
\|\mathcal E_D\|_F\,
\|\widehat A\|_{\op}\,
\|M_f\|_F\,
\|H_D^{-1}\|_{\op}.
\]
Lemma \ref{lem:pca_expansion_dml} yields
\begin{align*}
&\|\mathcal E_D\|_F\|M_f\|_F
+\|\mathcal E_G\|_F\|M_g\|_F\\
&\quad=
O_p\!\left(
\left\{\sqrt{\frac TN}(1+\eta_g)+q_{NT}\right\}\eta
+
\left\{\sqrt{\frac TN}(1+\eta_m)+q_{NT}\right\}\eta_g
\right)
=o_p(1).
\end{align*}
The last equality follows from the two one-sided rates, the joint rate
\(\sqrt{T/N}\,\eta_g\eta=o(1)\), and
\(q_{NT}(1+\eta_g+\eta_m)=o(1)\). It does not use
\(\eta_g,\eta_m=o(1)\). Therefore the preceding trace is \(o_p(1)\), and the
same argument gives
\[
\left|
\operatorname{tr}\!\left(
\mathcal E_GH_G^{-1}M_g'\widehat A
\right)
\right|
\le
\|\mathcal E_G\|_F
\|H_G^{-1}\|_{\op}\|M_g\|_F\|\widehat A\|_{\op}
=o_p(1).
\]
Substitution into \eqref{eq:R4} proves the proposition.
\end{proof}
\begin{proposition}\label{proposition:num_variance}
    Under the assumptions of Theorem \ref{thm:var-consistency}, \[\frac{1}{Tn_3}
\sum_{i\in\mathcal I_3^{(l)}}
\bigl(\hat v_{D,i}'\hat A(\hat v_{Y,i}-\hat{\theta}^{(l)}\hat{v}_{D,i})\bigr)^2\xrightarrow{p}\Omega_0.
\]
\end{proposition}
\begin{proof}
Set
\[
\widehat Z_i
=
\widehat v_{D,i}'\widehat A
(\widehat v_{Y,i}-\widehat\theta^{(l)}\widehat v_{D,i}),
\qquad
Z_i=\nu_i'A_0\varepsilon_i.
\]
The score expansion gives
\begin{align*}
\widehat Z_i-Z_i
&=
\nu_i'\Delta_A\varepsilon_i
+\bar C_{D,i}'G_D^{0\prime}\widehat A\varepsilon_i
+\nu_i'\widehat A G^0\bar C_i
+\bar C_{D,i}'G_D^{0\prime}\widehat A G^0\bar C_i\\
&\quad
-\delta_i^{g\prime}\widehat A u_i^*
-v_{D,i}^{*\prime}\widehat A\delta_i
+\delta_i^{g\prime}\widehat A\delta_i
-(\widehat\theta^{(l)}-\theta_0)
 \widehat v_{D,i}'\widehat A\widehat v_{D,i}\\
&=S_{1i}+\cdots+S_{7i}.
\end{align*}
We show that
\begin{equation}\label{eq:variance-score-L2}
\frac1{n_3T}\sum_{i\in\mathcal I_3^{(l)}}
(\widehat Z_i-Z_i)^2=o_p(1).
\end{equation}

For the first three terms, the conditional covariance bounds and the fact that
\(\widehat A\) is \(\mathcal G_l\)-measurable give
\[
E(S_{1i}^2\mid\mathcal G_l)
\le C\|\Delta_A\|_F^2,
\]
\[
E(S_{2i}^2\mid\mathcal G_l)
=
\bar C_{D,i}'G_D^{0\prime}\Delta_A
\Sigma_{\varepsilon,i}\Delta_A'G_D^0\bar C_{D,i},
\]
and
\[
E(S_{3i}^2\mid\mathcal G_l)
=
\bar C_i'G^{0\prime}\widehat A'
\Sigma_{\nu,i}\widehat A G^0\bar C_i.
\]
Here \(G_D^{0\prime}\Delta_A=G_D^{0\prime}\widehat A\). Using
\[
G_D^{0\prime}\widehat A
=H_D^{-\prime}\mathcal E_D'\widehat A,
\qquad
\widehat A G^0=\widehat A\mathcal E_GH_G^{-1},
\]
Lemmas \ref{lem:pca_expansion_dml} and \ref{lem:DeltaA_rate}, and conditional
Markov's inequality, we obtain
\[
\frac1{n_3T}\sum_i(S_{1i}^2+S_{2i}^2+S_{3i}^2)=o_p(1).
\]

The same identities imply
\[
S_{4i}
=
\bar C_{D,i}'H_D^{-\prime}
\mathcal E_D'\widehat A\mathcal E_G
H_G^{-1}\bar C_i.
\]
The fold-average fourth moments of the augmented loadings are bounded in
probability, the alignments are \(O_p(1)\), and the factor dimensions are fixed.
Hence Lemma \ref{lem:double_projection_bound_sharp}
gives
\[
\frac1{n_3T}\sum_i S_{4i}^2
\le
O_p(1)\frac{1}{T}
\|\mathcal E_D'\widehat A\mathcal E_G\|_F^2
=o_p(1).
\]

Decompose the nuisance component as
\[
S_{5i}
=
-\delta_i^{g\prime}\widehat A\varepsilon_i
-\nu_i'\widehat A\delta_i
-\delta_i^{g\prime}\widehat A G^0\bar C_i
-\bar C_{D,i}'G_D^{0\prime}\widehat A\delta_i.
\]
For its first two terms, conditional second moments are bounded, respectively,
by \(C\|\delta_i^g\|_2^2\) and \(C\|\delta_i\|_2^2\). After averaging,
Proposition~\ref{proposition:1st_stage} makes their contribution
\(O_p\{\bar r_{g}^2+(\bar r_{m}+\bar r_{g})^2\}=o_p(1)\).
For the last two,
Cauchy--Schwarz, the preceding projection identities, and the two
loading-weighted squared prediction rates in Assumption
\ref{ass:rate-main} give
\begin{align*}
\frac1{n_3T}\sum_i
(\delta_i^{g\prime}\widehat A G^0\bar C_i)^2
&\le
O_p(1)\|\mathcal E_G\|_F^2
\frac1{n_3T}\sum_i\|\delta_i^g\|_2^2\|\bar C_i\|_2^2\\
&=O_p\!\left(\eta_g^2\|\mathcal E_G\|_F^2\right)
=O_p\!\left\{
\frac TN\eta_g^2(1+\eta_m)^2+q_{NT}^2\eta_g^2
\right\}=o_p(1),
\end{align*}
and
\begin{align*}
\frac1{n_3T}\sum_i
(\bar C_{D,i}'G_D^{0\prime}\widehat A\delta_i)^2
&\le
O_p(1)\|\mathcal E_D\|_F^2
\frac1{n_3T}\sum_i\|\delta_i\|_2^2\|\bar C_{D,i}\|_2^2\\
&=O_p\!\left(\eta^2\|\mathcal E_D\|_F^2\right)
=O_p\!\left\{
\frac TN\eta^2(1+\eta_g)^2+q_{NT}^2\eta^2
\right\}=o_p(1).
\end{align*}
The last equalities use the two one-sided rates, the square of
\(\sqrt{T/N}\,\eta_g\eta=o(1)\), and
\(q_{NT}(1+\eta_g+\eta_m)=o(1)\). Thus
\[
\frac1{n_3T}\sum_iS_{5i}^2=o_p(1).
\]
Moreover,
\[
\frac1{n_3T}\sum_iS_{6i}^2
\le
\frac1{n_3T}
\left(\sum_i\|\delta_i^g\|_2^2\right)
\left(\sum_i\|\delta_i\|_2^2\right)
=o_p(1)
\]
by Proposition \ref{proposition:1st_stage} and its product-rate condition.

The splitwise expansion used in the proof of Theorem \ref{thm:CLT} gives
\(\widehat\theta^{(l)}-\theta_0=O_p((NT)^{-1/2})\). In addition,
Assumptions \ref{ass:factor_fold}, \ref{ass:training_main}(c), and
\ref{ass:rate-main} imply
\[
\frac1{n_3T^2}\sum_{i\in\mathcal I_3^{(l)}}
\|\widehat v_{D,i}\|_2^4=O_p(1).
\]
To see this, use
\(\widehat v_{D,i}=G_D^0\bar C_{D,i}+\nu_i-\delta_i^g\),
\(\|G_D^0\|_{\op}^2=O_p(T)\), the fold-average fourth-moment bound for
\(\bar C_{D,i}\), and Assumption \ref{ass:training_main}(c). Conditional on
the realized disturbance paths, the random-fold sampling argument transfers
the latter full-sample average bound to \(\mathcal I_3^{(l)}\).
It follows that
\[
\frac1{n_3T}\sum_iS_{7i}^2
\le
O_p((NT)^{-1})
\frac1{n_3T}\sum_i\|\widehat v_{D,i}\|_2^4
=O_p(N^{-1})=o_p(1).
\]
This proves \eqref{eq:variance-score-L2}.

It remains to establish the oracle second-moment law of large numbers. Let
\(W_i=Z_i\) and
\(q=1+\min\{\delta/2,1\}\in(1,2]\). Conditional on \(\mathcal G_l\), the
variables \(W_i^2-E(W_i^2\mid\mathcal G_l)\) are independent and mean zero.
Let \(a=2q/(2+\delta)\le1\). Conditional centering, Lyapunov's inequality,
and concavity give
\begin{align*}
\sum_{i\in\mathcal I_3^{(l)}}
E\!\left[
\left|W_i^2-E(W_i^2\mid\mathcal G_l)\right|^q
\middle|\mathcal G_l
\right]
&\le C\sum_i E(|W_i|^{2q}\mid\mathcal G_l)\\
&\le C n_3^{1-a}
\left\{\sum_i E(|W_i|^{2+\delta}\mid\mathcal G_l)\right\}^{a}.
\end{align*}
The fold-average Lyapunov bound in Assumption \ref{ass:eval_main}(b) therefore
implies
\[
\sum_{i\in\mathcal I_3^{(l)}}
E\!\left[
\left|W_i^2-E(W_i^2\mid\mathcal G_l)\right|^q
\middle|\mathcal G_l
\right]
\le C n_3T^q
\]
almost surely. The von Bahr--Esseen inequality consequently yields
\[
E\!\left[
\left|
\frac1{n_3T}\sum_i
\{W_i^2-E(W_i^2\mid\mathcal G_l)\}
\right|^q
\middle|\mathcal G_l
\right]
\le Cn_3^{1-q}=o_p(1).
\]
Therefore, Assumption \ref{ass:eval_main}(c) implies
\[
\frac1{n_3T}\sum_i Z_i^2\xrightarrow{p}\Omega_0.
\]
Finally, \eqref{eq:variance-score-L2}, the preceding convergence, and
Cauchy--Schwarz imply
\[
\left|
\frac1{n_3T}\sum_i(\widehat Z_i^2-Z_i^2)
\right|=o_p(1),
\]
which proves the proposition.
\end{proof}

\noindent\textbf{Proof of Theorem \ref{thm:rank_consistency_hl}}
\begin{proof}
We prove the result for the treatment equation. At the final nuclear-norm
step, the matrix supplied to singular-value thresholding is
\[
D-\widehat g^{(nn)}(X)=\Theta_D+\widetilde V_D.
\]
Because the terminal objective defining \(\widehat\Theta_D\) in
Section~\ref{subsec:rank-selection} is the proximal problem for the nuclear
norm,
\[
s_j(\widehat\Theta_D)
=
\{s_j(\Theta_D+\widetilde V_D)-\rho_D\}_+,
\qquad
j=1,\ldots,\min\{N,T\}.
\]
Assumption~\ref{assp:rank_consistency}(a) implies that, with probability approaching one,
there are constants \(0<c_D<C_D<\infty\) such that
\[
c_D\sqrt{NT}
\le s_{r_D}(\Theta_D)
\le s_1(\Theta_D)
\le C_D\sqrt{NT},
\qquad
s_j(\Theta_D)=0\quad(j>r_D).
\]

For \(j>r_D\), Weyl's inequality and the soft-thresholding formula give
\[
s_j(\widehat\Theta_D)
\le
\{\|\widetilde V_D\|_{\op}-\rho_D\}_+
\le \|\widetilde V_D\|_{\op}
=O_p(\rho_D).
\]
For \(j\le r_D\), the same inequality gives
\[
s_j(\widehat\Theta_D)
\ge
s_{r_D}(\Theta_D)-\|\widetilde V_D\|_{\op}-\rho_D
\ge c_D\sqrt{NT}-O_p(\rho_D).
\]
Since \(\rho_D/\sqrt{NT}\to_p0\), the \(r_D\) signal singular values of
\(\widehat\Theta_D\) are bounded below by a constant multiple of
\(\sqrt{NT}\) with probability approaching one. Moreover,
\[
s_1(\widehat\Theta_D)
\le s_1(\Theta_D+\widetilde V_D)
\le s_1(\Theta_D)+\|\widetilde V_D\|_{\op}
=O_p(\sqrt{NT}).
\]
Together with
\(s_1(\widehat\Theta_D)\ge s_{r_D}(\widehat\Theta_D)\), this proves
\(s_1(\widehat\Theta_D)\asymp_p\sqrt{NT}\).

Let
\[
\tau_D
=
\rho_D^{1/2-\mu}
s_1(\widehat\Theta_D)^{1/2+\mu}
\]
be the selection threshold. Because \(\mu<1/2\),
\[
\frac{\tau_D}{\sqrt{NT}}
=
\left(\frac{\rho_D}{\sqrt{NT}}\right)^{1/2-\mu}
\left(\frac{s_1(\widehat\Theta_D)}{\sqrt{NT}}\right)^{1/2+\mu}
=o_p(1).
\]
Thus every signal singular value exceeds \(\tau_D\) with probability
approaching one. For the noise singular values, \(\mu>-1/2\) and
\(s_1(\widehat\Theta_D)\asymp_p\sqrt{NT}\) imply
\[
\frac{\rho_D}{\tau_D}
=
\left\{
\frac{\rho_D}{s_1(\widehat\Theta_D)}
\right\}^{1/2+\mu}
=o_p(1).
\]
Therefore
\[
\max_{j>r_D}\frac{s_j(\widehat\Theta_D)}{\tau_D}=o_p(1).
\]
The strict threshold rule defining \(\widehat r_D\) in
Section~\ref{subsec:rank-selection} consequently
selects all and only the \(r_D\) signal singular values with probability
approaching one.
Hence \(P(\widehat r_D=r_D)\to1\).

The reduced-form proof is identical, replacing
\((\Theta_D,\widetilde V_D,\rho_D,r_D)\) by
\((\Theta_Y,\widetilde\Xi_Y,\rho_Y,r_Y)\).
\end{proof}
\noindent\textbf{Proof of Corollary \ref{cor:estimated-ranks}}
\begin{proof}
Construct the fixed-rank and estimated-rank procedures using the same folds,
initializations, auxiliary learner randomization, and tie-breaking conventions.
Let \(\hat J_l^{\,0}\) denote the splitwise denominator computed with the true
ranks. Define
\[
\mathcal F_{NT}
=
\bigcap_{l=1}^6\{\hat J_l^{\,0}\ne0\}.
\]
Let
\begin{align*}
\mathcal E_{NT}&=\{\hat r_D=r_D,\ \hat r_Y=r_Y\},\\
\mathcal E_{NT}^{\circ}&=\mathcal E_{NT}\cap\mathcal F_{NT}.
\end{align*}
The fixed positive true ranks are admissible for all training-fold PCAs for all
sufficiently large \(N,T\). Theorem \ref{thm:rank_consistency_hl}, Proposition
\ref{proposition:denominator}, and the fact that there are only six splits give
\[
P\!\left((\mathcal E_{NT}^{\circ})^c\right)
\le
P(\mathcal E_{NT}^c)
+
\sum_{l=1}^6P(\hat J_l^{\,0}=0)
\to0.
\]
On the event \(\mathcal E_{NT}^{\circ}\), the estimated-rank procedure coincides with the oracle-rank procedure. Hence
\[
\hat\theta_{\hat r}=\hat\theta,
\qquad
\widehat\sigma_{\theta,\widehat r}^2=\hat\sigma_\theta^2
\qquad\text{on }\mathcal E_{NT}^{\circ}.
\]

For asymptotic normality, let
\[
Z_{NT}^*=\sqrt{NT}\,(\hat\theta_{\hat r}-\theta_0),
\qquad
Z_{NT}=\sqrt{NT}\,(\hat\theta-\theta_0).
\]
Then for any \(x\in\mathbb R\),
\begin{align*}
P(Z_{NT}^*\le x)
&=
P(Z_{NT}^*\le x,\mathcal E_{NT}^{\circ})
+
P(Z_{NT}^*\le x,(\mathcal E_{NT}^{\circ})^c) \\
&=
P(Z_{NT}\le x,\mathcal E_{NT}^{\circ})
+
P(Z_{NT}^*\le x,(\mathcal E_{NT}^{\circ})^c).
\end{align*}
Therefore,
\[
\bigl|P(Z_{NT}^*\le x)-P(Z_{NT}\le x)\bigr|
\le
P((\mathcal E_{NT}^{\circ})^c)\to 0.
\]
By Theorem \ref{thm:CLT},
\[
Z_{NT}\xrightarrow{d} N(0,\sigma_\theta^2),
\]
and hence
\[
Z_{NT}^*\xrightarrow{d} N(0,\sigma_\theta^2).
\]

For variance consistency,
\[
\bigl|\widehat\sigma_{\theta,\widehat r}^2-\hat\sigma_\theta^2\bigr|
=
0
\qquad\text{on }\mathcal E_{NT}^{\circ},
\]
so for any \(\varepsilon>0\),
\[
P\!\left(
\bigl|\widehat\sigma_{\theta,\widehat r}^2-\hat\sigma_\theta^2\bigr|>\varepsilon
\right)
\le
P((\mathcal E_{NT}^{\circ})^c)\to 0.
\]
Thus
\[
\widehat\sigma_{\theta,\widehat r}^2-\hat\sigma_\theta^2=o_p(1).
\]
The desired result follows by Theorem~\ref{thm:var-consistency}.
\end{proof}
\section{Discussion of the First-Stage Rates}
\label{app:decomposition}

Assumption~\ref{ass:1st_stage} bounds the total conditional prediction errors
of the feasible first-stage procedures relative to \(g_0^*\) and \(m_0^*\),
including the effects of estimated factor projections.
Proposition~\ref{proposition:1st_stage} transfers these rates to the
evaluation fold and controls the nuisance-error remainder \(R_{1i}\).
Assumption~\ref{ass:rate-main} specifies loading-weighted conditions for
the training-fold factor expansions and the evaluation-fold
factor--nuisance interactions. Accordingly, \(\bar r_{g}\) and
\(\bar r_{m}\) describe ordinary prediction errors, while \(\eta_g\)
and \(\eta_m\) describe the corresponding loading-weighted quantities.

\paragraph{Lasso rate illustration.}
Consider a projected-Lasso implementation with a dictionary
\(Z_i=Z(X_i)\in\mathbb R^{T\times p}\), where \(p=p_{NT}\ge2\) is the
number of dictionary features. Suppose that \(g_0\) and \(m_0\) admit
approximately sparse representations
\[
g_0(X_i)=Z_i\beta_g^0+r_g(X_i),
\qquad
m_0(X_i)=Z_i\beta_m^0+r_m(X_i),
\]
where \(\beta_g^0,\beta_m^0\in\mathbb R^p\) have at most \(s_g\) and
\(s_m\) nonzero coordinates, respectively. The \(T\)-vector maps \(r_g\)
and \(r_m\) are the approximation errors. Let \(b_{g,NT}\) and
\(b_{m,NT}\) be nonnegative deterministic bounds on their normalized
conditional approximation norms \(\|r_g\|_{P,2}\) and \(\|r_m\|_{P,2}\).

Suppose the complete feasible Lasso procedure satisfies
Assumption~\ref{ass:1st_stage}(a)--(b) with vanishing deterministic rates
such that
\[
\bar r_{g}\lesssim
\sqrt{\frac{s_g\log p}{NT}}+b_{g,NT},
\qquad
\bar r_{m}\lesssim
\sqrt{\frac{s_m\log p}{NT}}+b_{m,NT}.
\]
Since \(n_3\asymp N\), substituting these bounds into
Assumption~\ref{ass:1st_stage}(c) gives the sufficient product-rate
restriction
\[
\begin{aligned}
&\sqrt{NT}
\left(\sqrt{\frac{s_g\log p}{NT}}+b_{g,NT}\right)\\
&\qquad\times
\left(\sqrt{\frac{s_m\log p}{NT}}+b_{m,NT}
      +\sqrt{\frac{s_g\log p}{NT}}+b_{g,NT}\right)=o(1).
\end{aligned}
\]
In the exactly sparse case, \(r_g=r_m=0\), so the approximation bounds
can be taken as \(b_{g,NT}=b_{m,NT}=0\). The sufficient restriction for
part~(c) then simplifies to
\[
\bigl(\sqrt{s_gs_m}+s_g\bigr)\log p=o(\sqrt{NT}).
\]

The prediction, loading-weighted, and residual conditions in
Assumptions~\ref{ass:1st_stage}, \ref{ass:rate-main}, and
\ref{ass:training_main}(d) jointly restrict the complete feasible fitting
procedure and the admissible growth of \(N\) and \(T\).

\paragraph{Connection to rank selection.}
For the terminal learners used by Algorithm~\ref{alg:rank_estimation},
the rank-selection residual conditions concern prediction errors relative
to the unprojected functions \(g_0\) and \(m_0\). Recall that
\[
\widetilde V_D
=
V-\{\hat g^{(nn)}(X)-g_0(X)\},
\qquad
\widetilde \Xi_Y
=
\Xi-\{\hat m^{(nn)}(X)-m_0(X)\}.
\]
By the triangle inequality and \(\|\cdot\|_{\op}\le \|\cdot\|_F\),
\[
\|\widetilde V_D\|_{\op}
\le
\|V\|_{\op}
+
\|\hat g^{(nn)}(X)-g_0(X)\|_F,
\]
and
\[
\|\widetilde \Xi_Y\|_{\op}
\le
\|\Xi\|_{\op}
+
\|\hat m^{(nn)}(X)-m_0(X)\|_F.
\]
The following conditions imply that both effective residuals are
\(o_p(\sqrt{NT})\):
\[
\|V\|_{\op}=o_p(\sqrt{NT}),
\qquad
\frac{1}{NT}\|\hat g^{(nn)}(X)-g_0(X)\|_F^2=o_p(1),
\]
and
\[
\|\Xi\|_{\op}=o_p(\sqrt{NT}),
\qquad
\frac{1}{NT}\|\hat m^{(nn)}(X)-m_0(X)\|_F^2=o_p(1).
\]
Thus
\[
\|\widetilde V_D\|_{\op}=o_p(\sqrt{NT}),
\qquad
\|\widetilde\Xi_Y\|_{\op}=o_p(\sqrt{NT}).
\]

These conditions separate the effective residuals from the factor-signal
scale. Assumption~\ref{assp:rank_consistency}(b) further requires
\[
\|\widetilde V_D\|_{\op}=O_p(\rho_D),
\qquad
\|\widetilde\Xi_Y\|_{\op}=O_p(\rho_Y).
\]
The terminal fits and tuning sequences used by
Algorithm~\ref{alg:rank_estimation} are required to satisfy these bounds
jointly. These requirements concern the unprojected targets \(g_0\) and
\(m_0\), whereas Assumption~\ref{ass:1st_stage} concerns the projected
first-stage targets used for inference.

\section{Additional Simulation Details and Evidence}
\label{app:additional-sim}

\paragraph{First-stage fitting.}
For both Lasso and Elastic Net, the \(T\times90\) series dictionary for unit
\(i\) contains the coordinatewise first, second, and third powers of the
original covariates, without cross-products. At each factor-augmented update,
the current time-series residual-maker is applied to both the \(T\)-vector
pseudo-outcome and every column of the dictionary. The projected arrays are
then stacked over the relevant training units before the learner is refit.
This implements the projected targets \(g_0^*\) and \(m_0^*\).

For both Lasso and Elastic Net, we select the penalty coefficient by three-fold cross-validation over a common grid of 31 candidate values ranging from \(10^{-3}/\sqrt{NT}\) to \(1\). The Elastic Net mixing parameter is fixed at \(\omega=1/2\).

Writing \(\widetilde y\) and \(\widetilde Z\) for the stacked training
response and dictionary, and \(n_{\mathrm{tr}}\) for their number of rows,
the regularized regression minimizes over \(b\in\R^{\widetilde p}\)
\[
\frac{1}{2n_{\mathrm{tr}}}
\|\widetilde y-\widetilde Zb\|_2^2
+\alpha\left\{
\omega\|b\|_1+\frac{1-\omega}{2}\|b\|_2^2
\right\},
\]
with \(\omega=1\) for Lasso and \(\omega=1/2\) for Elastic Net. The
regressions have no additional intercept, and dictionary columns are not
standardized. In factor-augmented regressions, the response and dictionary
are projected as described above, whereas standard DML uses unprojected data. The OLS first-stage benchmark uses the same dictionary and factor-projection
steps, replacing the regularized fit with unpenalized least squares.
Fold-specific point estimates and variance estimates are averaged over the
six role assignments for the three-fold specifications and over the two
assignments for the two-fold specifications.

\paragraph{Rank selection.}
In this simulation study, we select the treatment and reduced-form outcome ranks
separately on the full sample, alternating nuclear-norm soft thresholding
with penalized learner updates.
When the selected pair equals the true pair \((3,4)\), selected-rank FA-DML reduces to   known-rank FA-DML, and we use the known-rank estimates for that replication.

\begin{figure}[!htbp]
    \centering
    \includegraphics[width=\linewidth]{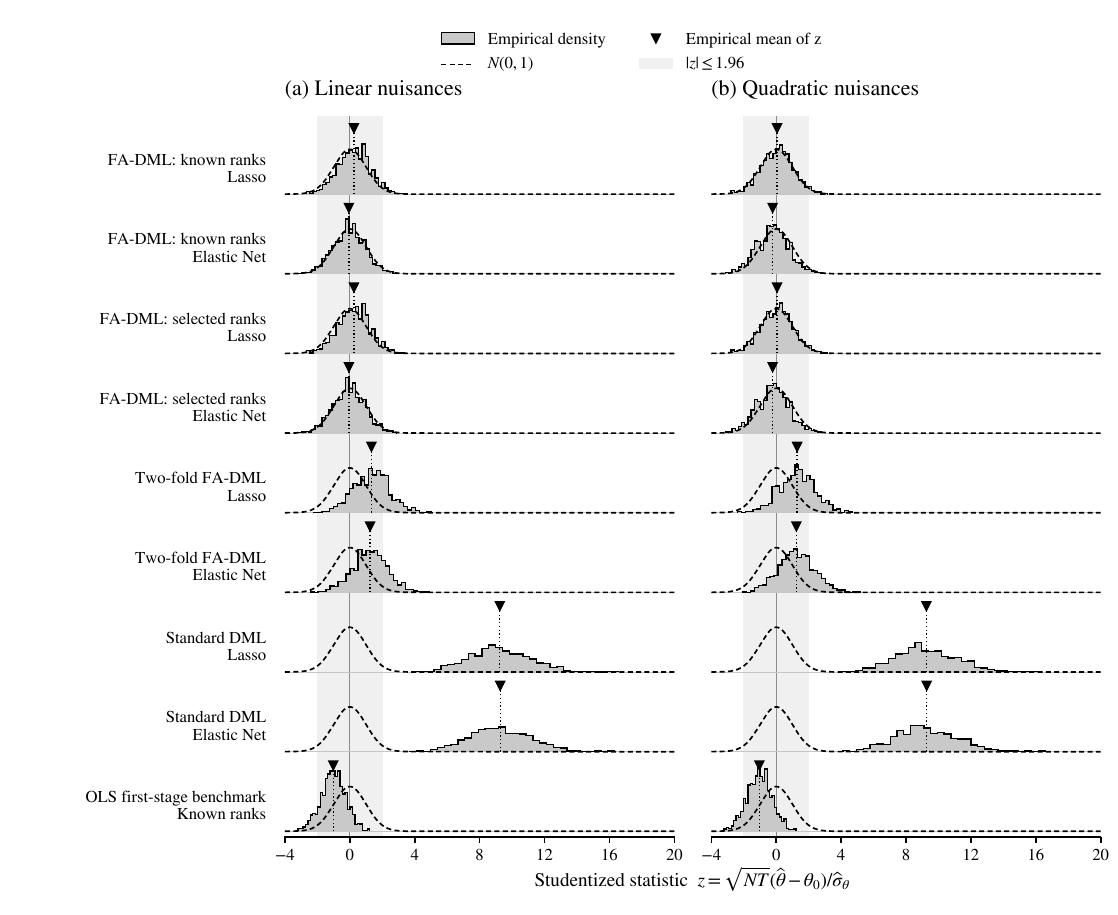}
    \begingroup
    \setstretch{1.0}
    \footnotesize
    \caption{Empirical distributions of
    \(z=\sqrt{NT}(\hat\theta-\theta_0)/\hat\sigma_\theta\) at
    \((N,T)=(100,100)\), with \(\theta_0=0.5\), based on
    \(1{,}000\) replications, where \(\hat\sigma_\theta/\sqrt{NT}\) is the estimated
    standard error. Panels (a) and (b) use linear and quadratic nuisance
    functions. Rows identify the estimator and learner. Factor-augmented
    estimators use \((r_D,r_Y)=(3,4)\), except selected-rank FA-DML, which
    selects both ranks once per replication on the full sample and holds them
    fixed across folds. Standard DML omits factor augmentation. The OLS
    first-stage benchmark retains three-fold cross-fitting and factor
    augmentation and is reported once per panel. Histograms have \(30\) bins.
    Dashed curves show the standard normal density, triangles and dotted lines mark the mean of \(z\), and  shading denotes \(|z|\le1.96\). Horizontal scales are identical, and heights above each row's baseline use a common density
    scale.}
    \label{fig:sim_clt_full}
    \endgroup
\end{figure}

\paragraph{Additional distribution comparisons.}
Figure~\ref{fig:sim_clt_full} supplements
Figure~\ref{fig:sim_studentized_main} by including the Elastic Net
specifications. For specifications appearing in both figures,
the histogram bins are identical. All distributions use common horizontal
and density scales without smoothing.

\clearpage
\section{Additional Evidence for the Empirical Applications}
\label{app:additional_empi}

This appendix reports factor-dimension sensitivity for the minimum-wage
application and presents the branching-deregulation application. The sensitivity grid contains the \(15\) pairs with \(2\le r_D\le r_Y\le6\),
for two learners and two control dictionaries. The inference results in Section~\ref{Section_thm} apply to the true or consistently selected ranks.

\subsection{Minimum-Wage Estimates across Factor Dimensions}
\label{subsection:robustness_rank_miniwage}

For the minimum-wage application, Algorithm~\ref{alg:rank_estimation} is
applied to the original dictionary with cross-validated Lasso,
\(\rho_D=\rho_Y=0.25\), and \(\mu=0.001\), and it selects \((2,3)\). The penalty level differs from the simulation design because the nuclear-norm penalty scales with the residual matrices.

Table~\ref{tab:ams_results_appendix} extends the diagonal comparison in
Table~\ref{tab:ams_results} to \(2\le r_D\le r_Y\le6\). The data,
control dictionaries, and first-stage settings are those of
Section~\ref{sec:ams}. All \(60\) estimates are negative, ranging from \(-0.1652\) to \(-0.1003\), and
every interval excludes zero. For a given pair and dictionary, the Lasso and random-forest estimates differ by at most \(0.0116\). Magnitudes depend more on \(r_D\) than on \(r_Y\).

\begin{table}[htbp!]
  \centering
  \captionsetup{font={footnotesize,stretch=1},justification=justified,
      singlelinecheck=false,width=\textwidth,skip=6pt}
  \fontsize{8}{12}\selectfont
    \begin{tabular}{lcccc}
    \toprule
          & \multicolumn{2}{c}{Original controls ($p=4$)} & \multicolumn{2}{c}{Expanded controls ($p=32$)} \\
    \cmidrule(lr){2-3}\cmidrule(lr){4-5}
          & Lasso & Random Forest & Lasso & Random Forest \\
    \midrule
    \multicolumn{5}{c}{$r_D = 2$} \\
    $r_Y=2$ & $-0.1155$ & $-0.1201$ & $-0.1152$ & $-0.1177$ \\
          & $(-0.199,\,-0.032)$ & $(-0.203,\,-0.037)$ & $(-0.195,\,-0.035)$ & $(-0.203,\,-0.033)$ \\
    $r_Y=3^{\dagger}$ & $-0.1103$ & $-0.1167$ & $-0.1088$ & $-0.1120$ \\
          & $(-0.187,\,-0.034)$ & $(-0.196,\,-0.037)$ & $(-0.185,\,-0.032)$ & $(-0.191,\,-0.033)$ \\
    $r_Y=4$ & $-0.1126$ & $-0.1141$ & $-0.1095$ & $-0.1153$ \\
          & $(-0.195,\,-0.030)$ & $(-0.197,\,-0.032)$ & $(-0.193,\,-0.026)$ & $(-0.199,\,-0.031)$ \\
    $r_Y=5$ & $-0.1060$ & $-0.1054$ & $-0.1105$ & $-0.1058$ \\
          & $(-0.188,\,-0.024)$ & $(-0.189,\,-0.022)$ & $(-0.194,\,-0.027)$ & $(-0.189,\,-0.023)$ \\
    $r_Y=6$ & $-0.1029$ & $-0.1028$ & $-0.1003$ & $-0.1031$ \\
          & $(-0.188,\,-0.018)$ & $(-0.195,\,-0.011)$ & $(-0.189,\,-0.012)$ & $(-0.192,\,-0.014)$ \\
    \midrule
    \multicolumn{5}{c}{$r_D = 3$} \\
    $r_Y=3$ & $-0.1398$ & $-0.1382$ & $-0.1296$ & $-0.1405$ \\
          & $(-0.234,\,-0.045)$ & $(-0.235,\,-0.041)$ & $(-0.222,\,-0.037)$ & $(-0.237,\,-0.044)$ \\
    $r_Y=4$ & $-0.1362$ & $-0.1356$ & $-0.1303$ & $-0.1368$ \\
          & $(-0.237,\,-0.036)$ & $(-0.237,\,-0.035)$ & $(-0.233,\,-0.027)$ & $(-0.238,\,-0.035)$ \\
    $r_Y=5$ & $-0.1379$ & $-0.1359$ & $-0.1285$ & $-0.1401$ \\
          & $(-0.238,\,-0.038)$ & $(-0.232,\,-0.040)$ & $(-0.223,\,-0.034)$ & $(-0.238,\,-0.042)$ \\
    $r_Y=6$ & $-0.1331$ & $-0.1303$ & $-0.1250$ & $-0.1299$ \\
          & $(-0.232,\,-0.034)$ & $(-0.227,\,-0.034)$ & $(-0.223,\,-0.027)$ & $(-0.227,\,-0.033)$ \\
    \midrule
    \multicolumn{5}{c}{$r_D = 4$} \\
    $r_Y=4$ & $-0.1612$ & $-0.1608$ & $-0.1520$ & $-0.1610$ \\
          & $(-0.280,\,-0.043)$ & $(-0.279,\,-0.043)$ & $(-0.264,\,-0.040)$ & $(-0.279,\,-0.043)$ \\
    $r_Y=5$ & $-0.1504$ & $-0.1510$ & $-0.1477$ & $-0.1510$ \\
          & $(-0.262,\,-0.039)$ & $(-0.263,\,-0.039)$ & $(-0.259,\,-0.036)$ & $(-0.265,\,-0.037)$ \\
    $r_Y=6$ & $-0.1500$ & $-0.1498$ & $-0.1417$ & $-0.1494$ \\
          & $(-0.270,\,-0.030)$ & $(-0.270,\,-0.030)$ & $(-0.255,\,-0.028)$ & $(-0.268,\,-0.030)$ \\
    \midrule
    \multicolumn{5}{c}{$r_D = 5$} \\
    $r_Y=5$ & $-0.1484$ & $-0.1496$ & $-0.1487$ & $-0.1486$ \\
          & $(-0.271,\,-0.026)$ & $(-0.276,\,-0.023)$ & $(-0.267,\,-0.030)$ & $(-0.274,\,-0.023)$ \\
    $r_Y=6$ & $-0.1489$ & $-0.1517$ & $-0.1526$ & $-0.1518$ \\
          & $(-0.266,\,-0.032)$ & $(-0.268,\,-0.036)$ & $(-0.278,\,-0.027)$ & $(-0.271,\,-0.032)$ \\
    \midrule
    \multicolumn{5}{c}{$r_D = 6$} \\
    $r_Y=6$ & $-0.1590$ & $-0.1606$ & $-0.1652$ & $-0.1600$ \\
          & $(-0.284,\,-0.034)$ & $(-0.287,\,-0.034)$ & $(-0.305,\,-0.025)$ & $(-0.286,\,-0.034)$ \\
    \bottomrule
    \end{tabular}
    \caption[Minimum-wage coefficients across factor dimensions.]{Minimum-wage coefficients across factor dimensions. Each cell reports the median FA-DML estimate over \(15\) random sample splits and, below it, the \(95\%\) Wald interval based on the median adjustment in Section~\ref{sec:ams}. Columns identify the learner and control dictionary. The dagger marks the selected pair \((2,3)\). Data and settings are as in Table~\ref{tab:ams_results}.
}
\label{tab:ams_results_appendix}
\end{table}

\subsection{Branching Deregulation and Income Inequality}
\label{app:banking}\label{sec:bll}
\citet{beck2010big} find that intrastate branching deregulation reduced
income inequality in a two-way fixed-effects specification. We ask whether this
conclusion survives flexible adjustment for observed covariates and interactive
state-time heterogeneity. As in the original study, the panel covers \(48\)
states and the District of Columbia from \(1976\) to \(2006\), excluding
Delaware and South Dakota, for \(N\times T=49\times31=1{,}519\)
observations.

The outcome is the logit transformation of the Gini coefficient,
\[
Y_{it}=\log\{g_{it}/(1-g_{it})\},
\qquad
D_{it}=\mathbf{1}\{t>t_i^*\},
\]
where \(g_{it}\) is the Gini coefficient and \(t_i^*\) is the year of
intrastate branching deregulation, so the treatment equals one from the year
after deregulation, as in \citet{beck2010big}. The parameter \(\theta_0\) is a
common effect on the log-odds of the Gini coefficient, and the specification
does not allow effects to vary by adoption cohort or event time. For the
binary treatment, the treatment equation is the factor-augmented linear
probability model of Section~\ref{sec_model}.

The original dictionary contains eight controls: growth in gross state
product (GSP) per capita, the shares of Black residents, high-school dropouts,
and female-headed households, the unemployment rate, log GSP per capita, log
population, and population dispersion. A degree-four polynomial expansion,
with nearly collinear terms removed, gives \(302\) controls. Interactive
effects allow common economic shocks to affect states differently.
Identification rests on Assumption~\ref{ass:identification}.

The benchmarks and the random-forest settings follow
Section~\ref{sec:ams}, and Lasso uses the self-normalized penalty of
\citet{belloni2012}. FA-DML follows Algorithm~\ref{algorithm_three_fold} with
the same \(15\)-split median adjustment. Algorithm~\ref{alg:rank_estimation}
selects \((r_D,r_Y)=(2,3)\), which we use as the primary specification for both
dictionaries and learners.
Table~\ref{tab:bll_results} also reports FA-DML with
\(r_D=r_Y=r\in\{2,\ldots,6\}\).

\begin{table}[htbp!]
  \centering
  \captionsetup{font={footnotesize,stretch=1},justification=justified,
      singlelinecheck=false,width=\textwidth,skip=6pt}
  \fontsize{8}{12}\selectfont
    \begin{tabular}{lccccccc}
    \toprule
          & \multicolumn{3}{c}{Original controls} & & \multicolumn{3}{c}{Expanded controls} \\
    \cmidrule(lr){2-4}\cmidrule(lr){6-8}
          & Estimate & s.e.  & 95\% CI                 & & Estimate & s.e.  & 95\% CI \\
    \midrule
        \multicolumn{8}{c}{Panel A. Benchmark estimators} \\
    OLS                & $-0.0277$ & $0.0127$ & $(-0.053,\,-0.003)$ & & $-0.0012$ & $0.0070$ & $(-0.015,\,+0.012)$ \\
    OLS (TWFE)         & $-0.0289$ & $0.0101$ & $(-0.049,\,-0.009)$ & & $-0.0208$ & $0.0075$ & $(-0.035,\,-0.006)$ \\
    DML                & $-0.0367$ & $0.0138$ & $(-0.064,\,-0.010)$ & & $-0.1184$ & $0.0708$ & $(-0.257,\,+0.020)$ \\
    \midrule
    \multicolumn{8}{c}{Panel B. OLS-IFE} \\
    $r=2$ & $+0.0024$ & $0.0057$ & $(-0.009,\,+0.014)$ & & $-0.0072$ & $0.0051$ & $(-0.017,\,+0.003)$ \\
    $r=3$ & $-0.0049$ & $0.0060$ & $(-0.017,\,+0.007)$ & & $-0.0097$ & $0.0055$ & $(-0.020,\,+0.001)$ \\
    $r=4$ & $-0.0028$ & $0.0062$ & $(-0.015,\,+0.009)$ & & $-0.0149$ & $0.0052$ & $(-0.025,\,-0.005)$ \\
    $r=5$ & $-0.0032$ & $0.0061$ & $(-0.015,\,+0.009)$ & & $-0.0088$ & $0.0052$ & $(-0.019,\,+0.001)$ \\
    $r=6$ & $-0.0059$ & $0.0058$ & $(-0.017,\,+0.005)$ & & $-0.0094$ & $0.0049$ & $(-0.019,\,+0.000)$ \\
    \midrule
    \multicolumn{8}{c}{Panel C. FA-DML (Lasso)} \\
    Selected $(2,3)^{\dagger}$ & $-0.0019$ & $0.0102$ & $(-0.022,\,+0.018)$ &  & $-0.0023$ & $0.0102$ & $(-0.022,\,+0.018)$ \\
    \addlinespace[2pt]
    $r=2$ & $-0.0001$ & $0.0116$ & $(-0.023,\,+0.023)$ &  & $-0.0001$ & $0.0116$ & $(-0.023,\,+0.023)$ \\
    $r=3$ & $+0.0058$ & $0.0120$ & $(-0.018,\,+0.029)$ &  & $+0.0058$ & $0.0116$ & $(-0.017,\,+0.029)$ \\
    $r=4$ & $-0.0021$ & $0.0118$ & $(-0.025,\,+0.021)$ &  & $-0.0021$ & $0.0118$ & $(-0.025,\,+0.021)$ \\
    $r=5$ & $-0.0059$ & $0.0119$ & $(-0.029,\,+0.017)$ &  & $-0.0059$ & $0.0119$ & $(-0.029,\,+0.017)$ \\
    $r=6$ & $-0.0063$ & $0.0126$ & $(-0.031,\,+0.018)$ &  & $-0.0063$ & $0.0126$ & $(-0.031,\,+0.018)$ \\
    \midrule
    \multicolumn{8}{c}{Panel D. FA-DML (Random Forest)} \\
    Selected $(2,3)^{\dagger}$ & $-0.0017$ & $0.0102$ & $(-0.022,\,+0.018)$ &  & $-0.0015$ & $0.0101$ & $(-0.021,\,+0.018)$ \\
    \addlinespace[2pt]
    $r=2$ & $+0.0007$ & $0.0112$ & $(-0.021,\,+0.023)$ &  & $+0.0002$ & $0.0107$ & $(-0.021,\,+0.021)$ \\
    $r=3$ & $+0.0061$ & $0.0117$ & $(-0.017,\,+0.029)$ &  & $+0.0060$ & $0.0120$ & $(-0.017,\,+0.030)$ \\
    $r=4$ & $-0.0024$ & $0.0117$ & $(-0.025,\,+0.021)$ &  & $-0.0020$ & $0.0117$ & $(-0.025,\,+0.021)$ \\
    $r=5$ & $-0.0062$ & $0.0120$ & $(-0.030,\,+0.017)$ &  & $-0.0069$ & $0.0118$ & $(-0.030,\,+0.016)$ \\
    $r=6$ & $-0.0062$ & $0.0126$ & $(-0.031,\,+0.019)$ &  & $-0.0055$ & $0.0124$ & $(-0.030,\,+0.019)$ \\
    \bottomrule
    \end{tabular}
    \caption[Branching deregulation and income inequality.]{Branching deregulation and income inequality.
    The outcome is the logit-transformed Gini coefficient and the treatment is an indicator
    for intrastate branching deregulation (\(N=49\), \(T=31\)). Entries are
    estimates, standard errors, and \(95\%\) Wald intervals with the original
    (\(p=8\)) or expanded (\(p=302\)) dictionary. OLS-IFE uses \(r\) outcome
    factors. The dagger marks the selected FA-DML pair \((r_D,r_Y)=(2,3)\), and
    the other FA-DML rows set \(r_D=r_Y=r\). FA-DML estimates use the
    \(15\)-split median adjustment. Standard errors are clustered by state, except for OLS-IFE.}
\label{tab:bll_results}
\end{table}

With the original controls, pooled OLS, TWFE, and standard DML all give
negative estimates with intervals excluding zero, and the TWFE estimate is
\(-0.0289\). With the expanded dictionary, TWFE remains negative at
\(-0.0208\), pooled OLS is close to zero, and standard DML is much less
precise. The benchmark results therefore depend on both the estimator and the
control set. The OLS-IFE intervals all include zero with the original
dictionary, and only \(r=4\) yields an interval excluding zero with the
expanded dictionary.

At the selected pair \((2,3)\), the Lasso and random-forest estimates
are \(-0.0019\) and \(-0.0017\) with the original controls and \(-0.0023\) and
\(-0.0015\) with the expanded controls, and all four intervals include zero.
Over all \(15\) pairs with \(2\le r_D\le r_Y\le6\), the estimates range from
\(-0.0069\) to \(0.0066\), and every interval includes zero. With the
self-normalized penalty, Lasso retains few controls, so its estimates are
nearly identical across the two dictionaries. Once interactive heterogeneity
is allowed, the estimated effect of deregulation is close to zero, so the TWFE
estimate is sensitive to how latent confounding is modeled. The
common-coefficient specification does not rule out cohort-specific or dynamic
effects.

\newpage

\end{document}